\RequirePackage{graphicx}
\documentclass[prd, superscriptaddress, nofootinbib, floatfix,preprintnumbers]{revtex4-2}   

\usepackage{xcolor}
\usepackage[subfigure]{graphfig}
\usepackage[normalem]{ulem} 
\usepackage{wrapfig}
\usepackage{amsmath}
\usepackage{amsfonts}
\usepackage{amssymb}
\usepackage{booktabs}
\usepackage{multirow}
\usepackage{slashed}
\usepackage{physics}
\usepackage{tabularx}
\usepackage{soul}
\usepackage{ulem}
\usepackage{hhline}
\usepackage{float}
\usepackage[colorlinks=true, linkcolor=blue, citecolor=blue, urlcolor=blue]{hyperref}
\usepackage{longtable}

\usepackage{nccmath}
\usepackage{esvect}

\newcommand\befs{\begin{figure*}}
\newcommand\eefs[1]{\label{fig:#1}\end{figure*}}
\newcommand\bef{\begin{figure}}
\newcommand\eef[1]{\label{fig:#1}\end{figure}}
\newcommand\beq{\begin{equation}}
\newcommand\eeq[1]{\label{#1}\end{equation}}
\newcommand\beqa{\begin{eqnarray}}
\newcommand\eeqa[1]{\label{#1}\end{eqnarray}}
\newcommand\bet{\begin{table}}
\newcommand\eet[1]{\label{tb:#1}\end{table}}
\newcommand\bets{\begin{table*}}
\newcommand\eets[1]{\label{tb:#1}\end{table*}}

\newcommand\eqn[1]{Eq.\ (\ref{eq:#1})}

\def\be{\begin{equation}}
\def\ee{\end{equation}}
\newcommand{\bea}{\begin{eqnarray}}
\newcommand{\eea}{\end{eqnarray}}
\newcommand{\ba}{\begin{align}}
\newcommand{\ea}{\end{align}}

\newcommand{\wt}{\widetilde}
\newcommand{\nn}{\nonumber}

\newcommand{\ms}{\overline{\rm{MS}}}

\begin{document}

\preprint{JLAB-THY-22-3663}

\widetext

\title{\Large  {Towards the determination of the gluon helicity distribution in the nucleon from lattice quantum chromodynamics}} 
\newcommand*{\Jlab}{Thomas Jefferson National Accelerator Facility, Newport News, Virginia, USA.}\affiliation{\Jlab}  
\newcommand*{\ORNL}{Oak Ridge National Laboratory, Oak Ridge, Tennessee, USA.}\affiliation{\ORNL} 
\newcommand*{\CU}{Department of Physics, Columbia University, New York City, New York, USA.}\affiliation{\CU}  
\newcommand*{\WM}{Department of Physics, William and Mary, Williamsburg, Virginia, USA.}\affiliation{\WM}       
\newcommand*{\BRAC}{BRAC University, 66 Mohakhali, Dhaka 1212, Bangladesh.}\affiliation{\BRAC}
\newcommand*{\ODU}{Department of Physics, Old Dominion University, Norfolk, Virginia, USA.}\affiliation{\ODU}
\newcommand*{\CNRS}{Aix Marseille Univ, Universit\'e de Toulon, CNRS, CPT, Marseille, France.}\affiliation{\CNRS}

\author{Colin Egerer}\affiliation{\Jlab}
\author{B\'alint Jo\'o}\affiliation{\ORNL}
\author{Joseph Karpie}\affiliation{\CU}
\author{Nikhil Karthik}\affiliation{\Jlab}\affiliation{\WM}
\author{Tanjib Khan}\affiliation{\BRAC}
\author{Christopher J. Monahan}\affiliation{\Jlab}\affiliation{\WM}
\author{Wayne Morris}\affiliation{\Jlab}\affiliation{\ODU}
\author{Kostas Orginos}\affiliation{\Jlab}\affiliation{\WM}
\author{Anatoly  Radyushkin}\affiliation{\Jlab}\affiliation{\ODU}
\author{David G. Richards}\affiliation{\Jlab} 
\author{Eloy Romero}\affiliation{\Jlab}
\author{Raza Sabbir Sufian}\affiliation{\Jlab}\affiliation{\WM}
\author{Savvas Zafeiropoulos}\affiliation{\CNRS}
\collaboration{On behalf of the \textit{HadStruc Collaboration}}

\begin{abstract}
We present the first exploratory lattice quantum chromodynamics (QCD) calculation of the polarized gluon Ioffe-time pseudo-distribution in the nucleon. The Ioffe-time pseudo-distribution provides a frame-independent and gauge-invariant framework to determine the gluon helicity in the nucleon from first principles. We employ a high-statistics computation using a $32^3\times 64$ lattice ensemble characterized by a  $358$~MeV pion mass and a $0.094$~fm lattice spacing. We establish the pseudo-distribution approach as a feasible method to address the proton spin puzzle with successive improvements in statistical and systematic uncertainties anticipated in the future. Within the statistical precision of our data, we find a good comparison between the lattice determined polarized gluon Ioffe-time distribution and the corresponding expectations from the state-of-the-art global analyses. We find a hint for a nonzero gluon spin contribution to the proton spin from the model-independent extraction of the gluon helicity pseudo-distribution over a range of Ioffe-time, $\nu\lesssim 9$. 
\end{abstract}

\maketitle

\section{Introduction}\label{sec:intro}
An outstanding question in particle and nuclear physics  is how the spin of the proton arises from its constituents, quarks and gluons, and their interactions that are governed by quantum chromodynamics (QCD), the fundamental theory of strong interactions. The spin sum rules are central to addressing this question by breaking down the proton spin into the quark and gluon spin and angular momentum components. The Jaffe-Manohar decomposition~\cite{Jaffe:1989jz} provides one such spin sum rule,
\bea \label{eq:JMdecomp}
J = \frac{1}{2}\Delta \Sigma + L_q + L_G + \Delta G ,
\eea
where $\frac{1}{2}\Delta \Sigma$ is the quark spin contribution, $L_q$ and $L_G$ are quark and gluon orbital angular momenta, and $\Delta G$ is the gluon spin contribution. Such a decomposition is not unique, and Ji's spin decomposition~\cite{Ji:1996ek} offers a frame-independent and gauge invariant way to decompose the proton's spin into quark spin, quark orbital angular momentum and gluon angular momentum contributions. A naive expectation, based on intuition from the quark model, would be that the quark spin term provides the dominant contribution to the spin sum rules,  but a deep inelastic scattering (DIS) experiment, conducted by the European Muon Collaboration, featuring polarized muons scattering from polarized protons  found that the quark spin contribution to the proton spin is very small ($\Delta \Sigma (Q^2= 10\, {\rm GeV}^2) =0.060(47)(69)$~\cite{EuropeanMuon:1987isl, EuropeanMuon:1989yki}. This surprising result 
is the so-called ``proton spin crisis," later confirmed by modern global analyses of DIS experimental data that show that quarks contribute roughly $30\%$~\cite{deFlorian:2009vb, Nocera:2014gqa, Ethier:2017zbq} to the proton spin. These results lead to the natural question:  how much do gluons contribute to the proton spin budget or, equivalently, what are the sign and magnitude of the gluon helicity $\Delta G$ in the nucleon? Experimentally, one can access $\Delta G(Q^2)$ via the integrated Bjorken-$x$ dependent polarized gluon distribution, $\Delta g(x,Q^2)$, as
\bea
\Delta G(Q^2)=\int_0^1 \dd x\ \Delta g(x,Q^2) .
\eea
An analysis~\cite{deFlorian:2014yva} of high-statistics 2009 STAR~\cite{Djawotho:2013pga} and PHENIX~\cite{PHENIX:2014gbf} data showed evidence of nonzero gluon helicity in the proton.  At the scale $Q^2=10\, {\rm GeV}^2$, it was found in~\cite{deFlorian:2014yva} that the gluon helicity distribution $\Delta g(x,Q^2)$ is positive and nonzero in the momentum fraction region $0.05 \leq x \leq 0.2$. However, the distribution has a large uncertainty in the small $x$-region. In contrast, in a recent global analysis~\cite{Zhou:2022wzm} that relaxed the positivity constraints  on  the  quark  and  gluon  helicity  parton distribution functions (PDFs), it was found that the existing experimental data does not  imply that the gluon polarization in the nucleon must  be  positive. In order to provide high-precision measurements of the gluon helicity $\Delta G$,  several experiments are being carried out at the Relativistic Heavy Ion Collider (RHIC)~\cite{Bunce:2000uv,PHENIX:2014gbf}, HERMES~\cite{HERMES:2008abz}, JLab~\cite{Dudek:2012vr}, COMPASS~\cite{COMPASS:2018pup} and are planned for the future Electron-Ion Collider (EIC)~\cite{Accardi:2012qut} to better understand the origin of proton spin.

Lattice QCD offers a nonperturbative approach to compute each of the quark and gluon helicity and angular momentum contributions 
to the net proton spin\footnote{We refer the reader to Refs.~\cite{Alexandrou:2020sml,Wang:2021vqy} for the most recent nucleon spin decomposition using the Ji decomposition and quark orbital angular momentum calculations using both the Jaffe-Manohar and Ji decompositions in~\cite{Engelhardt:2020qtg}.}.
 However, direct lattice QCD calculations of $\Delta G$ are not straightforward, because the decomposition is derived in the light-cone frame with the choice of light-cone gauge, which cannot be obtained from a local matrix element in a lattice QCD calculation. To circumvent this problem, it was proposed in~\cite{Ji:2013fga} that the matrix element of the equal-time local operator $\vec{E}\times \vec{A}_{\rm phys}$, where  $\vec{A}_{\rm phys}$ is the gauge invariant part of the gauge potential $A_\mu$, coincides with the gluon helicity obtained from the nonlocal operator of the light-cone gluon helicity distribution in~\cite{Manohar:1990jx}. A lattice QCD calculation~\cite{Yang:2016plb} was performed using this formalism and obtained $\Delta G = 0.251(47)(16)$. However, the matching coefficient in the finite piece in the one-loop large momentum effective theory~\cite{Ji:2014gla}  was found to be quite large, indicating a  possible convergence problem for the perturbative series. Thus the gluon helicity $\Delta G$ obtained in~\cite{Yang:2016plb} is not free of a large matching systematic error. For the status of lattice QCD results related to the nucleon spin decomposition, we refer readers to a recent review~\cite{Liu:2021lke} and the references therein.

The recent proposal to use the pseudo-PDF approach~\cite{Radyushkin:2017cyf} to access $\Delta g(x,Q^2)$ (or equivalently, the 
corresponding Ioffe-time distribution~\cite{Braun:1994jq}) offers a gauge-invariant and frame-independent alternative way to study $\Delta G$.  The pseudo-PDF approach~\cite{Radyushkin:2017cyf} and the associated proper combination of matrix elements derived in~\cite{Balitsky:2021cwr} give access to the polarized gluon Ioffe-time distribution (ITD) and corresponding gluon helicity parton distribution function (PDF) in the nucleon. Following the convention for symbols of matrix elements from Ref.~\cite{Balitsky:2021cwr}, we calculate the gluon helicity Ioffe-time  reduced pseudo-distribution function (reduced pseudo-ITD), $\wt{\mathfrak{M}} (\nu,z^2)$~\cite{Radyushkin:2017cyf, Radyushkin:2017lvu, Radyushkin:2018cvn}, where we refer to $\nu\equiv-\left(p\cdot z\right)$ as the Ioffe-time~\cite{Ioffe:1969kf}. The related pseudo-PDF, $\wt{\mathcal{P}} (x, z^2)$ can be determined from the Fourier transform of the pseudo-ITD. The pseudo-PDF and the pseudo-ITD can be factorized into the PDF and perturbatively calculable kernels, similar to the factorization framework for experimental cross sections. Our calculation applies the reduced pseudo-ITD approach~\cite{Orginos:2017kos}, in particular the ratio proposed in~\cite{Balitsky:2021cwr}, for which the multiplicative  renormalization factors cancel.

We apply this recently established theoretical framework for the polarized gluon pseudo-PDF~\cite{Balitsky:2021cwr} in a numerical lattice QCD computation.
In comparison to the significant progress in lattice QCD calculations of quark structures of the nucleon and mesons in recent years~\cite{Lin:2018pvv,Alexandrou:2018pbm,Liang:2019frk,Karpie:2021pap,Sufian:2019bol,Sufian:2020vzb,Zhang:2020rsx, Izubuchi:2019lyk, Gao:2020ito,Alexandrou:2020qtt, Alexandrou:2020uyt, Fan:2020nzz,Lin:2020fsj,Zhang:2020dkn,Gao:2021dbh,Hua:2020gnw,Joo:2019jct,Joo:2019bzr,HadStruc:2021qdf,Gao:2022vyh,Bhattacharya:2021moj,Alexandrou:2021oih,Detmold:2021qln,Bali:2018spj,Bhat:2022zrw} using different approaches~\cite{Liu:1993cv,Detmold:2005gg,Braun:2007wv,Ji:2013dva, Ji:2014gla,Radyushkin:2017cyf,Ma:2014jla,Ma:2017pxb}, there have been only a few  attempts to calculate the $x$-dependent unpolarized gluon PDFs in the nucleon~\cite{Fan:2018dxu,Fan:2020cpa,HadStruc:2021wmh},  pion~\cite{Fan:2021bcr}, and kaon~\cite{Salas-Chavira:2021wui}. There have been no lattice QCD calculations of the $x$-dependent gluon helicity distribution in the nucleon. The current work closes this gap in the 
literature and provides the first look into the feasibility and the associated challenges in addressing $\Delta g(x,Q^2)$ and $\Delta G(Q^2)$ from
the pseudo-PDF approach.  For more details of the different methods for calculating $x$-dependent hadron structure, and related lattice QCD calculations, see recent reviews~\cite{Cichy:2018mum,Constantinou:2020hdm, Ji:2020ect,Constantinou:2022yye} and the references therein.

The rest of this paper is organized as follows. In Sec.~\ref{sec:theory}, we first discuss the theoretical framework for  the construction of  matrix elements and the reduced pseudo-ITD associated with the polarized gluon parton distribution in the nucleon. In Sec.~\ref{comp_frame}, we briefly describe the lattice QCD methodologies for the construction of the gluonic currents needed for the gluon helicity distribution, nucleon two-point correlators and our lattice setup for this calculation of gluonic matrix elements. Sec.~\ref{sec:rITDextra} describes the methodology we implement to calculate the reduced pseudo-ITD from the nucleon three-point correlators. In Sec.~\ref{sec:pdf_calc}, we extract the polarized gluon pseudo-ITD and discuss the potential of extracting the gluon helicity PDF from the reduced pseudo-ITD and compare our results with  phenomenological distributions. Sec.~\ref{conclusion} contains our concluding remarks and outlook.


\section{Theoretical Background of Polarized Gluon pseudo-Distribution }\label{sec:theory}

To access the polarized gluon PDF, one needs the matrix elements of two gluon field strength tensors, $G_{\mu\nu}$, connected by a Wilson line inserted between nucleon states of definite helicity.  Throughout this paper, we follow the notation of Ref.~\cite{Balitsky:2021cwr}, in which the authors used the {\it tilde}-symbol to denote the polarized gluonic matrix elements and associated distributions. 

We start with the matrix elements of two spatially separated gluon fields 
\bea \label{eq1:ME}
\wt{m}_{\mu\alpha;\lambda\beta}(z,p,s) = \bra{p,s}G_{\mu \alpha} (z) \, W[z, 0] \, \wt{G}_{\lambda \beta} (0) \ket{p,s}\, ,
\eea
where  $z_\mu$ is the separation between the gluon fields, $p_\mu$ is the four-momentum of the nucleon, and $s_\mu\equiv\overline{u}\left(p,s\right)\gamma_\mu\gamma_5u\left(p,s\right)$, a pseudovector normalized by $s^2=-m_p^2$, is the nucleon polarization vector and $m_p$ being the nucleon mass. Finally, the dual field to $G_{\lambda\beta}$ is defined as $\wt{G}_{\lambda\beta}=\frac{1}{2}\epsilon_{\lambda\beta\rho\gamma}G^{\rho\gamma}$, and $W[z, 0]$ is a straight-line Wilson line in the adjoint representation,
\bea
    W[x, y] = {\cal P}\mathrm{exp} \Big\{ ig_s \int_0^1 d\eta \, (x-y)^\mu \Tilde{A}_\mu \big(\eta x + (1-\eta)y \big) \Big\} \, ,
\eea
for the gauge field $A_\mu$, where ${\cal P}$ indicates that the integral is path-ordered.
The spin-dependent part of the matrix element is determined by the linear combination in Eq.~\eqref{eq2:ME} that is odd in the separation $z_\mu$. The matrix elements associated with the polarized gluon distribution are then written as
\bea\label{eq2:ME}
\wt{M}_{\mu\alpha;\lambda\beta}(z,p,s) = \wt{m}_{\mu\alpha;\lambda\beta}(z,p,s) - \wt{m}_{\mu\alpha;\lambda\beta}(-z,p,s)\, .
\eea
As shown in~\cite{Balitsky:2021cwr}, using Lorentz invariance and  taking into account the antisymmetric properties  of the gluon field strength tensor with respect to its indices, one can write  these matrix elements as  an expansion involving 
 two types of invariant amplitudes, differing in the contribution of spin vector $s^\mu$ to the Lorentz tensor structures. Of these,
$\wt{\mathcal{M}}_{sp}$, $\wt{\mathcal{M}}_{ps}$, $\wt{\mathcal{M}}_{sz}$, $\wt{\mathcal{M}}_{zs}$, $\wt{\mathcal{M}}_{pzps}$, $\wt{\mathcal{M}}_{pzsz}$ 
are accompanied by  tensor structures in which each one of the 
$\mu, \alpha; \lambda,  \beta$ indices is carried by the nucleon polarization vector $s_\mu$, the nucleon momentum $p_\mu$, the separation 
$z_\mu$, and  the metric tensor $g_{\mu \nu}$.  The invariant amplitudes of the second group, $\wt{\mathcal{M}}_{pp}$, $\wt{\mathcal{M}}_{zz}$, $\wt{\mathcal{M}}_{zp}$, $\wt{\mathcal{M}}_{pz}$, $\wt{\mathcal{M}}_{ppzz}$,$\wt{\mathcal{M}}_{gg}$ 
are accompanied by tensor structures in which the spin vector enters through the product  $s\cdot z$~\cite{Balitsky:2021cwr}. 
All invariant amplitudes are functions of the invariant interval $z^2$ and the Lorentz invariant $(p \cdot z) \equiv - \nu$~\cite{Braun:1994jq} called the Ioffe time due to its relation, up to a sign and normalization of the target's mass, with the original variable from DIS cross section analyses~\cite{Ioffe:1969kf}.  Ultimately, the invariant amplitudes of interest will be those that contribute in the contraction $g^{\alpha \lambda}\wt{M}_{\mu\alpha;\lambda\beta}$ with specific kinematics that define the polarized PDF. All others that cannot be removed will contribute to systematic errors that must be modeled or corrected for.

The light-cone polarized gluon distribution $\Delta g(x)$ is obtained from
\bea \label{eq:lcdelg}
    g^{\alpha \beta} \, \wt{M}_{+\alpha; \beta+} (z_-, p) = - 2 p_+ s_+ \, [\wt{\mathcal{M}}_{ps}^{(+)} (\nu, 0) + p_+ z_- \wt{\mathcal{M}}_{pp}(\nu,0)]\, ,
\eea
where $z$ is taken in the light-cone ``minus'' direction, $z = z_-$, $p_+$ is the momentum in the light-cone ``plus'' direction, and $\wt{\mathcal{M}}_{ps}^{(+)}=[\wt{\mathcal{M}}_{ps}+\wt{\mathcal{M}}_{sp}]$. The polarized gluon PDF can be determined by the Ioffe-time distribution
\bea \label{eq:lcITD}
-i \wt{\mathcal{I}}_p(\nu) \equiv \wt{\mathcal{M}}_{ps}^{(+)}(\nu) - \nu \wt{\mathcal{M}}_{pp}(\nu)\, ,
\eea
where
\bea
 \wt{\mathcal{I}}_p(\nu) =  \frac{i}{2} \int_{-1}^1 \dd  x \, e^{-ix \nu} \, x\,\Delta g(x) \, .
\eea
As noted in~\cite{Braun:1994jq}, with knowledge of the polarized gluon ITD $\wt{\mathcal{I}}_p\left(\nu\right)$, one can immediately obtain the gluon helicity contribution to the nucleon spin 
\bea \label{eq:deltaG}
\Delta G = \int_0^\infty \dd \nu~\wt{\mathcal{I}}_p(\nu)=\int_0^1 \dd x~\Delta g(x)\, .
\eea

The field-strength tensor $G_{\mu \alpha}$ is antisymmetric with respect to its indices and $g_{--} = 0$, so the left-hand side of Eq.~\eqref{eq:lcdelg} reduces to a summation over the transverse indices  $i,j =x,y$, perpendicular to the direction of separation between the two gluon fields. The combination of the matrix elements $\wt{M}_{0i;0i}$
and $\wt{M}_{ij;ij}$ can be written in terms of invariant amplitudes as
\bea \label{eq:pseudo_Ip}
    \wt{M}_{0i;0i}(z,p) + \wt{M}_{ij;ij}(z,p) = - 2 p_z 
    p_0 \wt{\mathcal{M}}_{sp}^{(+)} (\nu,z^2)+ 2
    p_0^3 z \wt{\mathcal{M}}_{pp} (\nu,z^2)\, , 
\eea
where  the nucleon boost is along the $3$rd  ($z$) direction, $p= \{p_0, 0_\perp, p_z \}$~\cite{Balitsky:2021cwr}. The polarization vector that we use is $s= \{p_z, 0_\perp, p_0 \}$, such that the requirement $s\cdot p=0$ is satisfied.

The particular combination in Eq.~\eqref{eq:pseudo_Ip} cancels the contamination terms coming from  invariant amplitudes other than $\wt{\mathcal{M}}_{sp}^{(+)}$ and $\wt{\mathcal{M}}_{pp}$ present in  Eq.~\eqref{eq:lcITD}. Still,  it involves a contamination term proportional to $\wt{\mathcal{M}}_{pp}$ that, in fact,  can be removed (as we discuss below). Therefore,  the  matrix element in Eq.~\eqref{eq:pseudo_Ip}, after removal of the  ultraviolet (UV) divergences discussed in the next paragraph, can be used to extract the 
invariant amplitude associated with the matrix elements relevant for the polarized gluon ITD and corresponding PDF.

The bilocal quark and gluon operators separated by a spacelike Wilson line 
[such as the operator in Eq.~\eqref{eq1:ME}]  have additional link-related UV divergences that are multiplicatively renormalizable
(see Refs.~\cite{Izubuchi:2018srq,Ji:2017oey,Green:2017xeu} for the quark case). In particular, various combinations of spatially separated gluon operators are shown to be multiplicatively renormalizable in~\cite{Zhang:2018diq,Li:2018tpe,Balitsky:2019krf,Balitsky:2021cwr}. For our calculation of the matrix elements corresponding to  the gluon helicity distribution, these UV divergences can be canceled by forming the following ratio proposed  in~\cite{Balitsky:2021cwr}:
\bea \label{eq:rITDdef}
\mathfrak{\wt{M}}(\nu,z^2)\equiv i \frac{[\wt{\cal M}_{00}(z,p_z)/p_z { p_0}]/Z_{\rm L}(z/a_L)}{{\cal M}_{00}(z,p_z=0)/m_p^2}\, ,
\eea
where we have defined $\wt{\cal M}_{00} (z,p_z)\equiv [\wt{M}_{0i;0i}(z,p_z) + \wt{M}_{ij;ij}(z,p_z)]$, and ${\cal M}_{00}(z,p_z) \equiv [M_{0i;i0}(z,p_z)+M_{ji;ij}(z,p_z)]$ is the spin averaged matrix element corresponding to the unpolarized gluon PDF~\cite{Balitsky:2019krf,HadStruc:2021wmh}. The factor $1/Z_{\rm L} (z_3/a_L)$ [$z_3\mapsto z$] determined in~\cite{Balitsky:2021cwr} cancels the UV  logarithmic vertex  anomalous dimension
of the $\wt{\cal M}_{00} $ matrix element. The factor $i$ in~\eqref{eq:rITDdef} is introduced in accordance with the definition of the ITD $-i \wt{\mathcal{I}}_p(\nu) \equiv \wt{\mathcal{M}}_{ps}^{(+)}(\nu) - \nu \wt{\mathcal{M}}_{pp}(\nu)$.  The ratio in Eq.~\eqref{eq:rITDdef} utilizes  the presence of the same linear UV divergence in $\wt{M}_{00}(z,p_z)$ and $M_{00}(z,p_z=0)$ related to the gluon link self energy  and cancels this common divergent factor. 
Still, this ratio in Eq.~\eqref{eq:rITDdef} preserves the logarithmic IR divergence at small $z$-separations that corresponds to the Dokshitzer-Gribov-Lipatov-Altarelli-Parisi  (DGLAP) evolution of the PDF~\cite{Gribov:1972ri,Altarelli:1977zs,Dokshitzer:1977sg}. The ratio in~\eqref{eq:rITDdef} is 
referred to as the reduced pseudo-ITD in the rest of the paper. 

As  mentioned above (and shown in~\cite{Balitsky:2021cwr}), the reduced pseudo-ITD~\eqref{eq:rITDdef} contains a contamination term that is not present in the definition of the light-cone gluon helicity ITD in~\eqref{eq:lcITD}.  Indeed, writing the right-hand side of Eq.~\eqref{eq:rITDdef} in terms of the invariant amplitudes 
 of Eq.~\eqref{eq:pseudo_Ip}  and using $z=\nu /p_z$
[which is valid when $z_\mu=(0,0,0,z)$],  we obtain,
\bea \label{eq:Ipform1}
\mathfrak{\wt{M}}(\nu,z^2) = \left [\wt{\mathcal{M}}_{sp}^{(+)}(\nu,z^2) - \nu \wt{\mathcal{M}}_{pp}(\nu,z^2) \right ] - \frac{m_p^2 z^2}{\nu}\wt{\mathcal{M}}_{pp}(\nu,z^2)\, , 
\eea
or, alternatively,  
\bea \label{eq:Ipform2}
\mathfrak{\wt{M}}(\nu,z^2) = \left [ \wt{\mathcal{M}}_{sp}^{(+)}(\nu,z^2)- \nu \wt{\mathcal{M}}_{pp}(\nu,z^2) \right ] -\frac{m_p^2}{p_z^2}  \nu \wt{\mathcal{M}}_{pp} (\nu,z^2) \, ,
\eea
where $m_p$ is the nucleon mass.

There are other combinations derived in~\cite{Balitsky:2021cwr} that  also contain the
invariant amplitudes $\wt{\mathcal{M}}_{pp}$ and $\wt{\mathcal{M}}_{sp}^{(+)}$, but  these combinations involve more contamination terms. In this work, our goal  therefore is to
calculate the matrix elements of the combination in Eq.~\eqref{eq:pseudo_Ip},  try to eliminate the ${\cal O}(m_p^2/p_z^2)$ contamination term present  in 
Eq.~\eqref{eq:Ipform2}, and extract  $[\wt{\mathcal{M}}_{ps}^{(+)}(\nu,z^2) - \nu \wt{\mathcal{M}}_{pp}(\nu,z^2)]$, necessary for
determining the gluon helicity distribution in the nucleon.

With the removal of the ${\cal O}(m_p^2/p_z^2)$  terms present in Eqs.~\eqref{eq:Ipform1} and \eqref{eq:Ipform2}, the resulting reduced pseudo-ITD $\mathfrak{\wt{M}}(\nu,z^2)$ can be related, up to power corrections, to the light-cone polarized gluon ITD $\wt{\mathcal{I}}_g (\nu, \mu^2)$ and singlet quark ITD $\wt{\mathcal{I}}_S (\nu, \mu^2)$ in the $\ms$ scheme through the following short distance factorization relation~\cite{Balitsky:2021cwr}:
\bea \label{eq:matching}
\wt{\mathfrak{M}} ( \nu, z^2 ) \langle x_g \rangle_{\mu^2} &= &   
\wt{\mathcal I}_p (\nu, \mu^2 )  -  \frac{\alpha_s N_c }{2\pi}   \int_0^1 \dd u\,  \wt{\mathcal I}_p (u\nu, \mu^2 ) \bigg\{ \ln\bigg(z^2 \mu^2 \frac{e^{2\gamma_E}}{4}\bigg) \nn \\
&&  \bigg( \bigg[\frac{2u^2} {\bar{u}} + 4u\bar{u}  \bigg]_+ - \bigg(\frac{1}{2}  + \frac{4}{3}  \frac{\langle x_S \rangle_{\mu^2}}
 {  \langle x_g \rangle_{\mu^2} } \bigg) \delta( \bar{u} ) \bigg) \nn \\
&+& 4 \bigg[\frac{u+\ln (1-u)}{\bar{u}}\bigg]_+ - \bigg( \frac{1} {\bar u} - \bar{u} \bigg)_+  -\frac{1}{2} \delta(\bar u) +2\bar uu \bigg\}      \nn \\
&-& \frac{ \alpha_s C_F}{2\pi}  \int_0^1 \dd u \,   \wt{\mathcal{I}}_S (u \nu,\mu^2 )  
 \bigg\{\ln \bigg(z^2 \mu^2 \frac{ e^{2 \gamma_E}} {4 } \bigg)  \wt {\mathcal B}_{gq} (u) + 2\bar uu    \bigg\}  + {\cal O}(\Lambda_{\rm QCD}^2 z^2)\, ,
\eea
where $z^2$ provides the hard scale in the one-loop perturbative matching formula~\cite{Balitsky:2021cwr}, $\Lambda_{\rm QCD}$ is the scale of QCD, $N_c=3$, 
  $\bar{u} \equiv (1 - u)$, $\gamma_E$ is the Euler–Mascheroni constant, and the plus-prescription is defined by
\bea
    \int_0^1 \dd u \, \Big[ f(u) \Big]_+ \; g(u) = \int_0^1 \dd u \, f(u) \, \Big[ g(u) - g(1) \Big]\, .
\eea
We note that for a complete implementation of the one-loop matching, one requires the calculation of the singlet quark Ioffe-time distribution.   In this proof-of-principle calculation, we exclude this quark singlet contribution. Even though the perturbative matching formula in Eq.~\eqref{eq:matching} has been derived up to the next-to-leading order, we will only perform the analysis at leading order, which is also consistent with excluding the quark singlet contribution. This is because the statistical uncertainty in the lattice QCD matrix elements is quite large, as we will see in the subsequent sections. Future studies with improved precision will require both the gluon and quark singlet $O(\alpha_s)$ contributions when they become statistically significant.

\section{Computational Framework}\label{comp_frame}

We construct the gluonic operators and compute the nucleon two- and three-point functions using the same methodologies and numerical techniques as in our previous work on the unpolarized gluon distribution. We therefore refer readers to our previous work in Ref.~\cite{HadStruc:2021wmh} for a more detailed description and briefly summarize our procedure in the following.

We perform our calculation on an isotropic ensemble with $(2+1)$ dynamical flavors of clover Wilson fermions with stout-link smearing~\cite{Morningstar:2003gk} of the gauge fields and a tree-level tadpole-improved Symanzik gauge action. The approximate lattice spacing is $a\sim 0.094$ fm and the pion mass is $m_\pi\sim 358$ MeV~\cite{lattices}. We use $64$ temporal sources over $1901$ gauge configurations, with each configuration separated by 10 hybrid Monte Carlo~\cite{Duane:1987de} trajectories. We take the two light quark flavors, $u$ and $d$, to be degenerate, the lattice spacing was determined using the $w_0$ scale~\cite{Borsanyi:2012zs}, and the strange quark mass is tuned by setting the quantity, $(2\, m^2_{\rm{K}^+} - m^2_{\pi^0})/m^2_{\Omega^-}$ equal to its physical value. We summarize the parameters of the ensemble in Table~\ref{tab:latt}.

\begin{table}
  \renewcommand{\arraystretch}{1.5}
  \setlength{\tabcolsep}{10pt}
  \begin{tabular}{cccccc}
  \toprule
  ID & $a$ (fm) & $m_{\pi}$ (MeV) & $L^3 \times N_t$ & $N_{\rm cfg}$ & $N_{\rm srcs}$\\
    \midrule
    $a094m358$ & 0.094(1) & 358(3) & $32^3 \times 64$ & 1901 & 64\\
    \bottomrule
    \end{tabular}
\caption{The parameters of the ensemble used in this work. Here, $N_{\rm cfg}$ is the number of gauge configurations.  The same number of configurations has been used for the determination of effective matrix elements associated with all the momenta.
\label{tab:latt}}
\end{table}


 On the lattice, the gluonic currents are constructed using the gradient flow~\cite{Luscher:2010iy,Luscher:2011bx,Luscher:2013cpa}.  The gradient flow exponentially suppresses the UV gauge field fluctuations, which physically corresponds to smearing out the original degrees of freedom in coordinate space, therefore improving  the signal-to-noise ratio for the gluon observables. In this work, we perform the calculation of the gluonic matrix elements for flow times $\tau/a^2$ = 1.0, 1.4, 1.8, 2.2, 2.6, 3.0, 3.4, and 3.8. Below $\tau/a^2=1.0$, the calculation is limited by poor signal-to-noise ratios, and more gauge configurations are essential for the calculation at lower values of flow time.


We calculate nucleon two-point correlators using the framework of distillation~\cite{Peardon:2009gh}, a low-rank approximation to the gauge-covariant Jacobi-smearing kernel. Within this framework,  the two-point correlator factorizes into distinct computational components, the so-called elementals and perambulators.  The elementals have a well-defined momentum and encode the structure of the interpolating operators, while the   perambulators  encode the propagation of the quarks within the distillation space and do not feature an  explicit momentum projection. We use an extended basis of interpolators to  facilitate a high-fidelity isolation of the nucleon ground state matrix elements, which is achieved numerically by solving a summed generalized eigenvalue problem (sGEVP)~\cite{Bulava:2011yz}. Within our operator basis we include  interpolators that feature  derivatives of first and second-order to capture the effect of  nonzero angular momenta between the quarks~\cite{Edwards:2011jj} and the breaking of parity when the nucleon interpolators are projected to nonzero momenta;  combinations corresponding to the commutation of two gauge-covariant derivatives acting on the same quark field are also considered. To minimize the  computational cost for the extended set of configurations compared to our previous work on  the unpolarized gluon distribution in~\cite{HadStruc:2021wmh}, we perform some tests to remove the interpolators that have minimal  contributions to the  gluonic matrix elements.
We compare the restricted and full basis of operators in Fig.~\ref{fig:opcomp} of the appendix~\ref{sec:interpol}. Fig.~\ref{fig:opcomp} shows that the restricted basis of interpolators reproduces the  gluonic matrix elements with similar accuracy  and control of excited states over a range of $z$ and $p_z$.  To reduce the cost of the construction of the nucleon two-point correlators, we therefore use the smaller basis of interpolators. 
This reduced basis of interpolators found to be  most relevant for this calculation is listed in Table~\ref{tab:interpolator}. 

\begin{table}
  \renewcommand{\arraystretch}{1.5}
  \setlength{\tabcolsep}{10pt}
  \begin{tabular}{ccc}
  \toprule
    Spatial momentum & Interpolators (Ref.~\cite{HadStruc:2021wmh}) & Interpolators (this work)\\
    \midrule
    $\overrightarrow{p} = \overrightarrow{0}$ & $N^{\;2} S_S\, \frac{1}{2}^+, \;\; N^{\;2} S_M\, \frac{1}{2}^+, \;\; N^{\;4} D_M\, \frac{1}{2}^+$, & $N^{\;2} S_S\, \frac{1}{2}^+, \;\; N^{\;4} P_M^{\ast}\, \frac{1}{2}^+,$\\
    & $N^{\;2} P_A\, \frac{1}{2}^+, \;\; N^{\;4} P_M^{\ast}\, \frac{1}{2}^+, \;\; N^{\;2} P_M^{\ast}\, \frac{1}{2}^+$ & $N^{\;2} P_M^{\ast}\, \frac{1}{2}^+$\\
    \midrule
    $\overrightarrow{p} \neq \overrightarrow{0}$ & $N^{\;2} P_M\, \frac{1}{2}^-, \;\; N^{\;2} P_M\, \frac{3}{2}^-, \;\; N^{\;4} P_M\, \frac{1}{2}^-$, & $N^{\;2} S_S\, \frac{1}{2}^+,$ \\ 
    & $N^{\;4} P_M\, \frac{3}{2}^-, \;\; N^{\;4} P_M\, \frac{5}{2}^-, \;\; N^{\;2} S_S\, \frac{1}{2}^+$, & $N^{\;4} P_M^{\ast}\, \frac{1}{2}^+,$\\
    & $N^{\;2} S_M\, \frac{1}{2}^+, \;\; N^{\;2} P_M^{\ast}\, \frac{1}{2}^+, \;\; N^{\;4} P_M^{\ast}\, \frac{1}{2}^+$ & $N^{\;2} P_M^{\ast}\, \frac{1}{2}^+$\\
    \bottomrule
  \end{tabular}
  \caption{Nucleon interpolators used in the calculation  classified according to the spectroscopic notation: $X^{\;2S+1}L_\pi J^P$ where $X$ is the nucleon, $N$; $S$ is the Dirac spin; $L=S, \; P, \; D, \dots$ is the orbital angular momentum of the continuum interpolator; $\pi = S, \; M \; \mathrm{or} \; A$ is the permutational symmetry of any derivatives; $J$ is the total angular momentum; and $P$ is the parity. The interpolators with an asterisk (*)  are hybrid in nature.   The sets of interpolators used in our previous work on the unpolarized gluon distribution in~\cite{HadStruc:2021wmh} are given in the middle column, and the sets of interpolators implemented in this calculation are given in the rightmost column.\label{tab:interpolator} }
\end{table}

We  apply momentum smearing~\cite{Bali:2016lva} to enhance the overlap of our nucleon interpolators onto the lowest-lying states at high momenta. Following  the procedure introduced in~\cite{Egerer:2020hnc}, the momentum smearing algorithm is realized  through computation of  a  ``phased" distillation space. This modified eigenvector space is obtained by applying spatially varying phases of the form $e^{i\vec{\zeta}\cdot\vec{x}}$ onto a precomputed Laplace eigenvector basis, where we use phases of the form
\bea
    \vv{\zeta} = 2 \cdot \frac{2 \pi}{L} \hat{z}\, .
\eea    

The largest momentum along the $z$-direction we access through this procedure is $p_z = n \times \frac{2 \pi}{La}$ with $n=6$ ( 2.46 GeV in physical units). The momentum smearing is applied for momenta $p_z > 3 \times \frac{2 \pi}{La}$. As in~\cite{HadStruc:2021wmh}, we will simply use the notation $p \equiv p_z$ to describe the nucleon boost along the $z$-direction in the rest of the paper.

\begin{figure}[!htb]
\center{\includegraphics[scale=0.7]{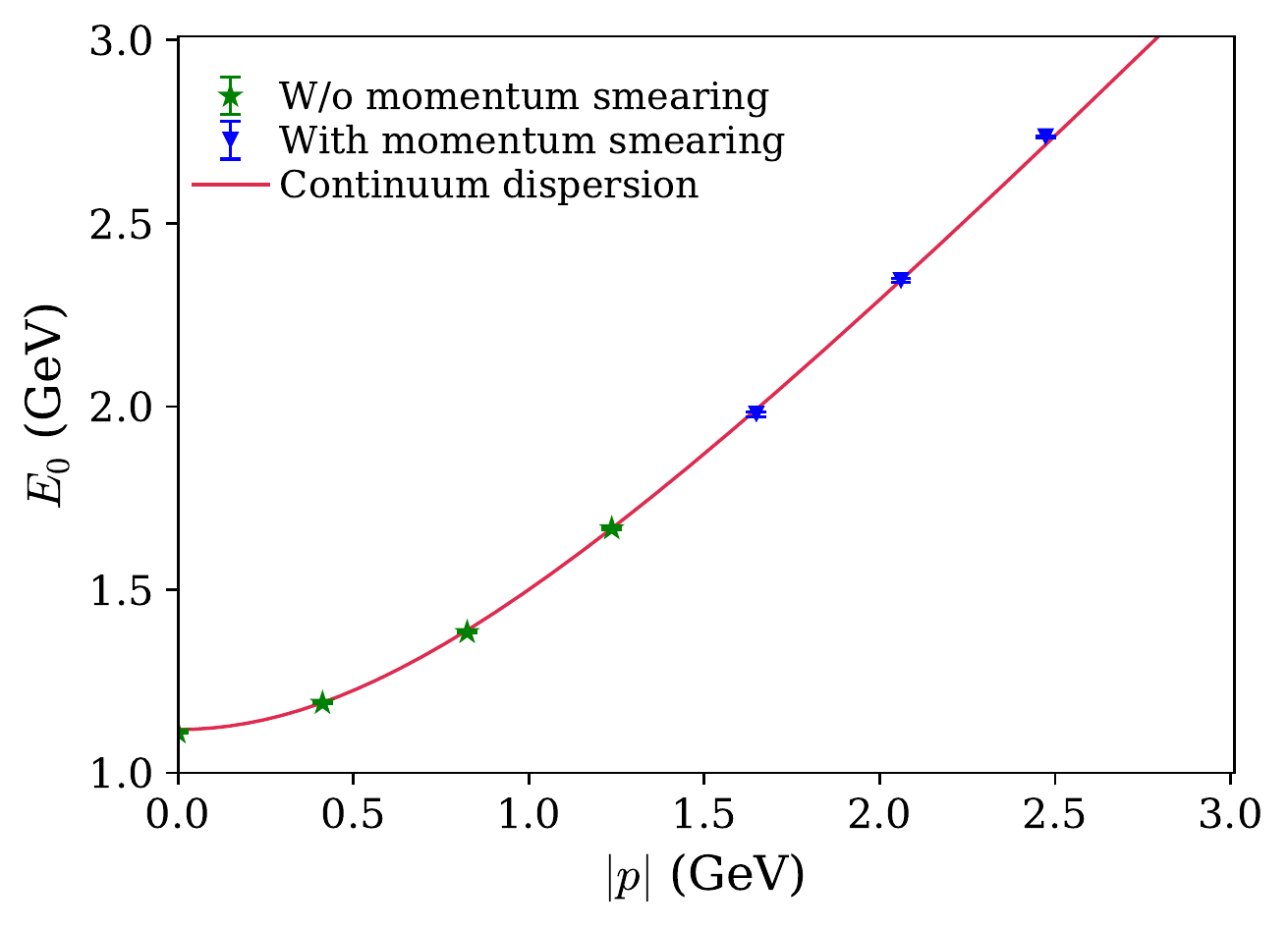}}\caption{The ground state nucleon dispersion relation on the ensemble $a094m358$, the solid line being the continuum dispersion relation. Energies without phasing are in green and energies with phasing are in blue. 
\label{fig:disp_plot}}
\end{figure}

For each momentum, we extract the   three lowest-lying principal correlators, and therefore the energy  states of the nucleon two-point correlators, by performing a variational analysis using the fitting procedure discussed in~\cite{Khan:2020ahz,HadStruc:2021wmh}. We plot the ground state nucleon energies  for the accessible spatial momenta in Fig.~\ref{fig:disp_plot} and compare with  expectations from the continuum dispersion relation. The resulting energies as a function of the momentum agree with the continuum dispersion relation within error, with a slight deviation at the highest momentum where $\mathcal{O}\left((ap)^2\right)$ errors are significant.

We construct the nucleon gluonic matrix elements relevant to the gluon helicity ITD using the combination of gluonic currents in Eq.~\eqref{eq:pseudo_Ip}, and numerically implement a  summed generalized eigenvalue problem (sGEVP)~\cite{Bulava:2011yz, Blossier:2009kd} to extract the ground-state matrix elements with high fidelity from the three-point correlators. { In the sGEVP analysis, the effective matrix elements are calculated for the value of source-sink separation, $t/a = 2$ to $14$. The $t_0/a$ value is the time slice with respect to which the orthogonality of the generalized eigenvectors is defined so that the excited-state contributions in the effective matrix elements would be minimum. We have used different  values of $t_0/a$ for each momentum and the chosen values of $t_0/a$ are: $t_0/a = 8,  9, 6, 7, 7, 7, 6$ for momenta $p_z = 0, 0.41, 0.82, 1.23, 1.64, 2.05, 2.46$ GeV, respectively. The solution to the sGEVP is marked by excited-state contamination  that decays as $ \big[ \, t \; \mathrm{exp}(-\Delta E \, t) \big]$ - a stronger suppression  than the $\big[ \, \mathrm{exp}(-\Delta E \, t/ 2) \big]$ decay in the GEVP method, where $\Delta E$ is the energy gap between the ground state and the lowest excited state not removed by the GEVP. In fact, for $\Delta E \, t \gg 1$, the suppression of the excited-state contamination becomes so significant that the sGEVP requires approximately half the total temporal separation for the same size of systematic corrections compared to the GEVP~\cite{Bulava:2011yz}. As we will see, the nucleon gluonic matrix elements for this gluon helicity distribution calculation are heavily influenced  by noise as the temporal separation is increased. For this reason,  we are reliant on the fits performed at small source-sink separations $t$. Solving the sGEVP where excited-state contributions decay faster is therefore crucial for this calculation.

For each flow time and momentum, we calculate the matrix elements for field separations, $z = a$ to $z =8a$, where $a$ is the lattice spacing. We construct the effective matrix element, $\wt{\mathcal{M}}^{\rm eff}(t, z, p, \tau)$ for each flow time, nucleon momentum, and field separation and perform fits of  the matrix elements using the functional form~\cite{Bulava:2011yz}:
\bea \label{eq:fiteq}
\wt{\mathcal{M}}^{\rm eff}(t) = A + B\,t \exp(-\Delta E\, t) \, .
\eea
Here, $\Delta E$ is the energy gap between the ground-state and an effective excited-state. Before performing a joint and correlated fit to the  gluonic matrix elements at a fixed value of $p\equiv p_z$ and $\tau/a^2$, we first determine the value of $\Delta E$  from fits to the matrix elements  for $z=a,2a,$ and $3a$; each resulting in almost identical values of $\Delta E$. We adopt $\Delta E$ determined from the fit to the matrix element for $z = 2a$ and use it as a prior for the joint fit. An alternative determination of the prior from $z=a$ or $3a$ matrix elements does not alter the outcome of the joint fit as we choose a flexible prior width of $3\times {\rm error}(\Delta E)$--in other words,  a prior width that is three times larger than the uncertainty obtained in the $\Delta E$ determination from the $z=2a$ matrix element fit. 
Our implicit assumption in our choice of prior is that for a given momentum  and flow time,  the nucleon gluonic matrix elements for all $z$-separations have the same excited-state contamination. 

We then perform a simultaneous and correlated fit to the matrix elements for $z \in [a,8a] = [0.094, 0.752]\, {\rm fm}$ using the fit-equation:
\bea \label{eq:fiteqsimul}
\wt{\mathcal{M}}_i^{\rm eff}(t) = A_i+ B_i\,t \exp(-\Delta E\, t) \, ,
\eea
where $i=1, 2, \cdots \, 8$ denoting the $z=a,2a, \cdots \, 8a$ data being fit and $\Delta E$ is assumed to be $z$-independent, and thus held fixed,  for matrix elements of a given  nucleon momentum and flow time. All  subsequent fits in this calculation are performed using the fitting package  {\tt XMBF}~\cite{XMBF}.

In Fig.~\ref{fig:Mefftauunpol}, we show sample fits to the gluonic matrix elements associated with the unpolarized gluon pseudo-ITD, needed to calculate the reduced pseudo-ITD~\eqref{eq:rITDdef}, for zero nucleon  momentum with $z=2a$ and $z=8a$, and two values of flow-time $\tau/a^2=1.4,3.0$. In Fig.~\ref{fig:Mefftau}, we illustrate the fits to the matrix elements for $\tau/a^2=1.4$ (upper row)  and for $\tau/a^2=3.0$ (bottom row)  for momenta $0.82$ GeV ($p = 2 \times \frac{2 \pi}{La}$), $1.46$ GeV ($p = 4 \times \frac{2 \pi}{La}$), and $2.46$ GeV ($p = 6 \times \frac{2 \pi}{La}$) and separations $z=2a, \,8a$. We list the fitted parameters in Table~\ref{tab:fitparams} of the appendix~\ref{appendix:fitparams}.  

\befs 
\centering

\includegraphics[scale=0.45]{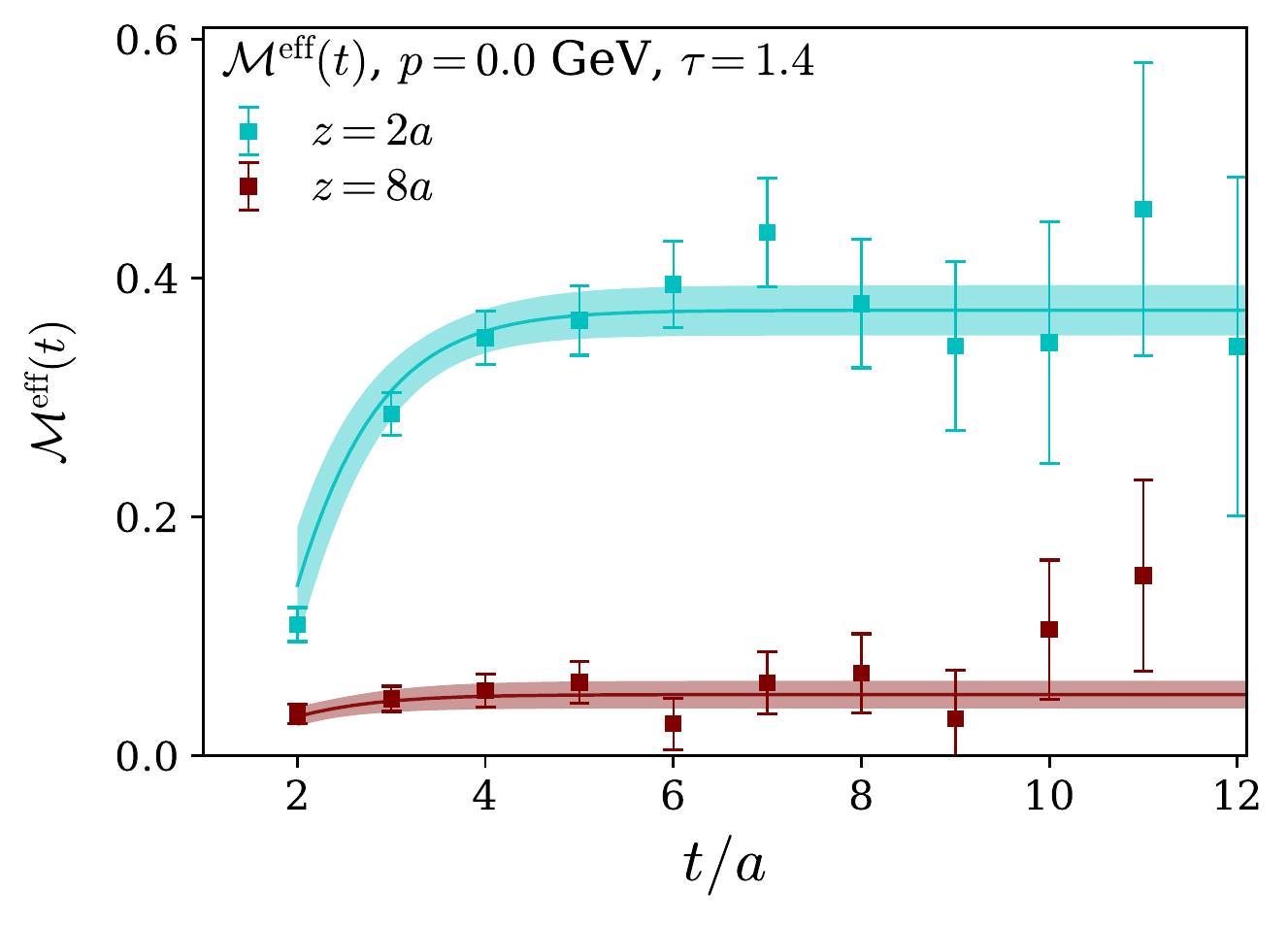}
\includegraphics[scale=0.45]{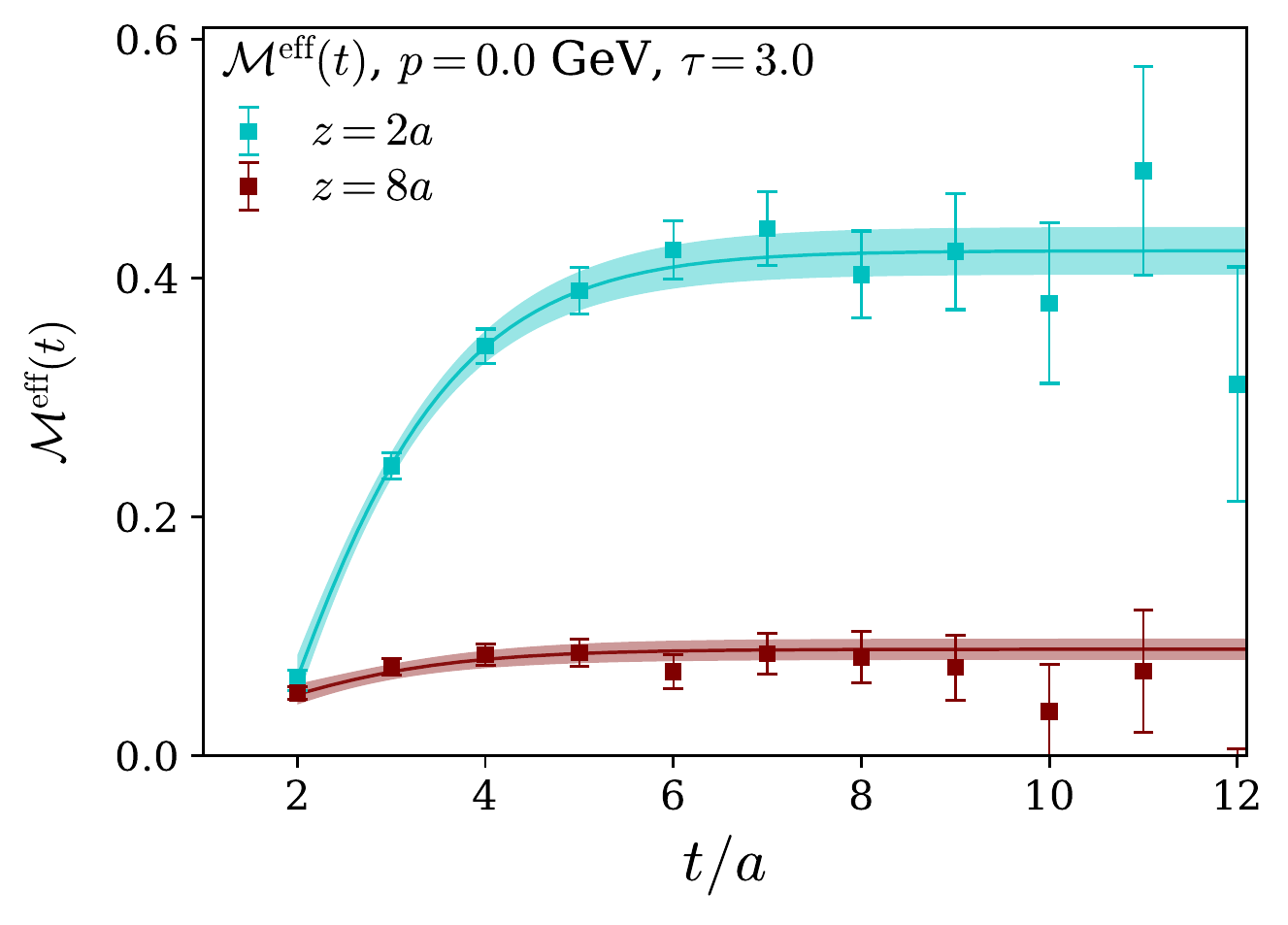}

\caption{\label{fig:Mefftauunpol} 
Extraction of the  zero-momentum  gluonic matrix elements associated with unpolarized gluon pseudo-ITD using the sGEVP method for different flow times 1.4 (left) and 3.0 (right). The bands are the fits described in the text for separations $z=2a$ and $8a$. We calculate the unpolarized $p_z=0$ matrix element on 1901 configurations, equal to the number of configurations used in this work to compute the polarized matrix elements. 
 }      
\eefs{mockdemocn}


\befs 
\centering

\includegraphics[scale=0.42]{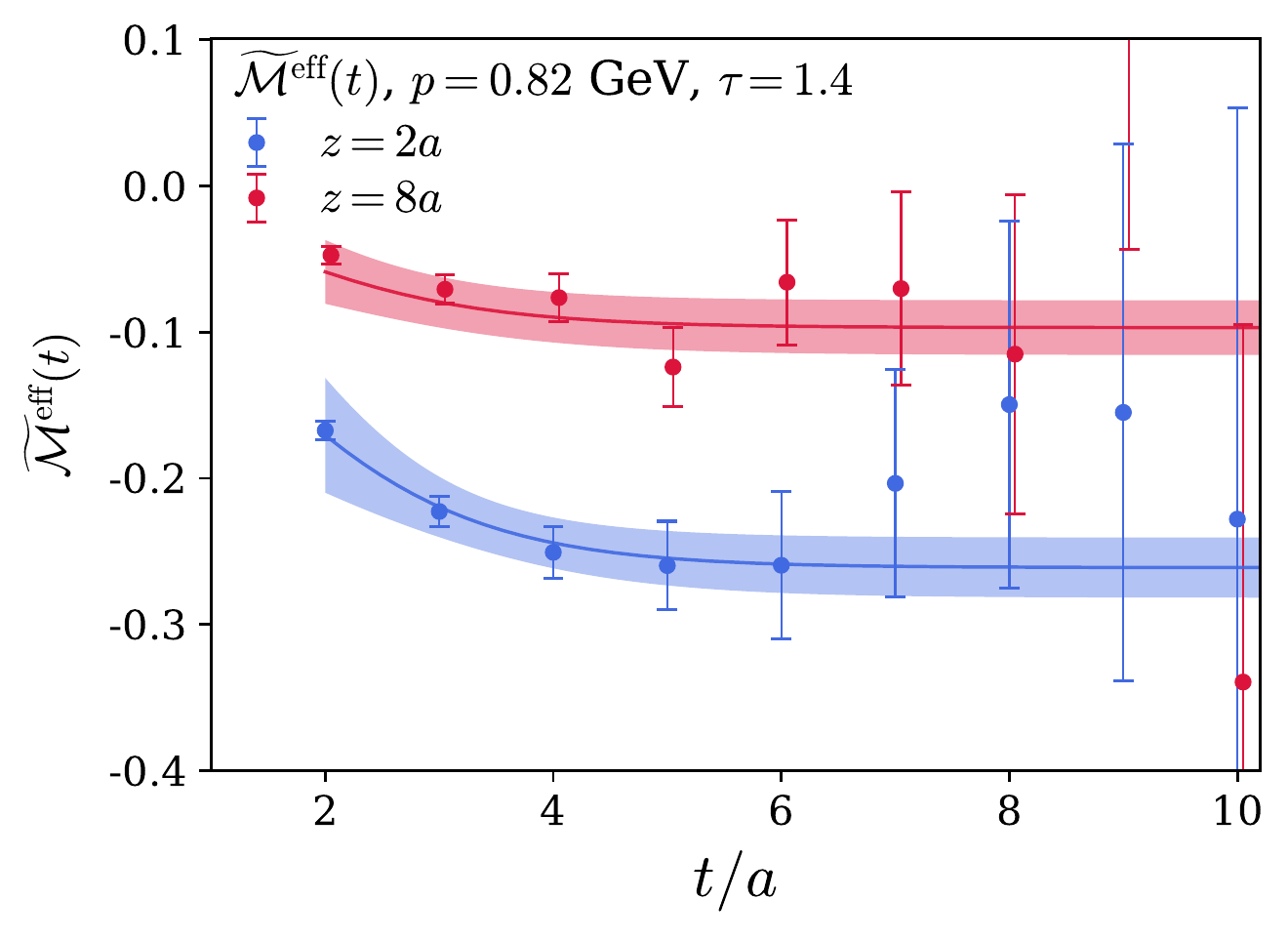}
\includegraphics[scale=0.42]{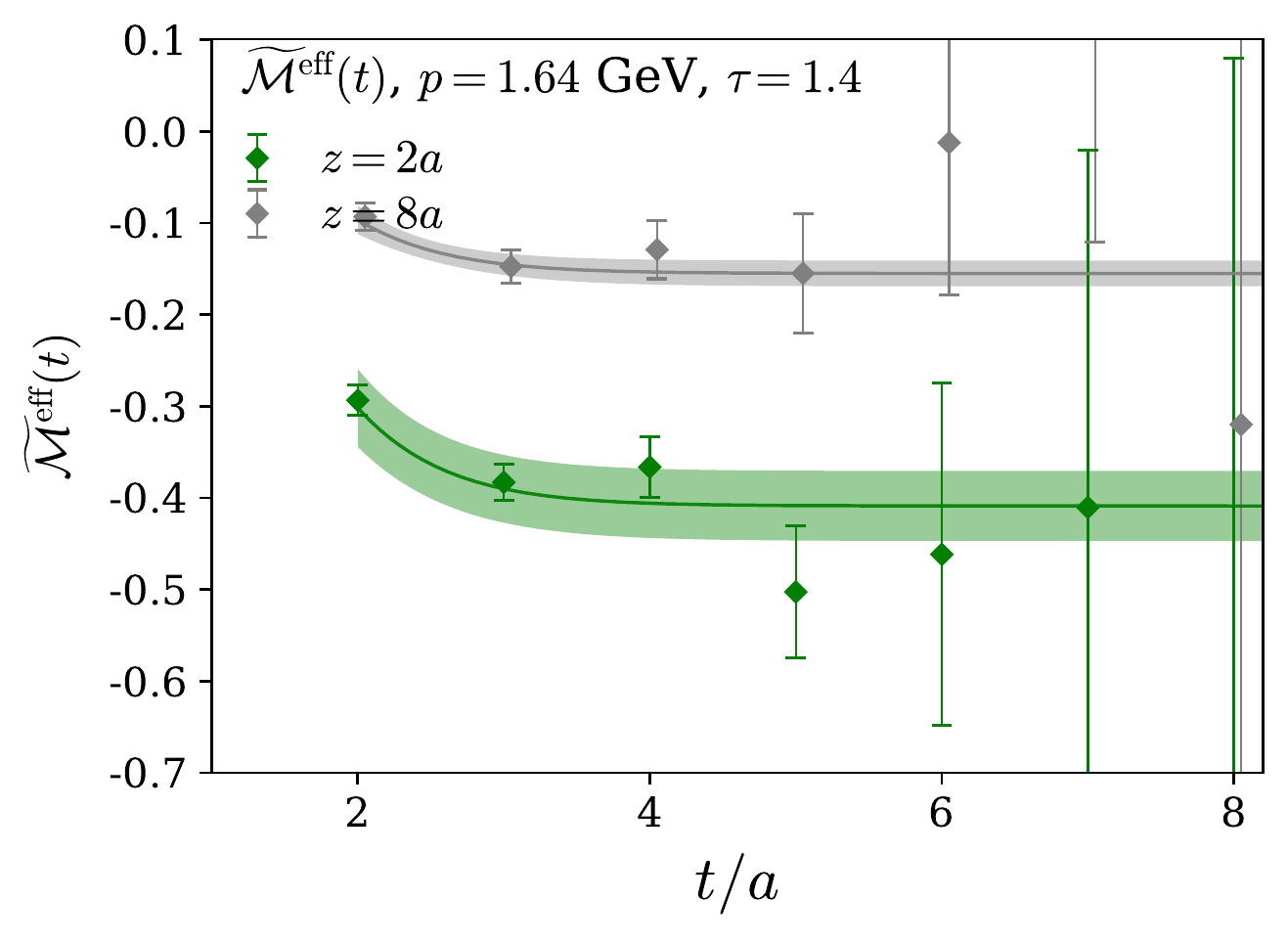}
\includegraphics[scale=0.42]{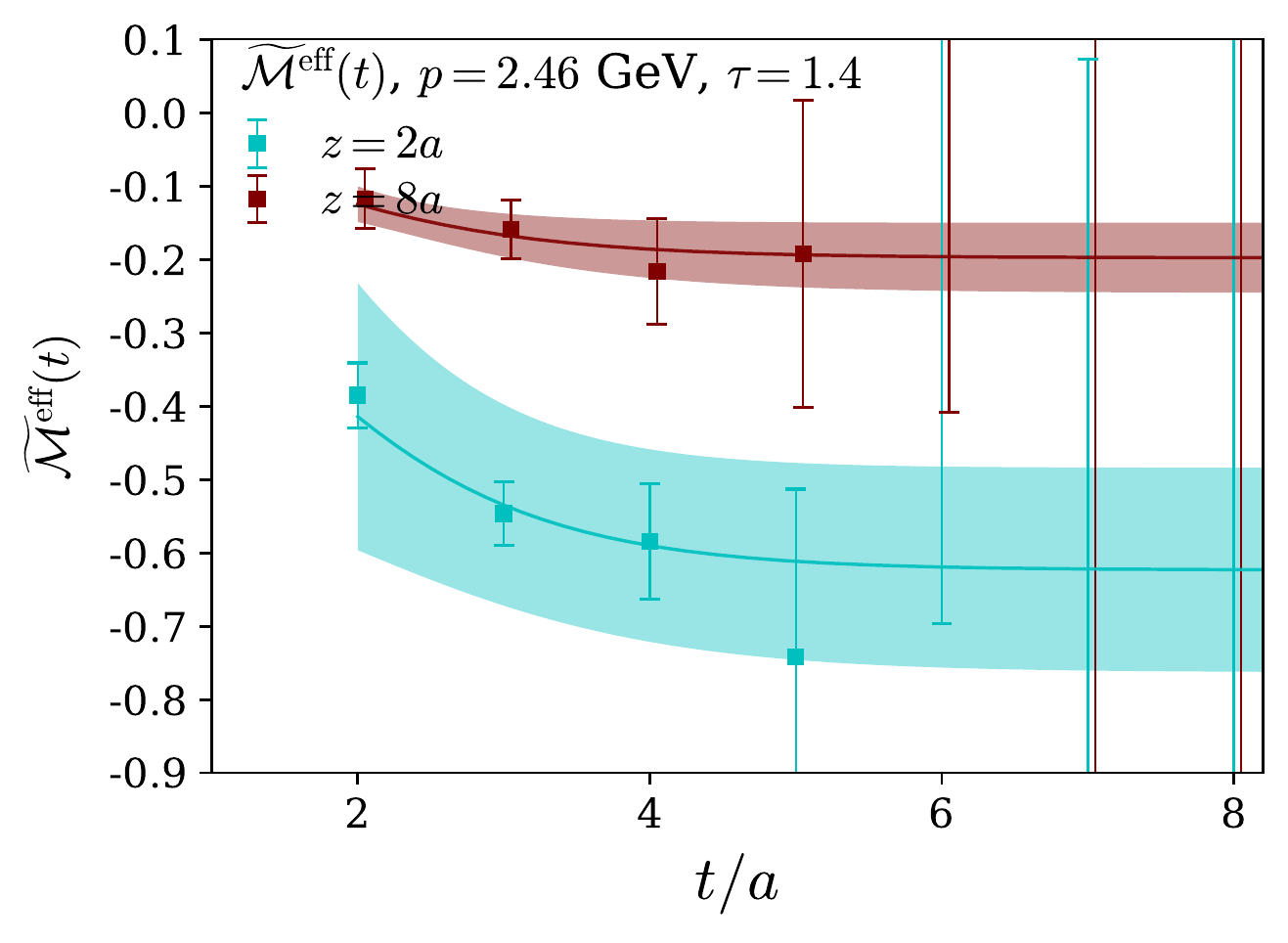}

\includegraphics[scale=0.42]{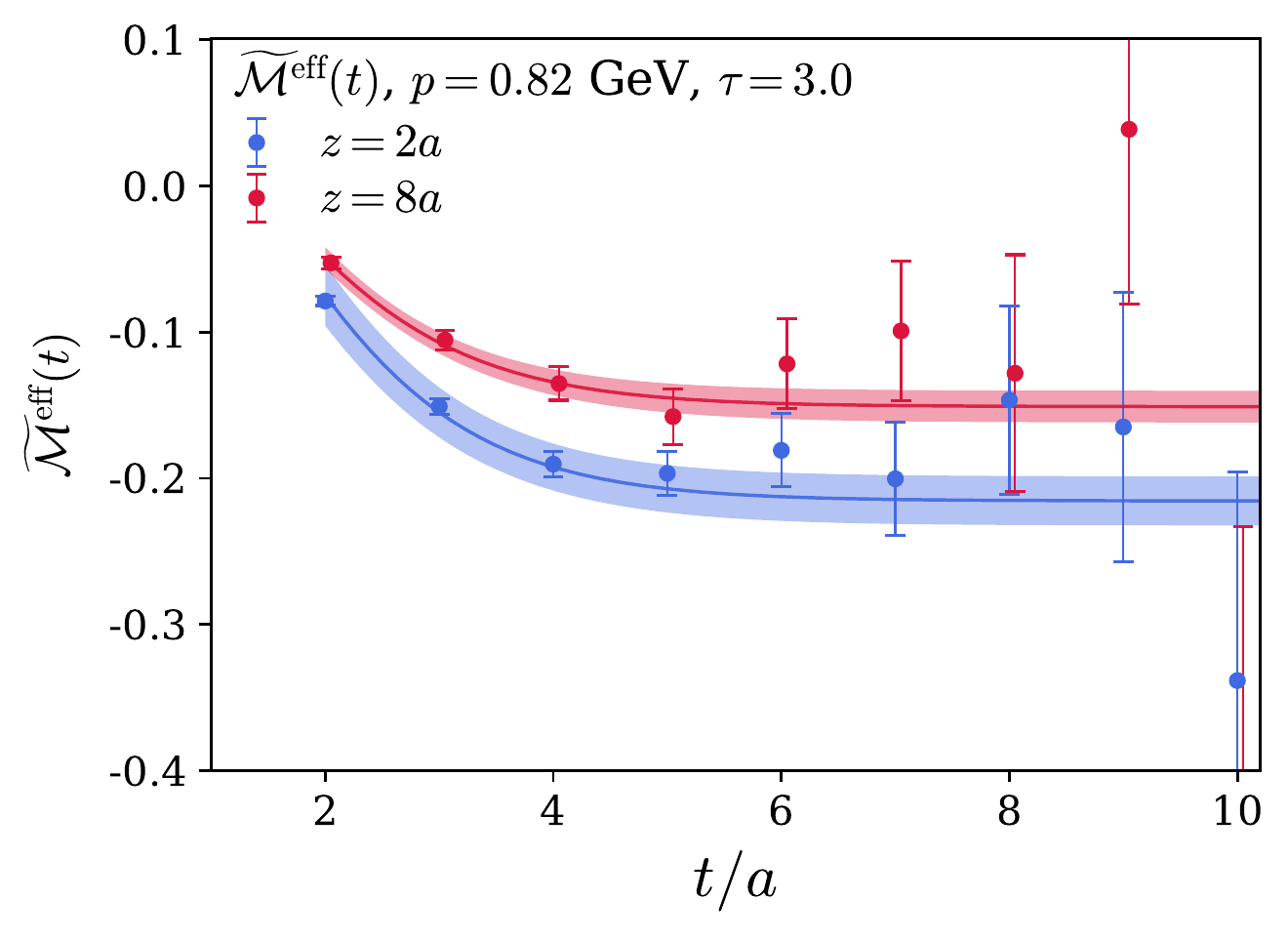}
\includegraphics[scale=0.42]{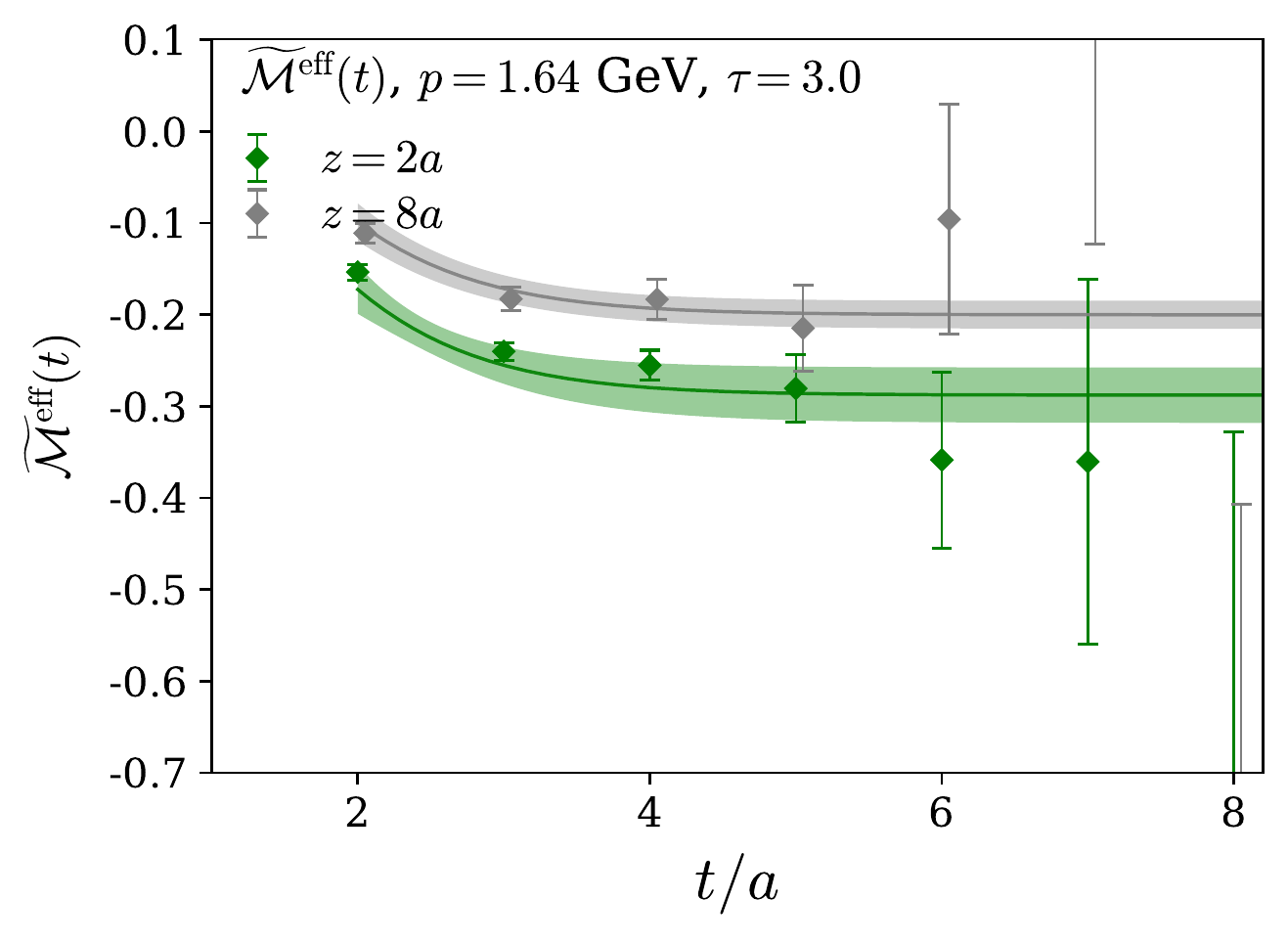}
\includegraphics[scale=0.42]{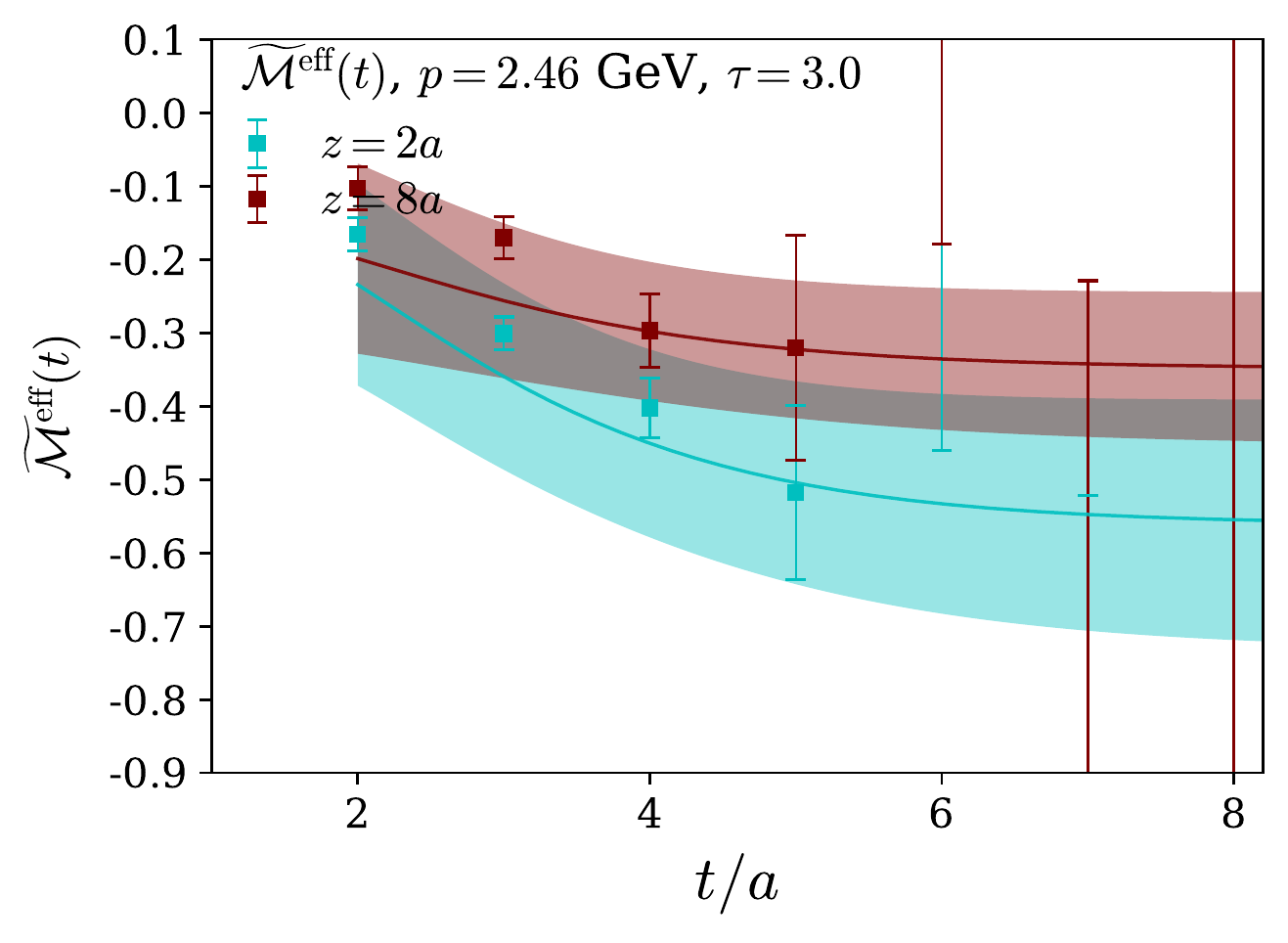}

\caption{\label{fig:Mefftau} 
Extraction of the matrix elements associated with polarized gluon pseudo-ITD using the sGEVP method for different flow times, nucleon momenta and field separations. The top and bottom rows contain the matrix elements for flow time $\tau/a^2$ = 1.4 and 3.0, respectively. In each row, the left, middle and right column illustrate  the matrix elements for $p =$  0.82 GeV, 1.64 GeV, and 2.46 GeV respectively at separations $z=2a, \, 8a$.  The largest source-sink separations used in the fits are indicated by the $t/a$-axis range of the effective matrix element plots.  Note that $p_z$ is written simply as $p$ in all the plots.
}      
\eefs{mockdemocn}

 Fig.~\ref{fig:Mefftauunpol} shows that the excited-state contribution to the unpolarized gluonic matrix elements dies out around source-sink separations of $t/a=4$, demarcated by a  plateau  in the effective matrix element signal. This rapid relaxation of the effective matrix element to one of ground-state dominance is a testament to our having identified a variationally-optimized interpolator through solution of the sGEVP induced by our chosen operator basis. From the $N\times N$ correlation matrix, the variational approach will create a series of $N$ orthogonal generalized eigenvectors that overlap significantly with the $N$ lowest energy states of a particular finite volume irrep, even if states are nearly degenerate. By using the generalized eigenvector associated with the ground state, the contributions of the $N-1$ lowest excited states are suppressed by this orthogonality.  This has proven successful in previous calculations of nucleon structure~\cite{Khan:2020ahz,Egerer:2018xgu}. In particular, it was shown in~\cite{Dudek:2012gj} that the optimized interpolators reduce the excited-state contributions at much earlier source-sink separations, allowing us to start the fit at significantly smaller time slices. Similar to other lattice calculations of gluonic matrix elements, this is  critically important for this calculation, as  the matrix elements at large momenta (see right panel of Fig.~\ref{fig:Mefftau}) lose any meaningful  signal around a source-sink separation $t = 6a$ = 0.564 fm. 

Here we point out some interesting features of the fits to the  matrix elements associated with the polarized gluon pseudo-ITD. From the left-side top and bottom panels in Fig.~\ref{fig:Mefftau}, we see for $p_z=0.82$ GeV the matrix elements  tend to fall on a plateau starting from $t/a=4$ and the fit bands describe the lattice data well. However, for some particularly large $z$-values the fit bands for both $\tau/a^2$ have a broader error band compared to the lattice data point at $t/a=2$. This is caused by the imposed constraint of fixed  $\Delta E$, which is assumed to be the same for matrix elements at a fixed nucleon momentum and flow time [see our discussion around Eq.~\ref{eq:fiteqsimul}].  This simultaneous and correlated fit, along with the  $\Delta E$ constraint for  all  $z\in[a,8a]$ matrix elements of  a fixed nucleon momentum and flow time, helps to  stabilize the fits for the noisier large $p_z$ and $z$ data points.  This fit, in particular, is helpful for fitting the largest $p$ data as shown on the right top and bottom panel in Fig.~\ref{fig:Mefftau}. 

 The right top and bottom panel in Fig.~\ref{fig:Mefftau} show that, although the lattice data points at earlier time slices become more precise with increasing $\tau/a^2$, the final fit bands for momenta $2.46$ GeV have larger error at $\tau/a^2=3.0$ compared to those at $\tau/a^2=1.4$. This is because the fall-off of the early time-slice data points for  $z=2a$ at  $\tau/a^2=3.0$ is more prominent within  uncertainties and does not exhibit  a plateau before $t/a=6$, at  which time the small  signal-to-noise ratio limits a robust determination of the plateau.  On the other hand, the $z=8a$ matrix elements are seen to reach a plateau around $t/a=4$ even at this large nucleon boost.  Since the simultaneous and correlated fits rely on a constancy constraint for $\Delta E$, the $z=8a$ matrix element fits, for example, are in fact advantageous as the clear plateau limits the flexibility of $\Delta E$ and hence dominates the remaining fits for matrix element at smaller $z$-values where there is no clearly observed plateau at a fixed nucleon momentum and flow time. By simultaneously analyzing multiple correlation functions and fixing the energy gaps to be the same, the matrix element and $\Delta E$ can be more precisely determined.

Compared to the analogous matrix elements of quark bilinears needed to resolve the quark PDFs, the gluonic bilinears, especially in the present polarized case, require considerably higher statistics in order to resolve the matrix elements at a commensurate level. As this work represents the first attempt to resolve the $x$-dependent gluon helicity distribution from lattice QCD, the number of measurements we identified was found to be sufficient to control the extraction of the polarized gluonic matrix elements for each $\lbrace p_z,z\rbrace$ and hence the polarized gluon reduced pseudo-ITD. A much larger sample size, with which matrix element plateaux at larger source-sink separations and at the largest nucleon momentum should become manifest,  will be required to rigorously estimate systematic effects present in these matrix element extractions. This will be the goal of a future calculation with much larger statistics and inclusion of the singlet quark distribution. To demonstrate the feasibility of a lattice calculation to determine the gluon helicity PDF, in the following sections we use the extracted matrix elements quoted only with statistical uncertainties and discuss the construction of the reduced pseudo-ITD.

\subsection{Reduced pseudo-ITD calculation and extrapolation to 
zero flow time }\label{sec:zero-flowtime}

With the bare matrix elements, we now calculate the reduced matrix elements and determine the polarized gluon reduced pseudo-ITD using the ratio in Eq.~\eqref{eq:rITDdef} for different flow times, nucleon momenta, and field separations. We present the reduced matrix elements for each value  of $\tau/a^2$ used in this work in Fig.~\ref{fig:ritdall}
(note that $p_z$ is written simply as $p$ in all the plots).

We observe that the data  strongly vary with the change of $p_z$,
indicating the importance of the ${\cal O}(m_p^2/p_z^2)$-type  contamination. In agreement with  Eq.~\eqref{eq:Ipform2},  the $p_z$-dependence of the data is less visible at large $p_z$, because the ${\cal O}(m_p^2/p_z^2)$ term  is suppressed for large $p_z$ at a fixed $\nu$.  Alternatively, for small $p_z$ the data as functions of $\nu$ strongly change when one changes $p_z$. Note, however, that the original matrix element,  given by   Eq.~\eqref{eq:pseudo_Ip}, 
 is  finite for $p_z=0$.  When the nucleon is at rest, this matrix element  is dictated entirely  by the  ${\cal O}(m_p^2)$ contribution
 and is equal  to $2m_p^3 z \wt{\cal M}_{pp}(\nu=0,z^2)$. 

\befs 
\centering

\includegraphics[scale=0.6]{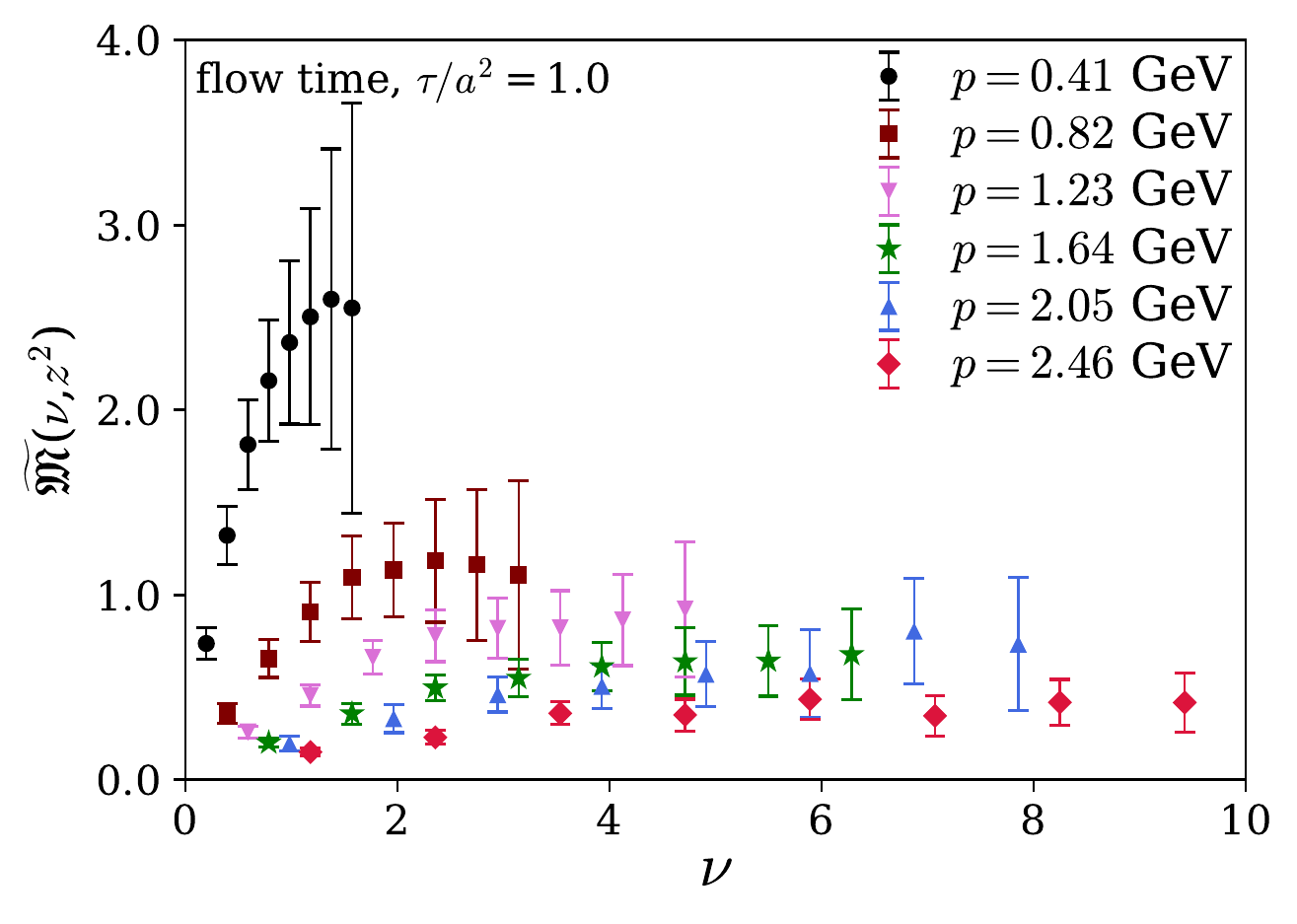}
\includegraphics[scale=0.6]{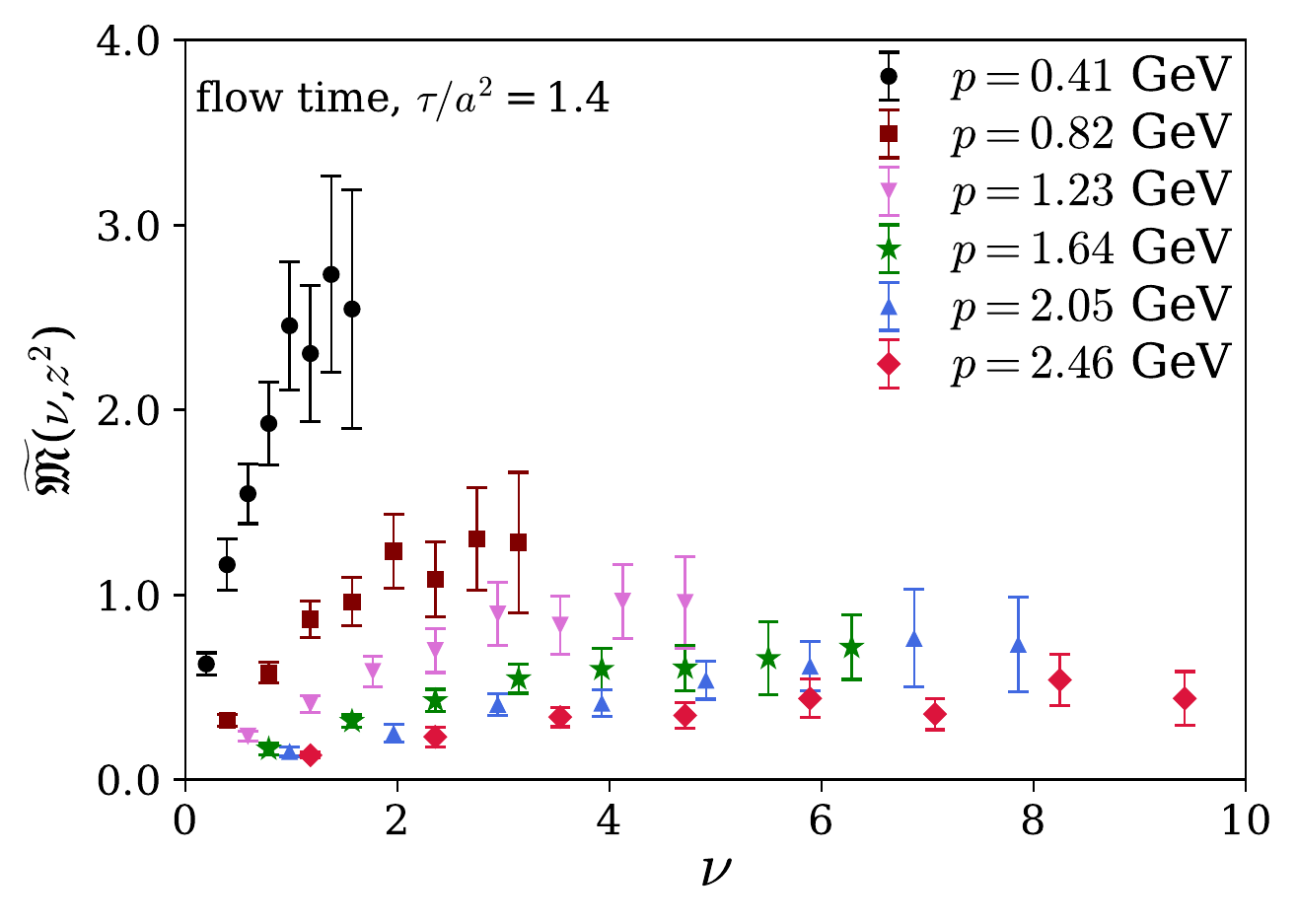}

\includegraphics[scale=0.6]{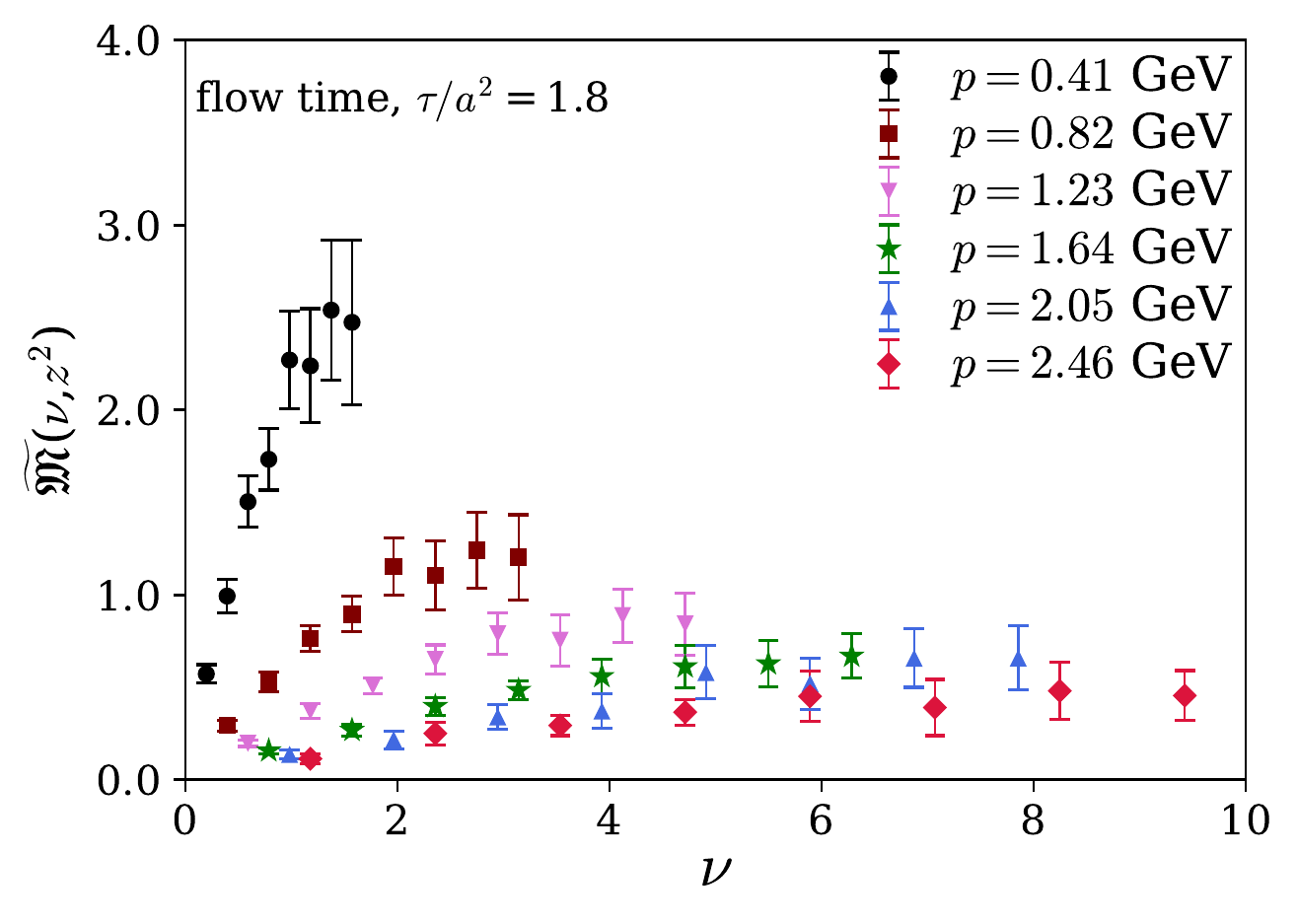}
\includegraphics[scale=0.6]{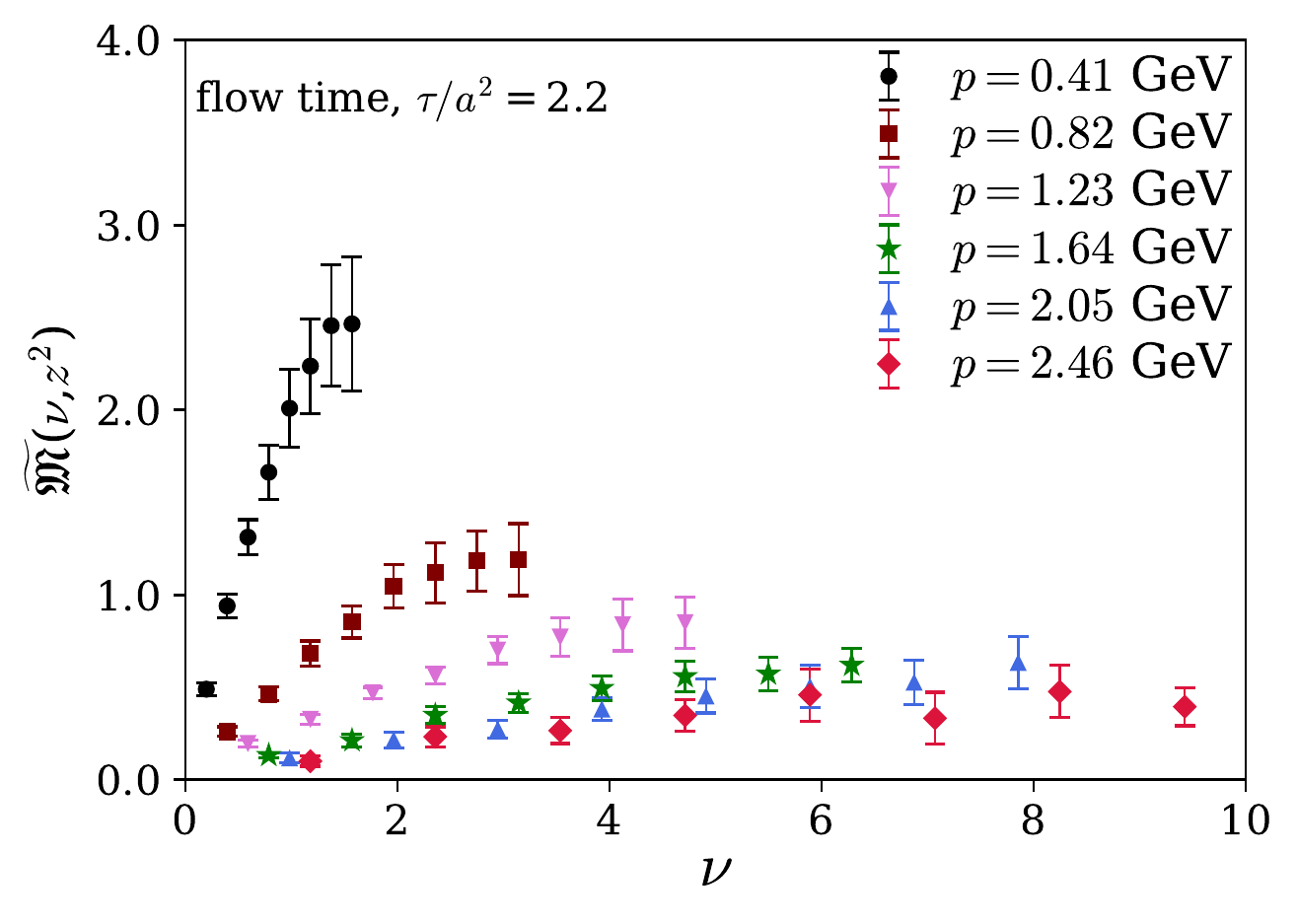}

\includegraphics[scale=0.6]{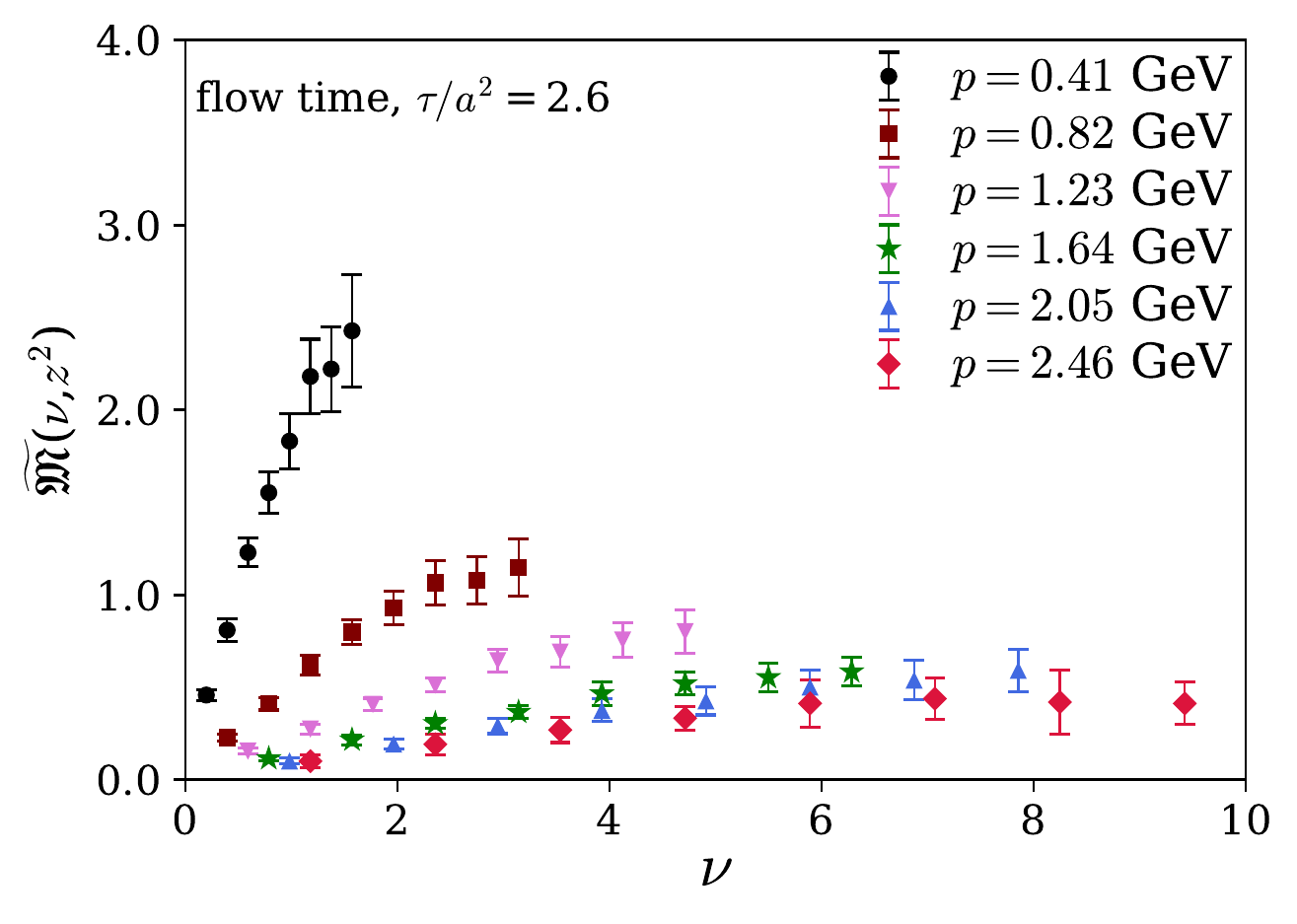}
\includegraphics[scale=0.6]{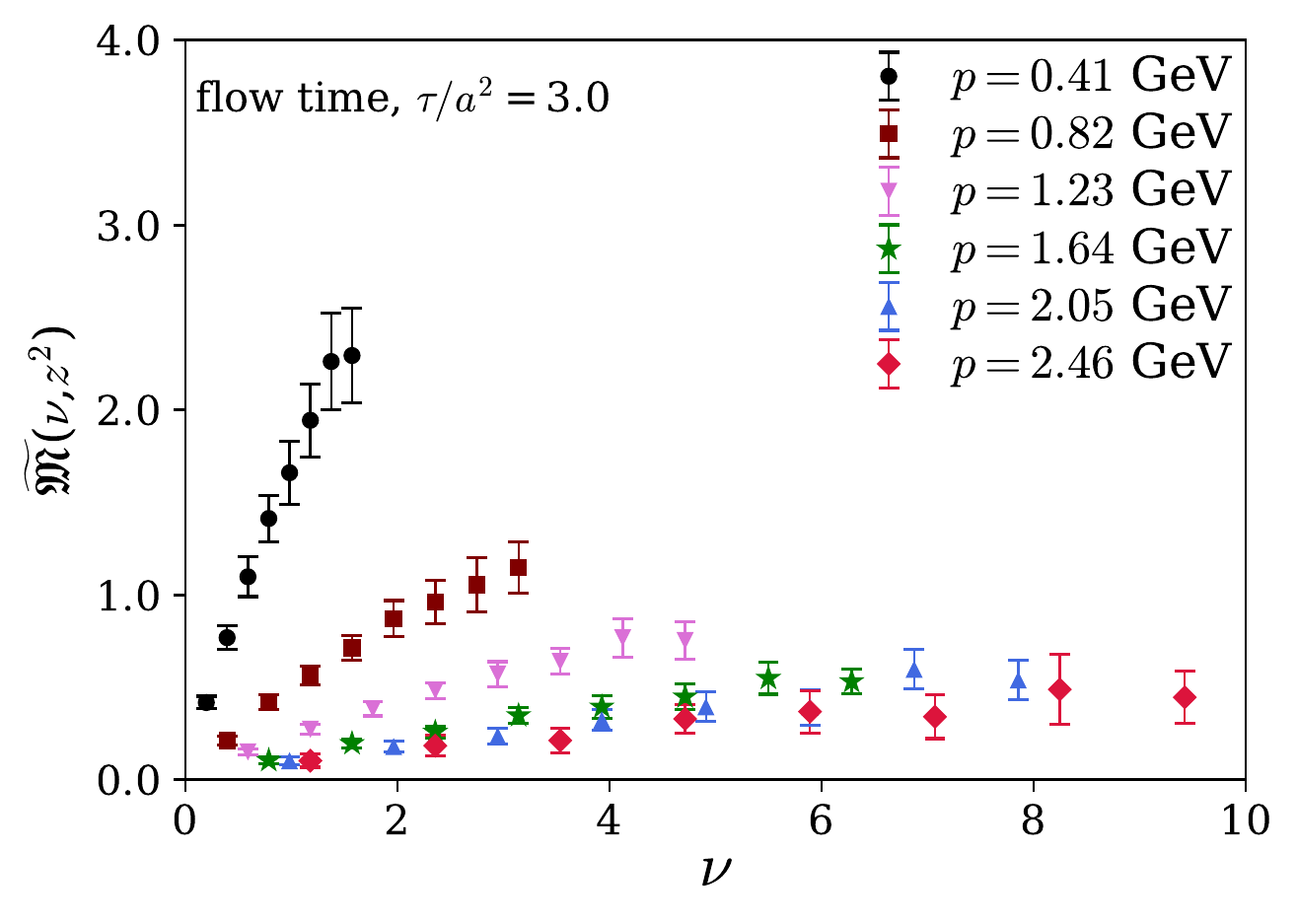}

\includegraphics[scale=0.6]{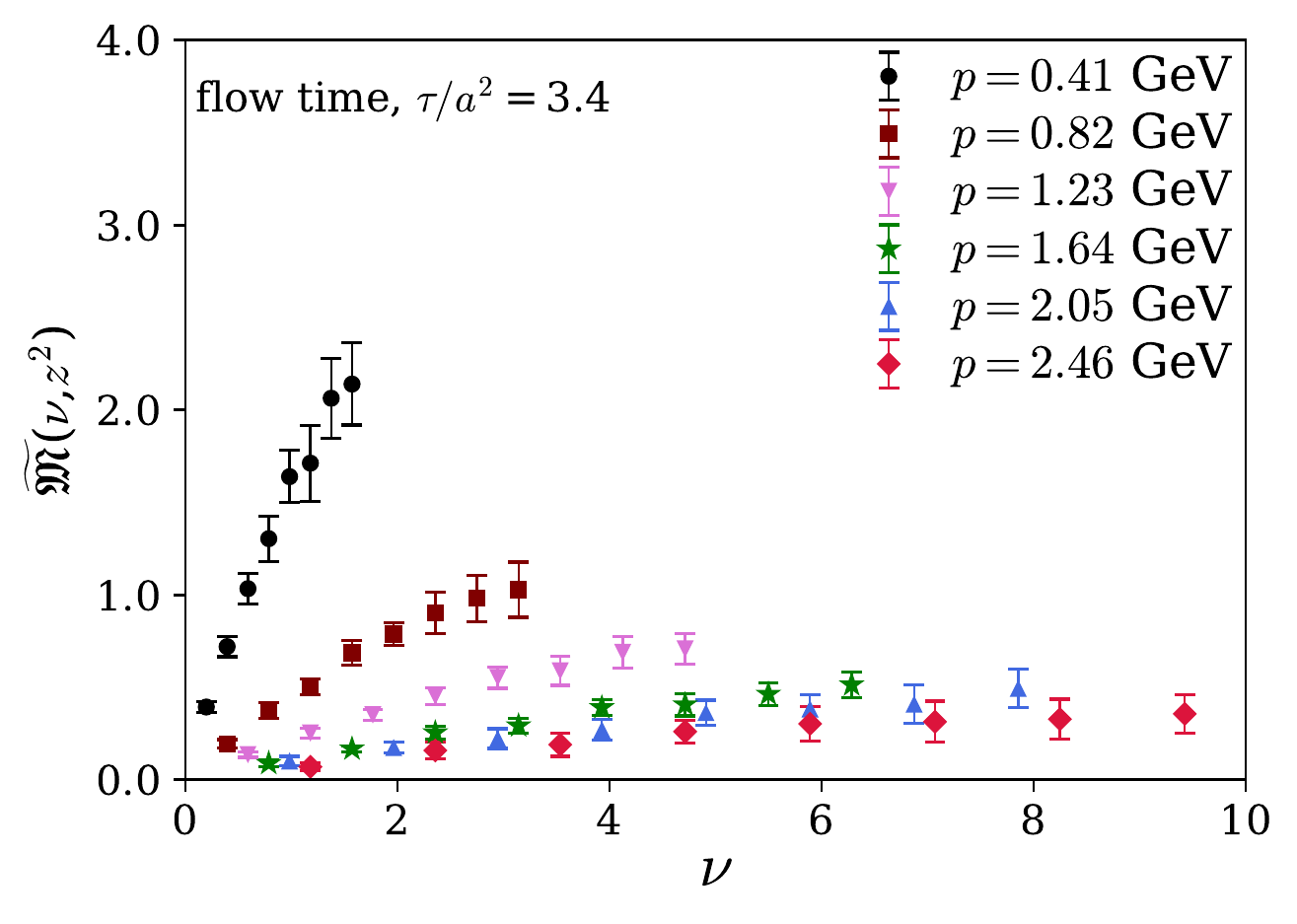}
\includegraphics[scale=0.6]{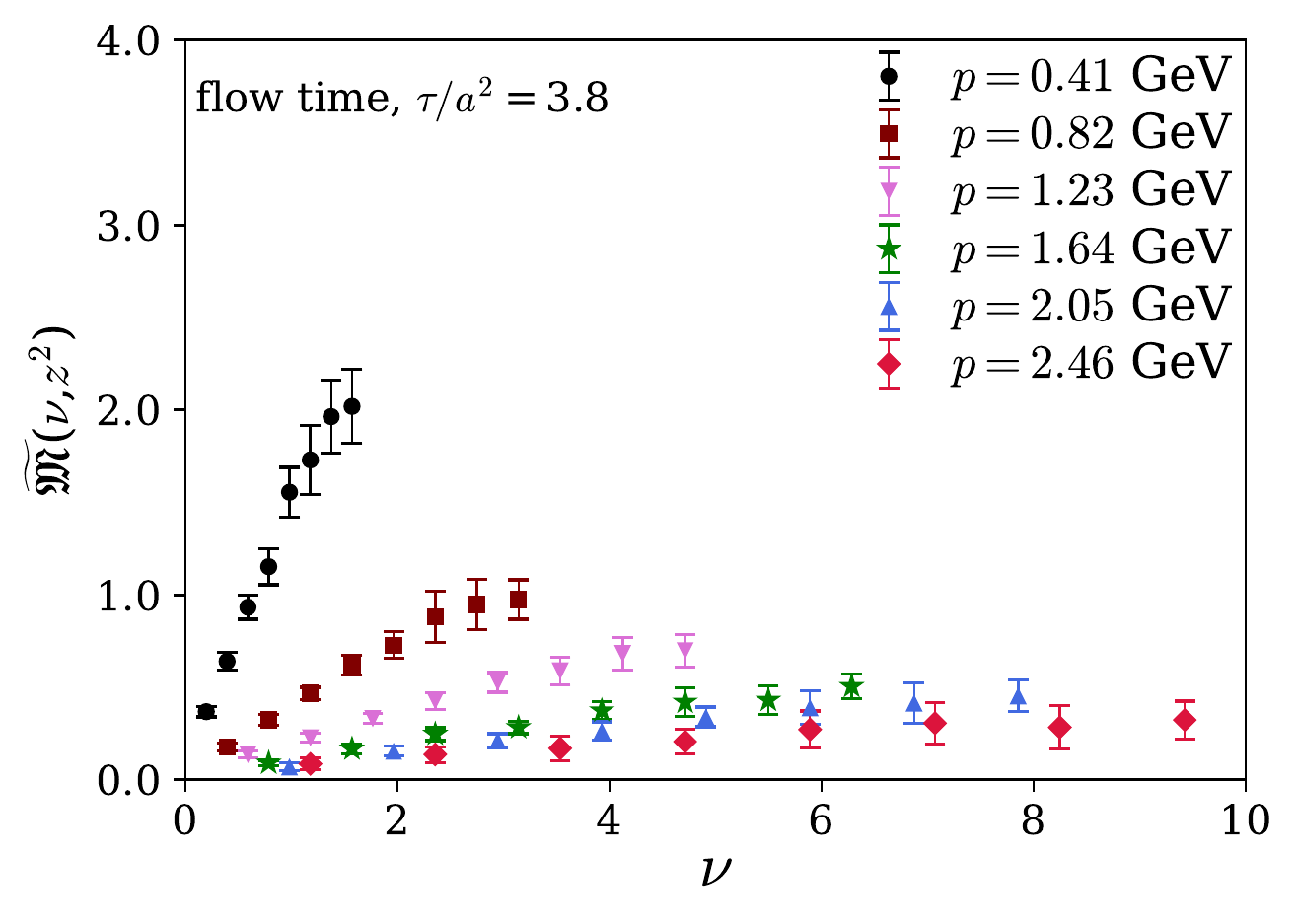}

\caption{\label{fig:ritdall} The reduced matrix elements, $\wt{\mathfrak{M}} (\nu, z^2)$ with respect to the Ioffe-time for different flow times. The reduced matrix elements are shown for increasing $\tau/a^2 =$ 1.0 to 3.8 traversing from the top-left to lower-right panels.}
\eefs{mockdemocn}

Next, we calculate the reduced pseudo-ITD from the reduced matrix elements at different flow times by extrapolating these matrix elements to zero flow time. As can be seen from Fig.~\ref{fig:tauextrapolation}, the flow-time dependence, 
at fixed values of the field separation and nucleon momentum, can be best described by a linear fit of the form: $\wt{\mathfrak{M}}(\tau/a^2) = c_0 + c_1 \tau$. Similar to the calculation of the unpolarized gluon distribution in~\cite{HadStruc:2021wmh}, the  addition of a term like $c_2 \tau^2$ turned out not having any contribution in the fit and we therefore use the linear fit form to determine the reduced pseudo-ITD matrix elements in the subsequent analysis. We list the values of the fitted parameters  in Table~\ref{tab:zft_reduced_mtx_elem} of appendix~\ref{zero_flow_time_reduced_mtx_elem} in a convenient  form  for reproducible analysis and presentation of our calculation in the subsequent section. Out of forty-eight such extrapolations, we demonstrate four arbitrary examples of such extrapolation in Fig.~\ref{fig:tauextrapolation}, and note that for all extrapolations we find $\chi^2/{\rm d.o.f.}< 1.0$.  Finally, we present the reduced pseudo-ITD in the zero flow time limit in Fig.~\ref{fig:pseudo-rITDfinal}.

\befs 
\centering

\includegraphics[scale=0.6]{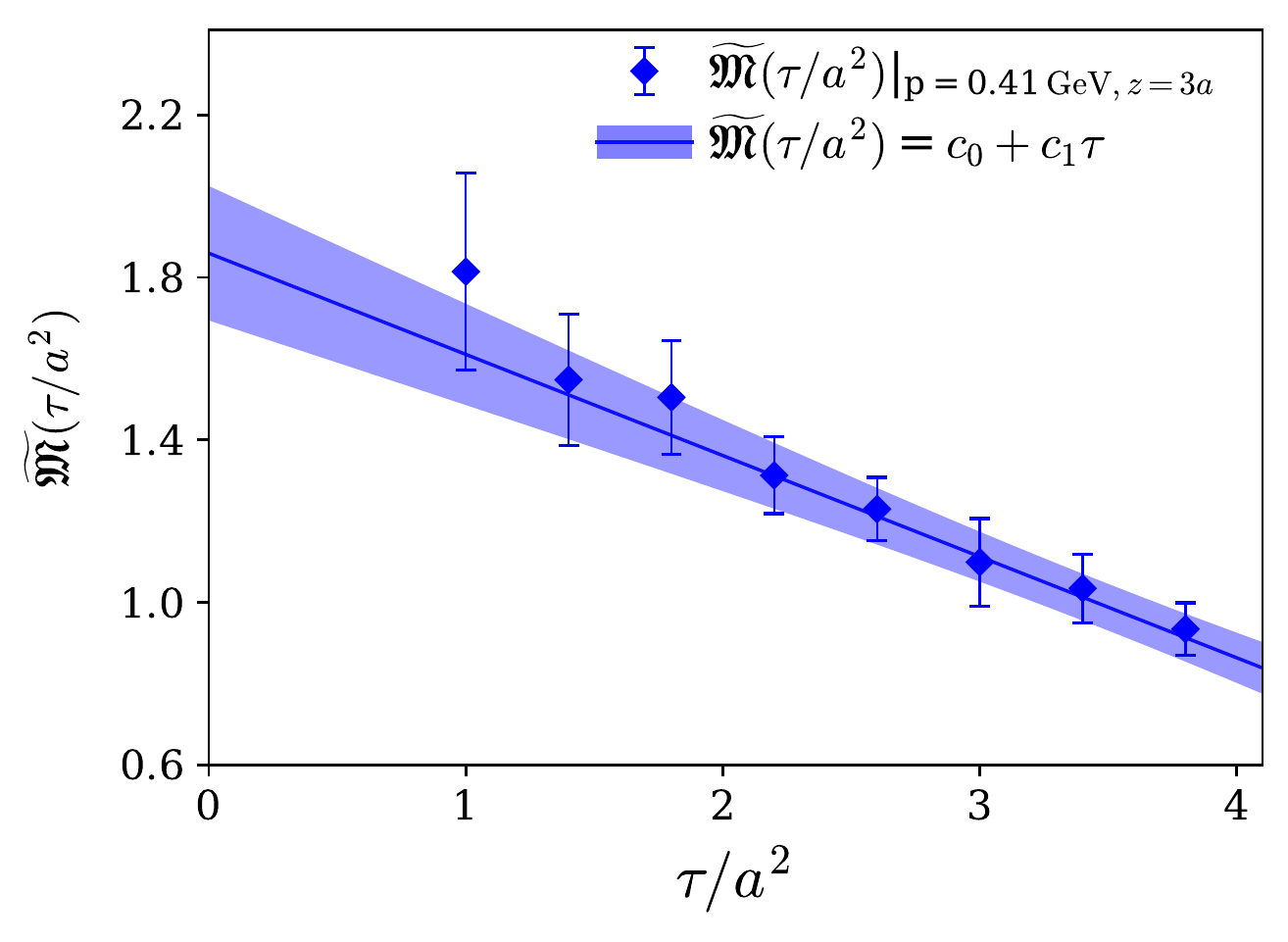}
\includegraphics[scale=0.6]{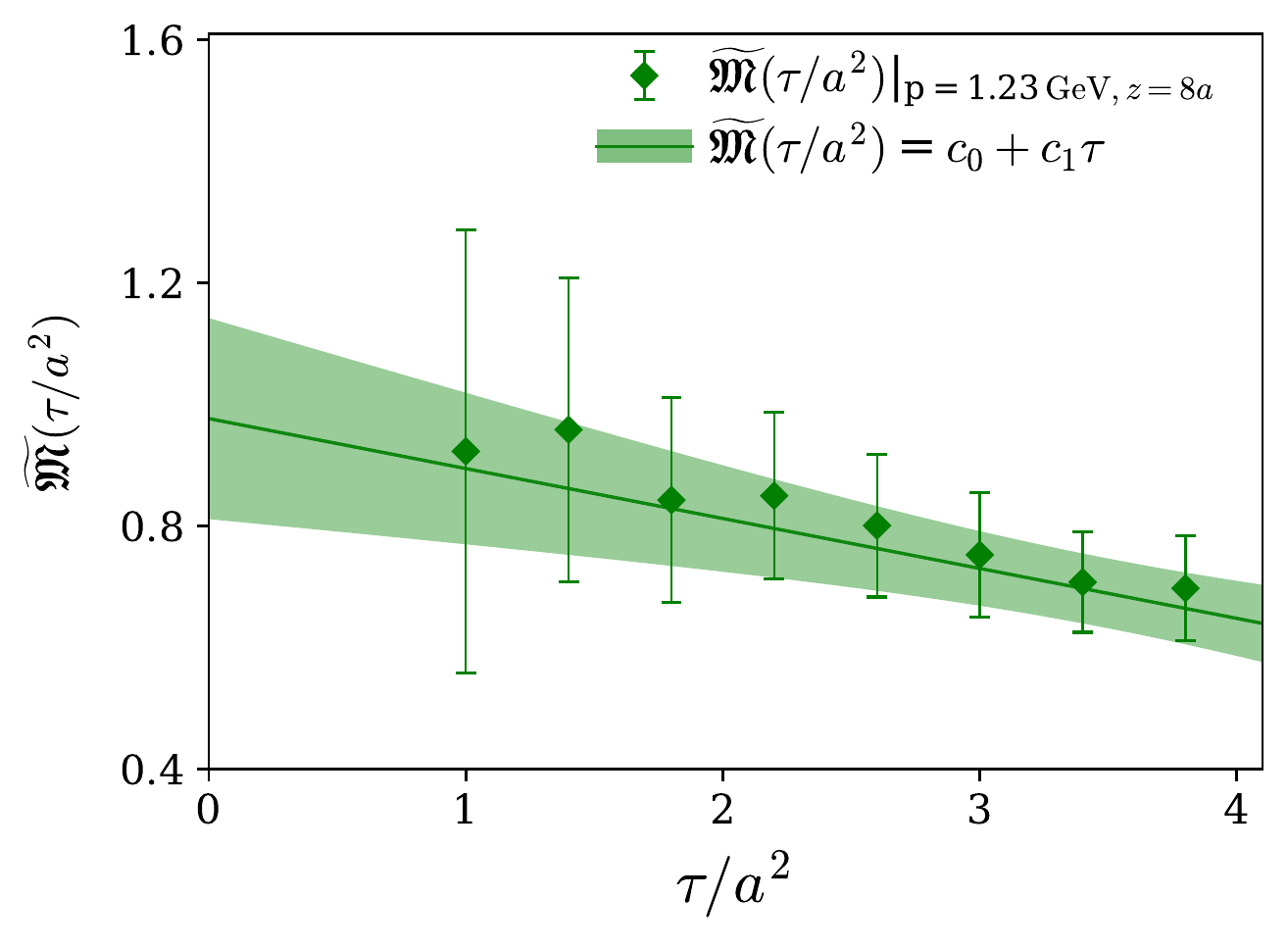}

\includegraphics[scale=0.6]{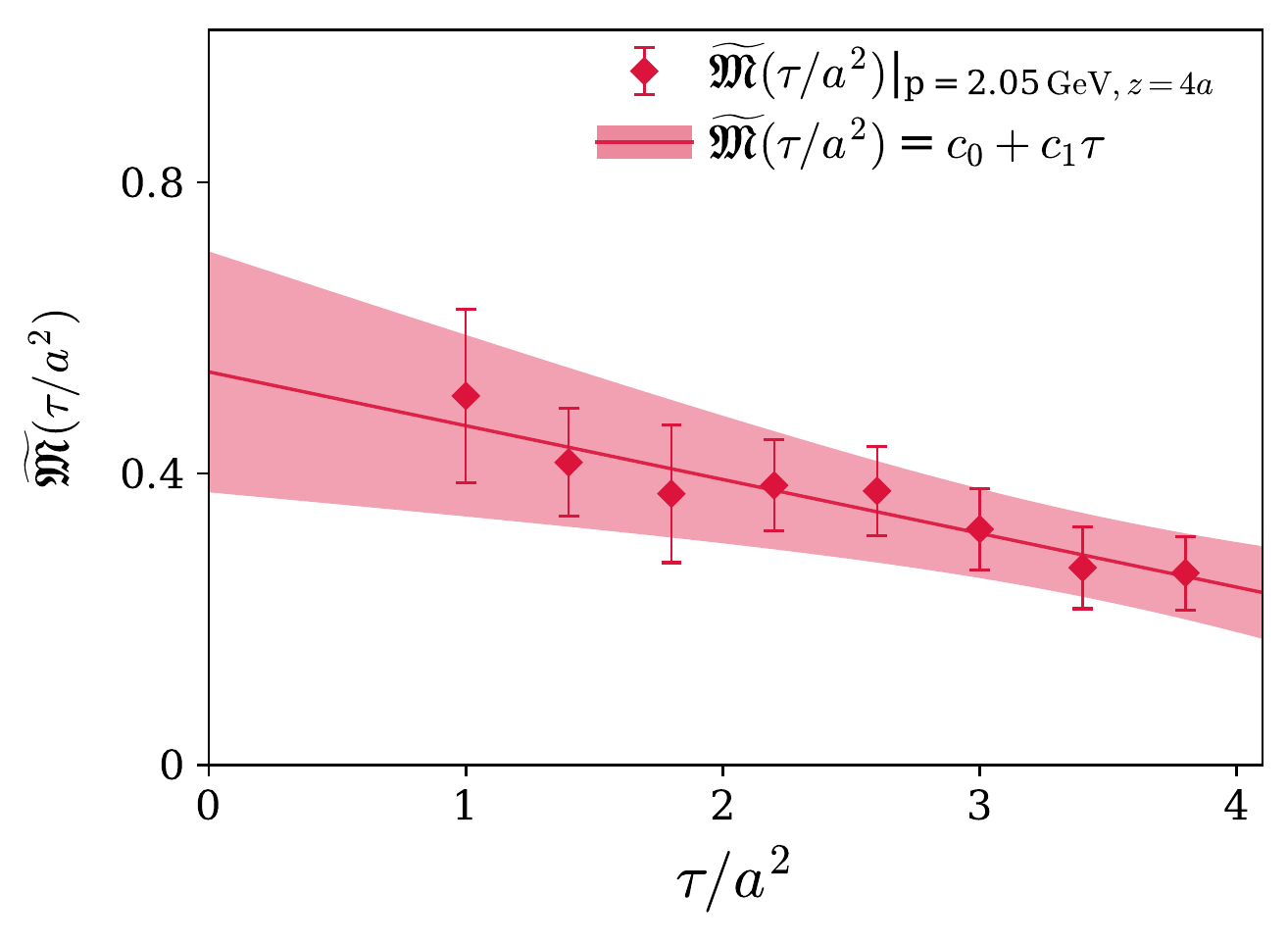}
\includegraphics[scale=0.6]{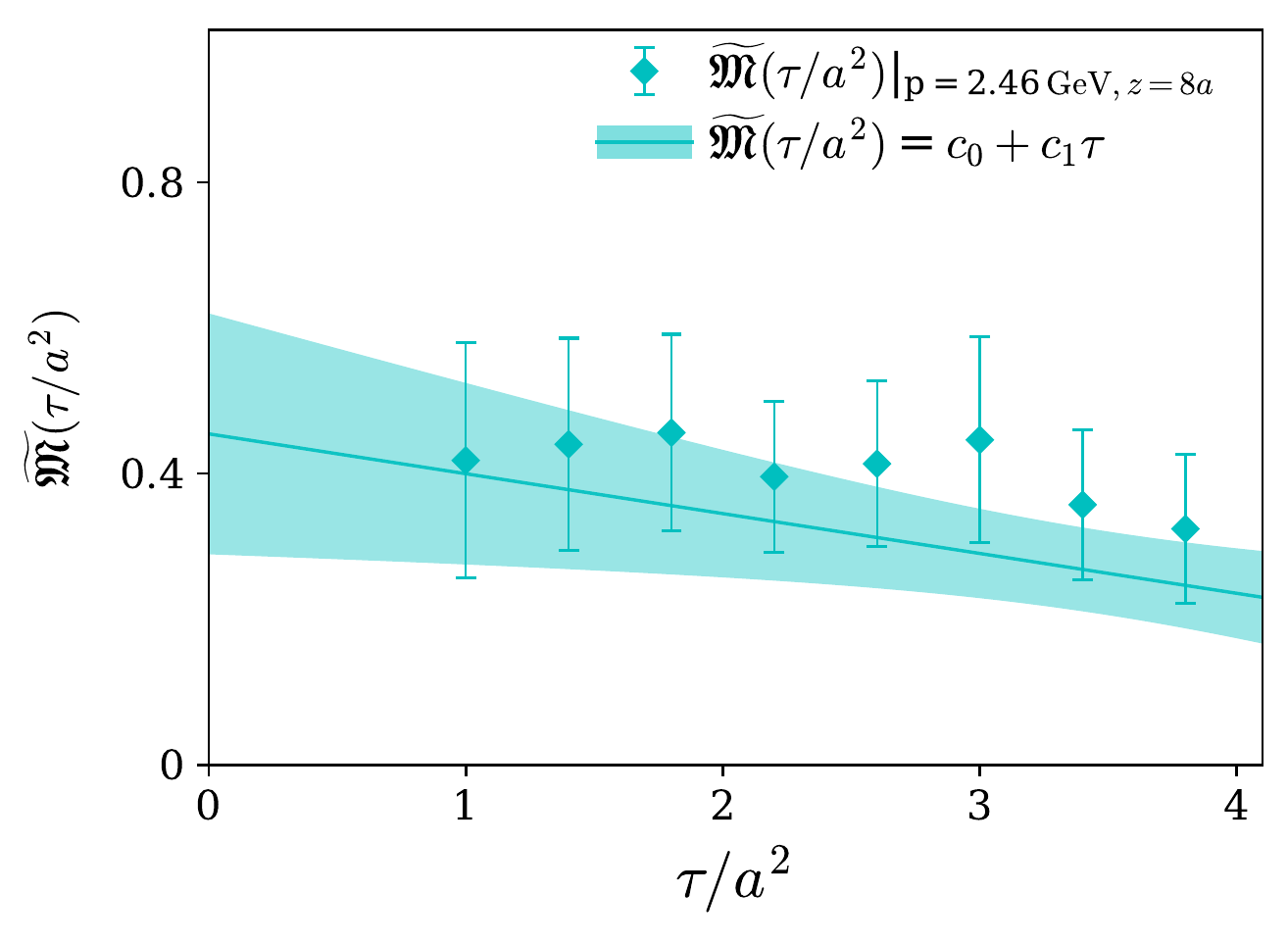}

\caption{\label{fig:tauextrapolation}  Arbitrary examples of reduced matrix elements $\wt{\mathfrak{M}} (\tau/a^2)$ extrapolated to the $\tau \to 0$ limit for different nucleon momenta and  field separations. The functional form used to fit the reduced matrix elements is $\wt{\mathfrak{M}} (\tau/a^2) = c_0 + c_1 \tau \,$. The top-left panel shows the fit for $p = 1 \times \frac{2 \pi}{a L}$ = 0.41 GeV and $z = 3a $. The top-right panel shows the fit for $p = 3 \times \frac{2 \pi}{a L}$ = 1.23 GeV and $z = 8a $. The bottom-left panel shows the fit for $p = 5 \times \frac{2 \pi}{a L}$ = 2.05 GeV and $z = 4a $. The bottom-right panel shows the fit for $p = 6 \times \frac{2 \pi}{a L}$ = 2.46 GeV and $z = 8a $.}
\eefs{mockdemocn}

\begin{figure}[!htb]
\center{\includegraphics[scale=0.7]{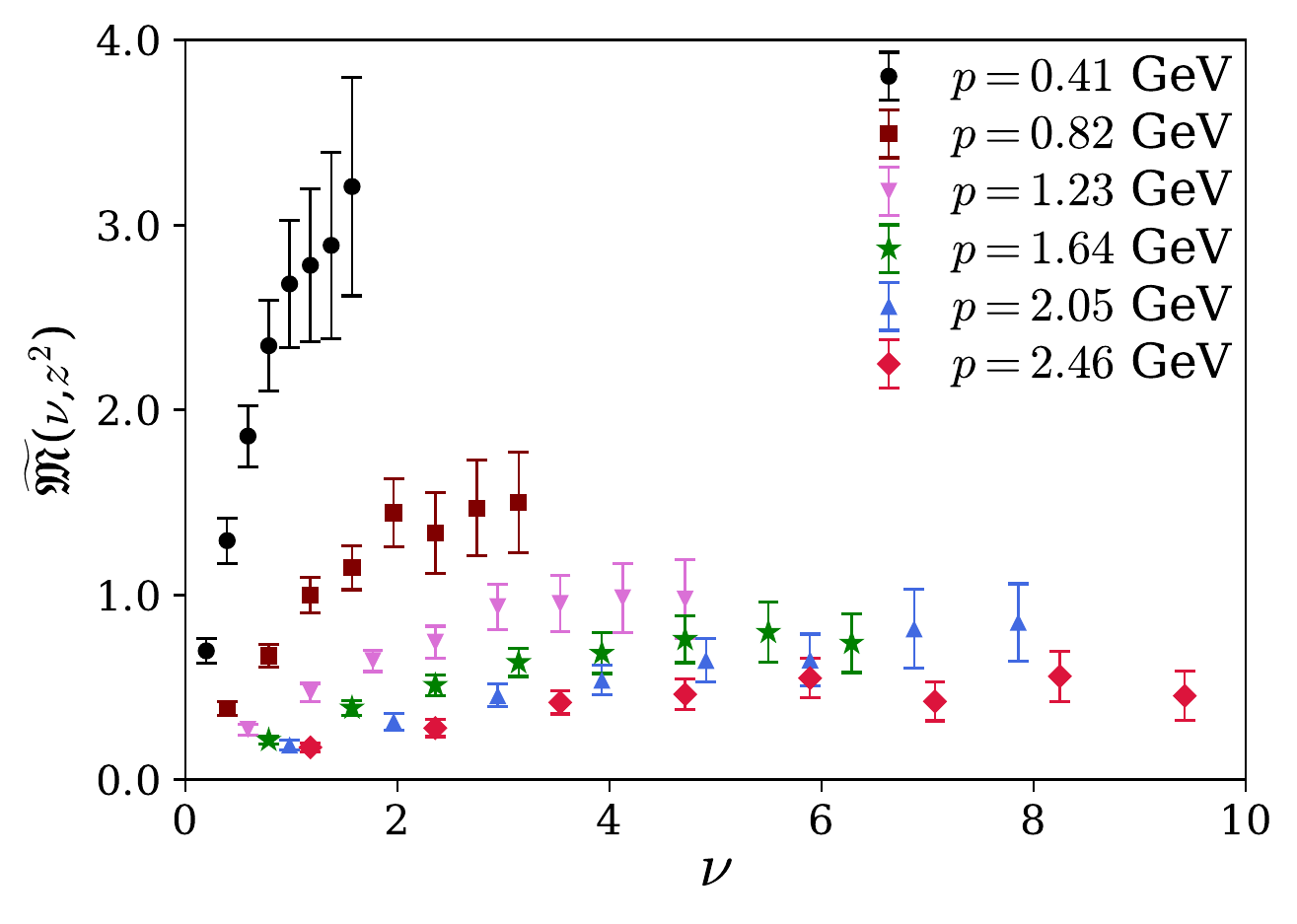}}
\caption{\label{fig:pseudo-rITDfinal} The reduced Ioffe-time pseudo-distribution $\wt{\mathfrak{M}} (\nu, z^2)$ in the zero flow-time limit.} 
\end{figure}



\section{ Isolating the gluon helicity  Ioffe-time pseudo-distribution} \label{sec:rITDextra}

In this section, we use two methods to correct for the ${\cal O}(m_p^2/p_z^2)$ term to  determine the  gluon helicity  Ioffe-time pseudo-distribution. As can be seen in Fig.~\ref{fig:pseudo-rITDfinal}, the ${\cal O}(m_p^2/p_z^2)$ terms give a substantial contribution to the matrix elements for the kinematics available to a lattice calculation. In Sec.~\ref{sec:methodI},  the $\nu$-dependence of  $[\wt{\mathcal{M}}_{sp}^+ -\nu  \wt{\mathcal{M}}_{pp}]$ and the correction from the ${\cal O}(m_p^2/p_z^2)$ contribution are modeled to isolate  the gluon helicity pseudo-distribution. In Sec.~\ref{sec:sub}, a rest-frame matrix element is subtracted to correct for the dominant ${\cal O}(m_p^2/p_z^2)$ contamination term.

\subsection{Method-I: 
fits using moments}\label{sec:methodI}

From Eq.~\eqref{eq:Ipform2}, we see that at a fixed value of $\nu$ and large $p_z$, we can approximate 
\bea \label{eq:limitITD}
\mathfrak{\wt{M}}(\nu,z^2)  \approx  [\wt{\mathcal{M}}_{sp}^+(\nu,z^2) -\nu  \wt{\mathcal{M}}_{pp}(\nu,z^2)] \ ,
\eea
and it is the $\nu$-dependence of  $[\wt{\mathcal{M}}_{sp}^+ -\nu  \wt{\mathcal{M}}_{pp}]$ that governs the $x$-dependent gluon helicity distribution. Therefore, our goal in this section is to estimate the size of the ${\cal O}(m_p^2/p_z^2)$ contamination and try to eliminate the contamination effects from the  matrix elements used to define the  reduced pseudo-ITD. 

In Eq.~\eqref{eq:Ipform2},  $\wt{\mathcal{M}}_{sp}^+$ is an odd function of $\nu$ and $\wt{\mathcal{M}}_{pp}$ is an even function of $\nu$. We therefore write these amplitudes in terms of odd and even moments that can describe the data in the accessible Ioffe-time region, and thereby parametrize the lattice data for the  reduced pseudo-ITD in Eq.~\eqref{eq:rITDdef} using the following fit form:
\bea \label{eq:ITDfit}
\wt{\mathfrak{M}}(\nu) = \sum_{i=0} \frac{(-1)^i}{(2i+1)!}a_i \nu^{2i+1} +  \nu \frac{m_p^2}{p_z^2} \sum_{j=0} \frac{(-1)^j}{(2j)!}b_j \nu^{2j} \ , 
\eea
where the coefficients $a_i$  are the Mellin moments of the pseudo-distribution related to $[ \wt{\mathcal{M}}_{sp}^+(\nu,z^2) -\nu  \wt{\mathcal{M}}_{pp}(\nu,z^2)]$, which can be related to the moments of the PDF~\cite{Karpie:2018zaz}, while the $b_i$ are those for $\wt{\mathcal{M}}_{pp}(\nu,z^2)$.

\begin{table*}
  \centering
  \setlength{\tabcolsep}{3pt}
  \renewcommand{\arraystretch}{1.5}
  \begin{tabular}{cccccccccc}
  \toprule
    Fit &  $a_0$ & $a_1$ & $b_0$ &  $b_1$ & ${\rm cov}[a_0,a_1]$ & ${\rm cov}[a_0,b_0]$ & ${\rm cov}[a_0,b_1]$ &  ${\rm cov}[a_1,b_0]$  &  ${\rm cov}[a_1,b_1]$  \\
    \midrule
    Fit-1 & $0.051(13)$ & $0.0058(11)$ & 0.362(22) & - & $7.34\times10^{-6}$ & $0.00019$ & - & 4.484231574  & - \\
    Fit-2 & 0.061(14) & 0.0043(20) & 0.371(31) & 0.0066(33) & $5.50\times 10^{-6}$ & -0.00032 & $4.41\times 10^{-6}$ & $-1.72\times 10^{-5}$ & $-5.44\times 10^{-6}$ \\
    \bottomrule
  \end{tabular}
\caption{Fitted parameters from the fits to the reduced pseudo-ITD through fits~\eqref{eq:ITDfit} using moments. The covariance among different fit parameters are listed as ${\rm cov}[a_i,b_j]$.  The two different fits are labeled by Fit-1 and Fit-2 as described in the text.}\label{tab:fitextrapol}
\end{table*}

We only have lattice data in a limited $\nu$-domain, so it is only possible to determine the first few moments with good accuracy in the fit of  Eq.~\eqref{eq:ITDfit}. 
 We perform two fits labeled 
 by ``Fit-1" and ``Fit-2,"   and obtain the fit parameters listed in Table~\ref{tab:fitextrapol}.
First,  we fit the  reduced pseudo-ITD lattice data for all the available $z$ and $p_z$ using $i=0,1$ and $j=0$ in the parametrization Eq.~\eqref{eq:ITDfit} and call this fit Fit-1. 
Table~\ref{tab:fitextrapol} demonstrates that the coefficient $b_0$ is much larger than $a_0$.
In this sense, the data are dominated by the ${\cal O}(m_p^2/p_z^2)$ contamination term. The  relative smallness of $a_0$ also means 
that $[\wt{\mathcal{M}}_{sp}^+ -\nu  \wt{\mathcal{M}}_{pp}]$
is a small difference, determined by the subtraction of much larger functions 
$\wt{\mathcal{M}}_{sp}^+ $ and $\nu  \wt{\mathcal{M}}_{pp}$, within the range of $\nu$ spanned by our data.

Next, we try to incorporate another moment $b_1$ in the fit using $j=0,1$,  and find that the second moment does not result in any significant value within error. We therefore use the fit parameters $a_0,a_1,b_0$ as the prior for the Fit-2 and obtain the fit results listed in Table~\ref{tab:fitextrapol}. The  smaller domain of Ioffe time data within the fixed $z$-range and in particular, smaller $p_z$-values are  not sufficient to constrain higher moments in the fits without other prior information. We compare the results of these two fits to the lattice data in the top panel of Fig.~\ref{fig:mpcfit} with $\chi^2/{\rm d.o.f.}$ shown. These two fits are consistent within uncertainty and reproduce the lattice data. We have excluded $z=a$ and $8a$ data from the fits to get a smaller $\chi^2/{\rm d.o.f.}$.   The $z=a$ matrix element potentially has significant discretization errors ${\cal O}(\frac az)$ while the $z=8a$ matrix elements could have significant higher twist contamination ${\cal O}(z^2 \Lambda_{\rm QCD}^2)$. The improved $\chi^2/{\rm d.o.f.}$ when these data are neglected support these possibilities. In future studies with increased precision and range of $\nu$, these systematic errors could be modeled as well. For this proof of principle study, the previously mentioned cuts on the data are made. The bottom panel of the figure shows the extrapolation of the lattice data in the limit of zero ${\cal O}(m_p^2/p_z^2)$ contamination term contribution within the fit parametrization. This is simply the fit band constructed using 
\bea \label{eq:ITDfit2}
\wt{\mathfrak{M}}(\nu) = \sum_{i=0} \frac{(-1)^i}{(2i+1)!}a_i \nu^{2i+1}
\eea
from the above fits. From both of the fits, it is seen that
the ${\cal O}(m_p^2/p_z^2)$ term  is dominated by the first moment $b_0$ of $\wt{\mathcal{M}}_{pp}$. While this might be a consequence of our inability to constrain higher moments and choice of the fit forms, this outcome will indeed become clearer  in the following subsection, where we correct for the target mass correction in a model-independent manner. From both of these methods, one might argue that within the Ioffe-time region $\nu \approx [0,9]$, the fit bands in the bottom panel of Fig.~\ref{fig:mpcfit} are good representations of the gluon helicity reduced pseudo-ITD. We note that to determine the normalization of the gluon PDF according to Eq.~\eqref{eq:matching}, we need to normalize the results by  the gluon momentum fraction.  The calculation of the gluon momentum fraction at finite flow time, however, must be matched to the $\overline{\rm MS}$-scheme. The flow time both controls the signal-to-noise ratio and characterizes the energy scale of the matrix elements. To control the noise-to-signal ratio, we use a flow time that corresponds to a non-perturbative scale. The gluon momentum fraction, evaluated at this scale, would require non-perturbative evolution to higher scales before it can be matched to the $\overline{\rm MS}$-scheme. This non-perturbative evolution is currently unknown and is a subject of future investigation. Therefore, we did not calculate the gluon momentum fraction on this particular ensemble in our previous work~\cite{HadStruc:2021wmh}. For the  overall normalization of the reduced pseudo-ITD, 
we take the result from~\cite{Alexandrou:2020sml}, which is $\langle x \rangle_g$=0.427(92), and apply this normalization to the ${\cal O}(m_p^2/p_z^2)$ 
contamination-corrected gluon helicity  pseudo-ITD  in Figs.~\ref{fig:mpcfit}. 

\befs 
\centering

\includegraphics[scale=0.6]{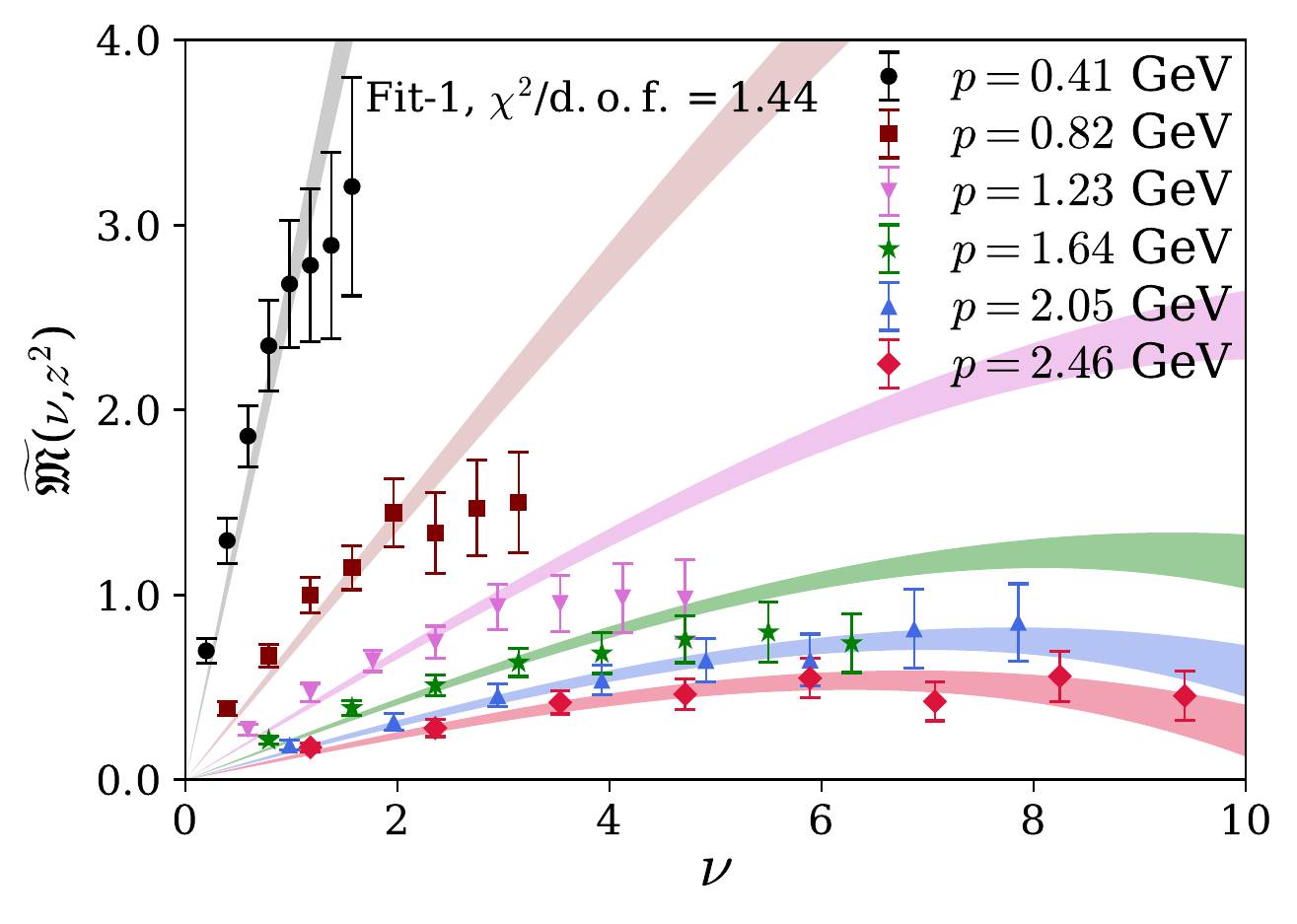}
\includegraphics[scale=0.6]{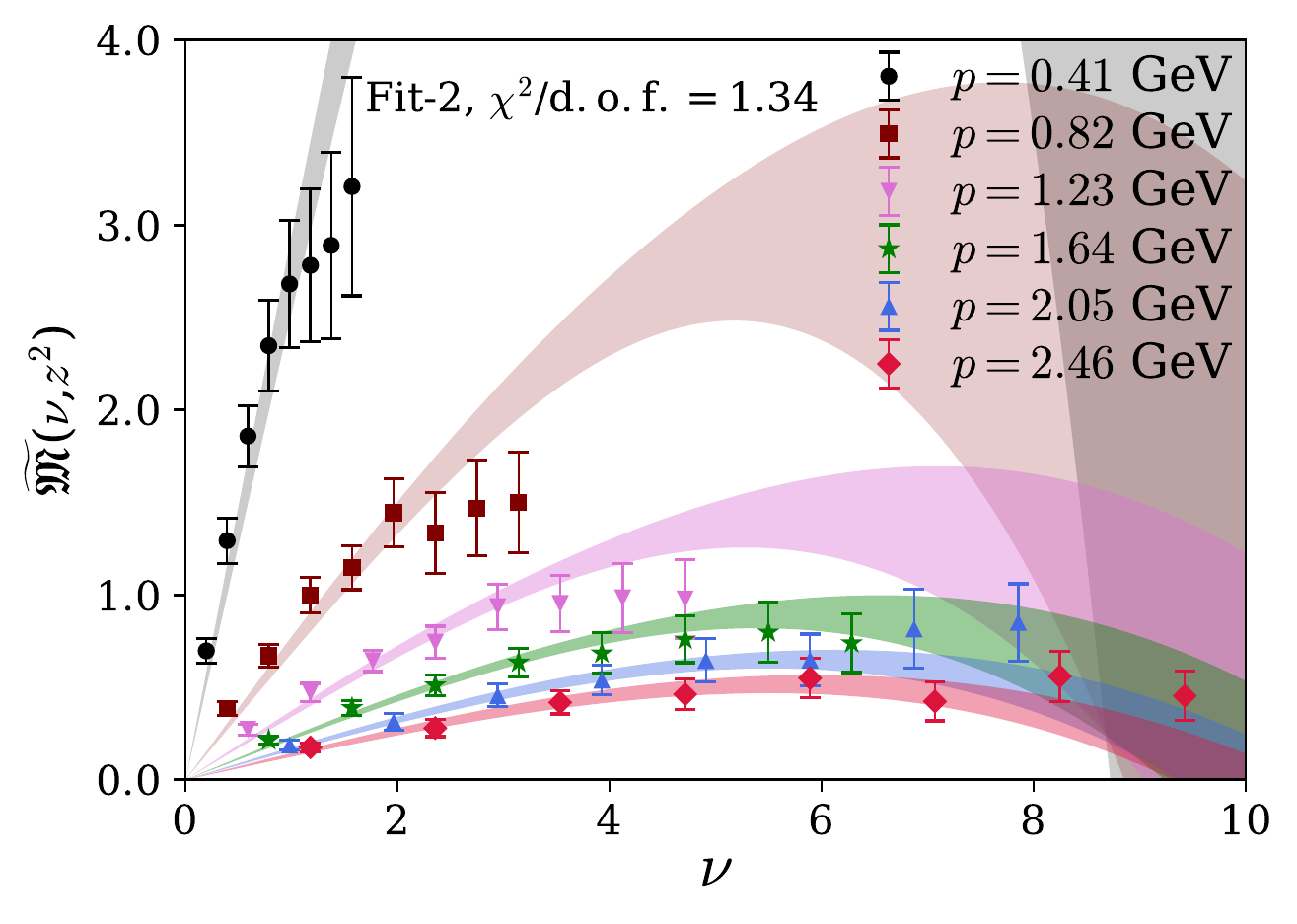}

\includegraphics[scale=0.6]{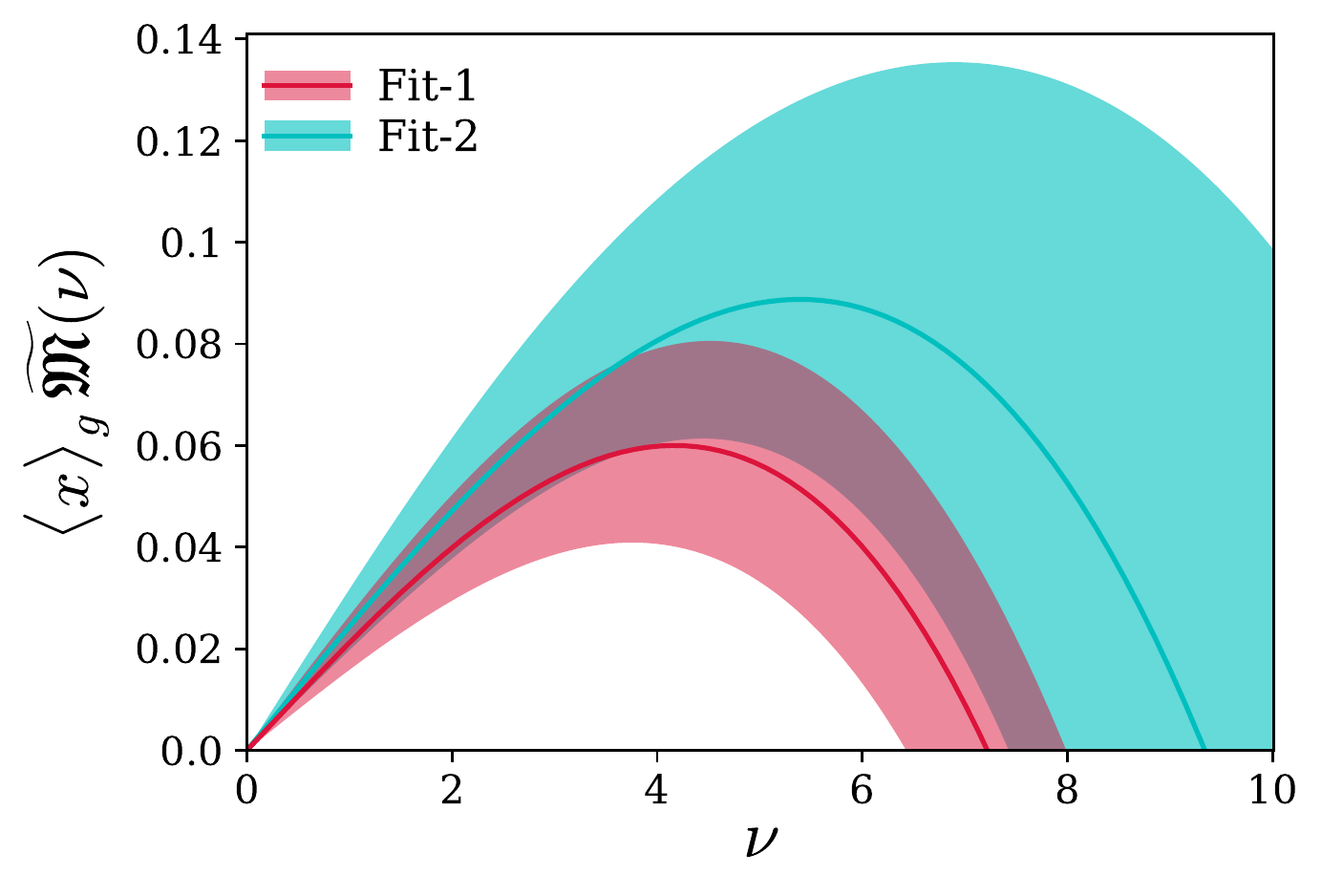}

\caption{\label{fig:mpcfit}  Simultaneous fit to the gluonic matrix elements at all momenta used in this calculation. The lattice data points in the upper panel are the reduced pseudo-ITD in the zero flow time limit and the fitted bands that describe the lattice data points are generated using the fit parameters listed in Table~\ref{tab:fitextrapol}. After correcting for the ${\cal O}(m_p^2/p_z^2)$ contamination term in the matrix element, the desired reduced pseudo-ITDs associated with the gluon helicity distribution from both fits are shown in the lower panel. For an appropriate comparison of the magnitude of these extrapolations, the fitted pseudo-ITD bands in the bottom panel are normalized by the gluon momentum fraction, $\langle x\rangle_g$ from~\cite{Alexandrou:2020sml}. }
\eefs{mockdemocn}

\subsection{Method-II:  subtraction using $p_z=0$ matrix element} \label{sec:sub}

As discussed, the ${\cal O}(m_p^2/p_z^2)$ contamination term   dominates the original matrix element for modest values of momenta accessible in this calculation. A natural consideration would be to eliminate this contamination, which would be possible  if we could  perform a separate 
measurement of the $ \wt{\mathcal{M}}_{pp}(\nu, z^2)$ invariant amplitude.
While we cannot do this, \eqn{pseudo_Ip} allows  access to  $\nu=0$ value of $ \wt{\mathcal{M}}_{pp}(\nu, z^2)$, i.e. to the first term in the 
Taylor expansion of $ \wt{\mathcal{M}}_{pp}(\nu, z^2)$ over $\nu$. 
Indeed, taking $p_z=0$ in \eqn{pseudo_Ip}, it follows 
\bea
\label{eq:pz0val}
 [\wt{M}_{0i;0i} + \wt{M}_{ij;ij}](z,p_z=0) = 2 m_p^3 z 
 \wt{\mathcal{M}}_{pp}(\nu=0, z^2).
\eea
Now, we define a ``subtracted'' matrix element 
\be
\label{eq:submel}
 [\wt{M}_{0i;0i} + \wt{M}_{ij;ij}]_{\rm sub}(z,p) \equiv [\wt{M}_{0i;0i} + \wt{M}_{ij;ij}](z,p) - \frac{p_0}{m_p} [\wt{M}_{0i;0i} + \wt{M}_{ij;ij}](z,0)\, , 
\ee
which vanishes for $p_z=0$. Dividing the subtracted matrix element  by $(-2p_zp_0)$ and introducing 
\bea
\label{eq:subcalm}
\wt{\mathcal M}_{\rm sub}(z,p) \equiv (-2p_0\, p_z)^{-1}\left[\wt{M}_{0i;0i} +\wt{M}_{ij;ij}\right]_{\rm sub}(z,p) \, , 
\eea
we derive  the representation 
\bea
\label{eq:subrep}
\wt{\mathcal M}_{\rm sub}(z,p_z) = & \widetilde{\mathcal{M}}_{sp}^{(+)}(\nu,z^2)  -\nu  \widetilde{\mathcal{M}}_{pp}   (\nu,z^2)
  -\nu \frac{ m_p^2  }{p_z^2} \left [  \widetilde{\mathcal{M}}_{pp}(\nu,z^2)    -  \wt{\mathcal{M}}_{pp}(\nu=0, z^2)
 \right ]   \ . 
\eea
Although the subtracted representation still contains an ${\cal O}(m_p^2/p_z^2)$ contamination term, it now contains $\mathcal{M}_{pp}$
in a subtracted $[\widetilde{\mathcal{M}}_{pp}(\nu,z^2)-\wt{\mathcal{M}}_{pp}(\nu=0, z^2)]$ form, the Taylor expansion (in $\nu$) of which starts with $\nu^2$ and is accompanied by the coefficient $b_1$. As Fit-2 suggests that this coefficient is very small, we  expect that the contamination term for  $\wt{\mathcal M}_{\rm sub}$ to be even smaller than that for  $\wt{\mathcal M}$.

Using the above prescription, we calculate $\wt{\mathcal M}_{\rm sub}(z,p_z)$ for each $z$ and $p_z$ for a given flow time and calculate the reduced pseudo-ITD using Eq.~\eqref{eq:rITDdef}. The results for $\wt{\mathcal M}_{\rm sub}(z,p_z)$ are shown as a function of flow-time in Fig.~\ref{fig:ritdallsub}. 

\befs 
\centering

\includegraphics[scale=0.6]{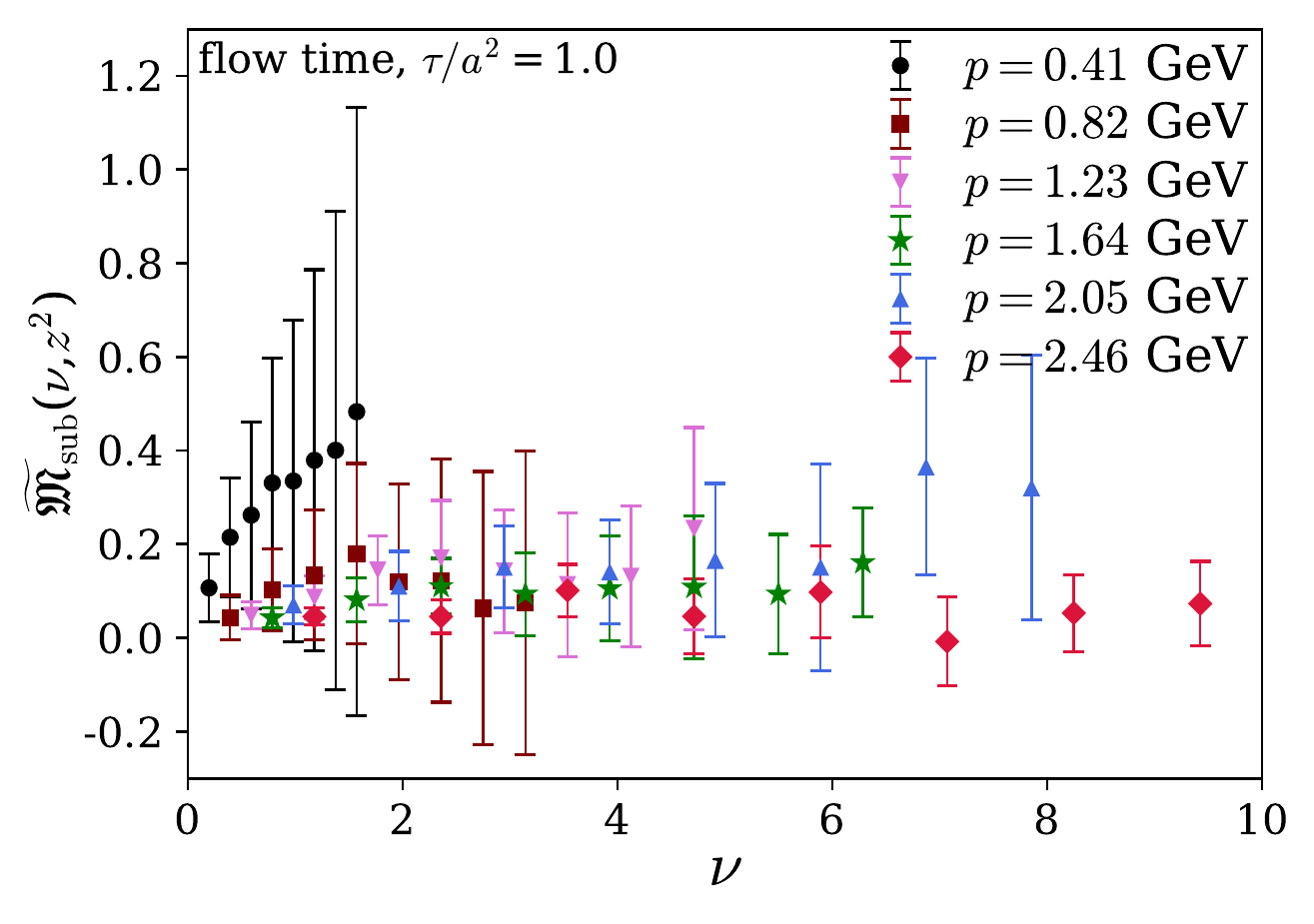}
\includegraphics[scale=0.6]{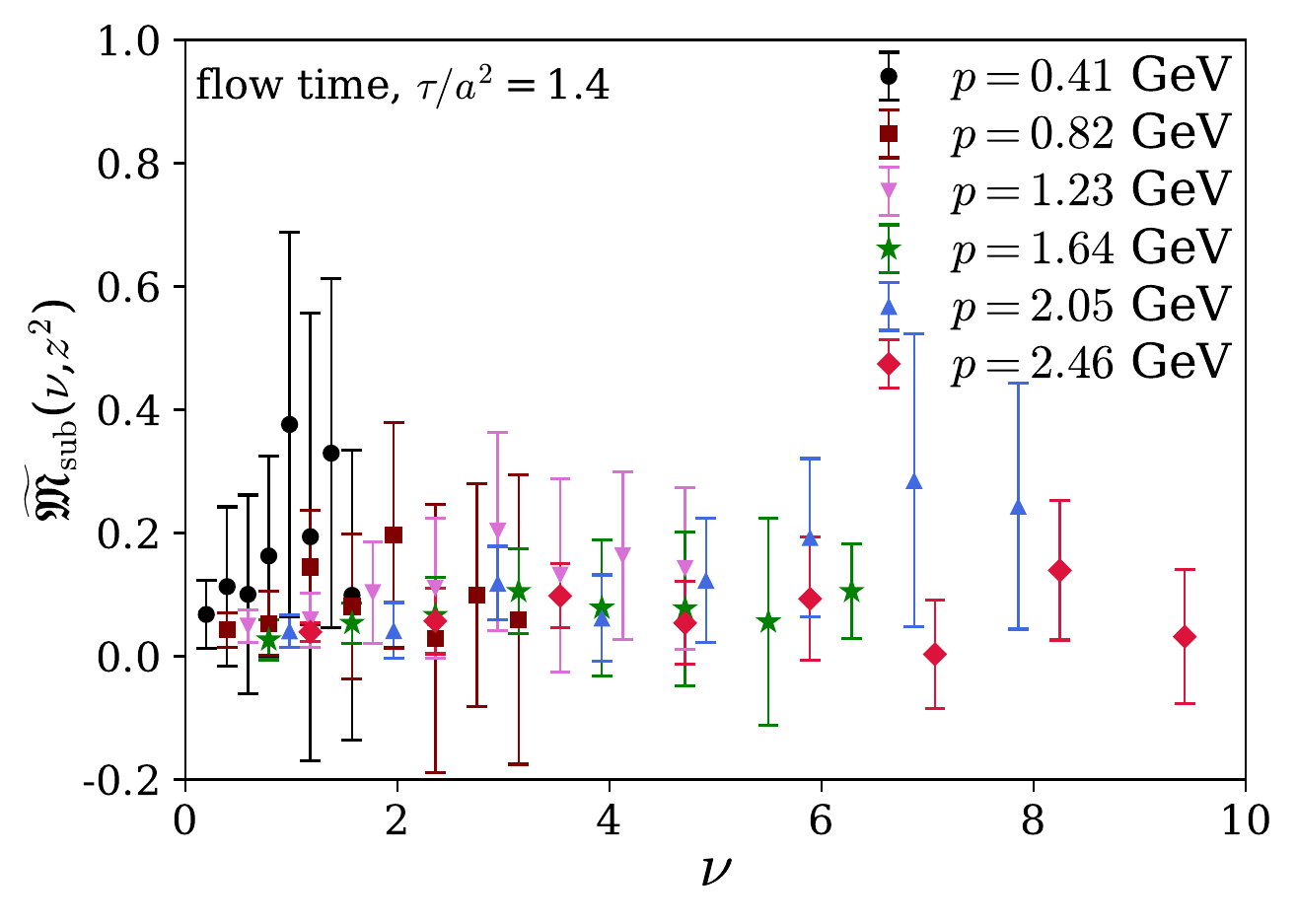}

\includegraphics[scale=0.6]{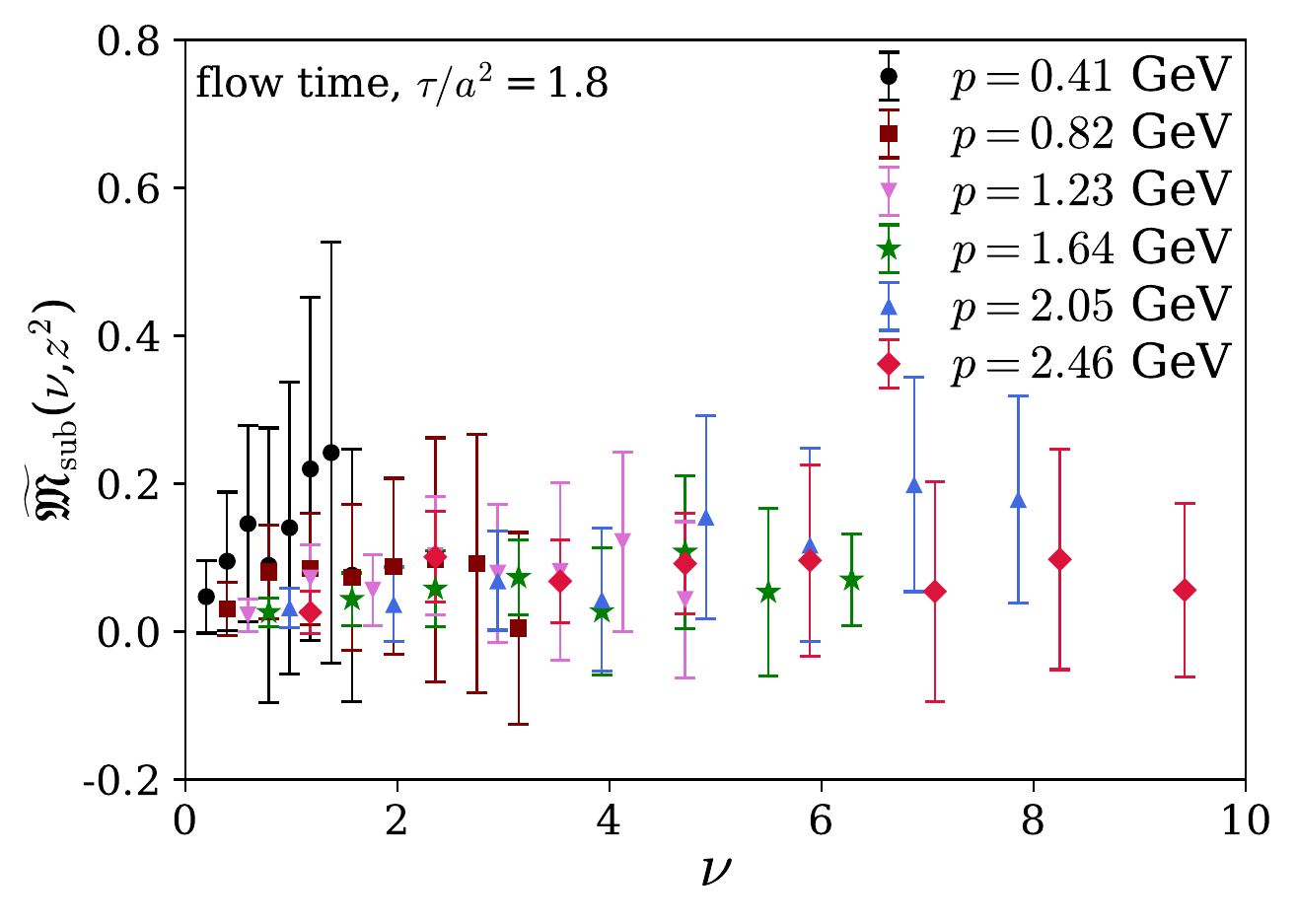}
\includegraphics[scale=0.6]{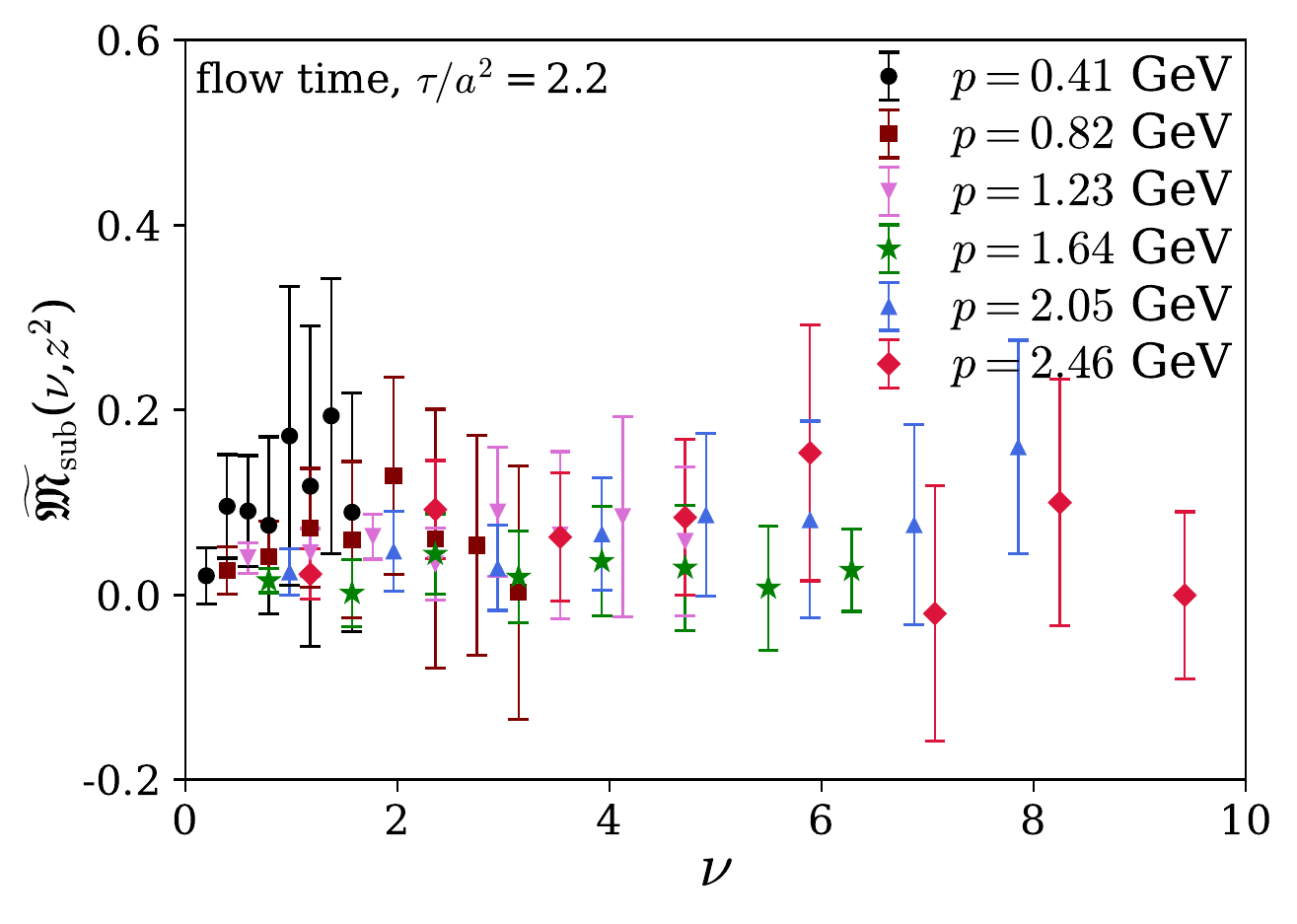}

\includegraphics[scale=0.6]{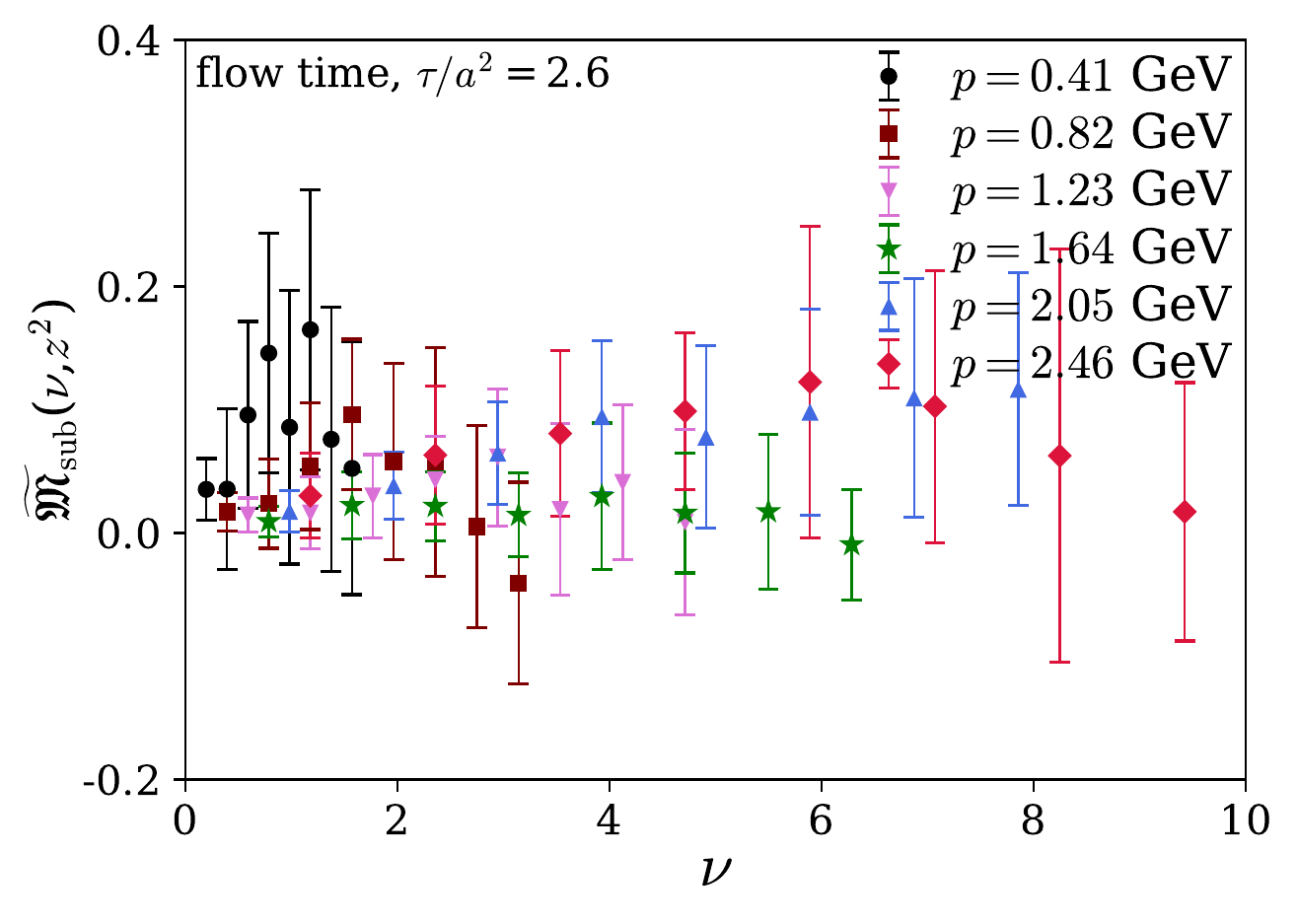}
\includegraphics[scale=0.6]{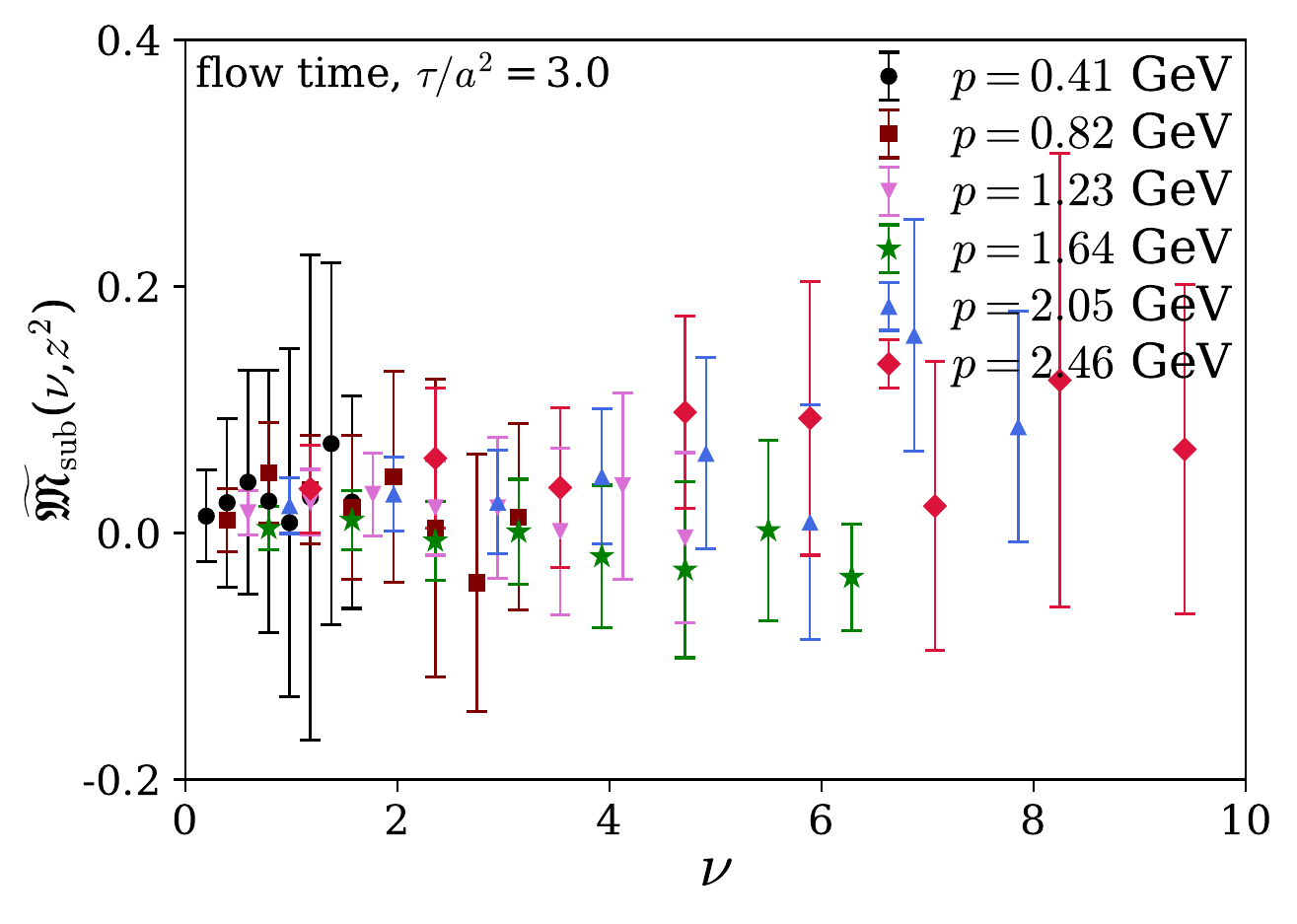}

\includegraphics[scale=0.6]{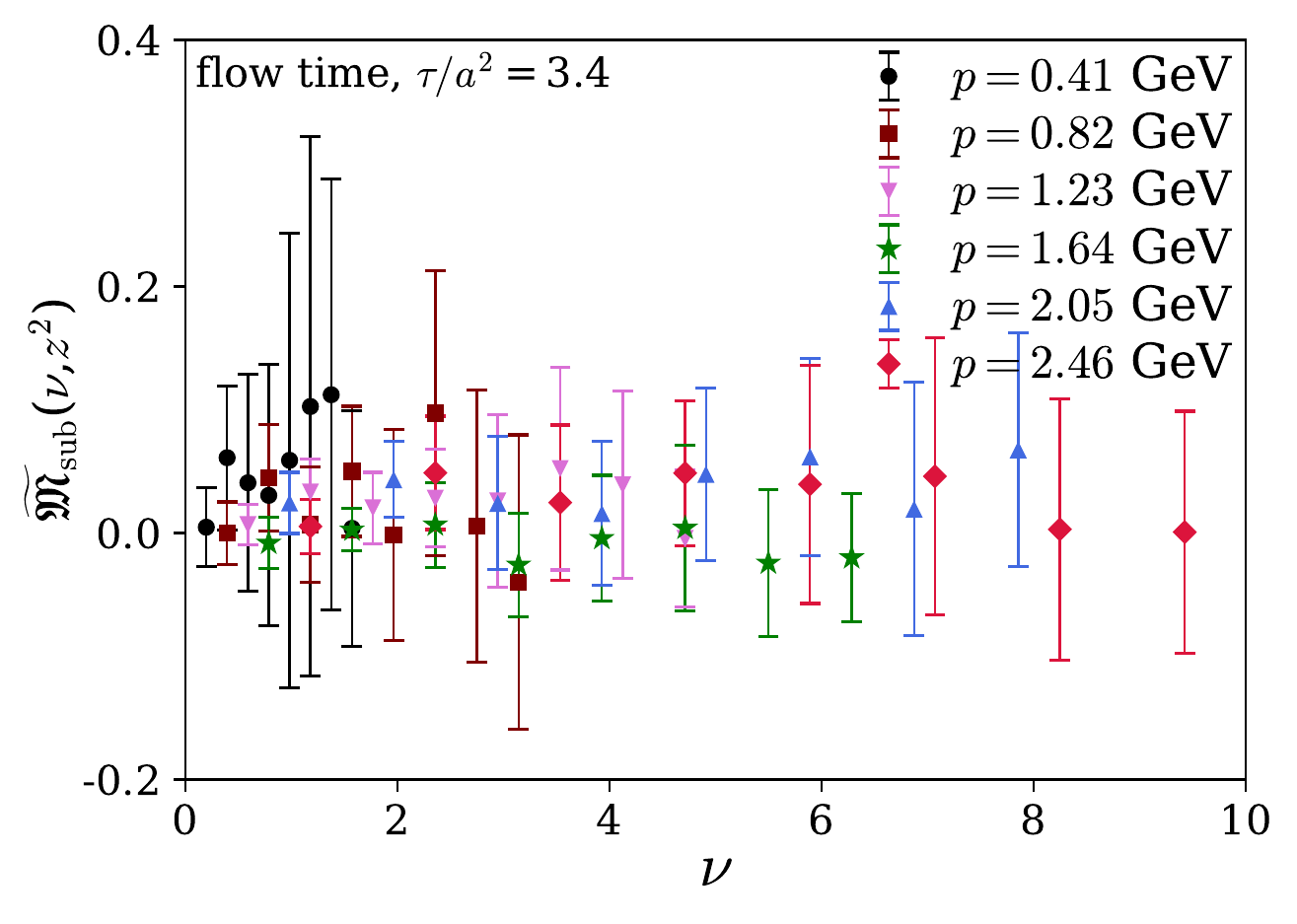}
\includegraphics[scale=0.6]{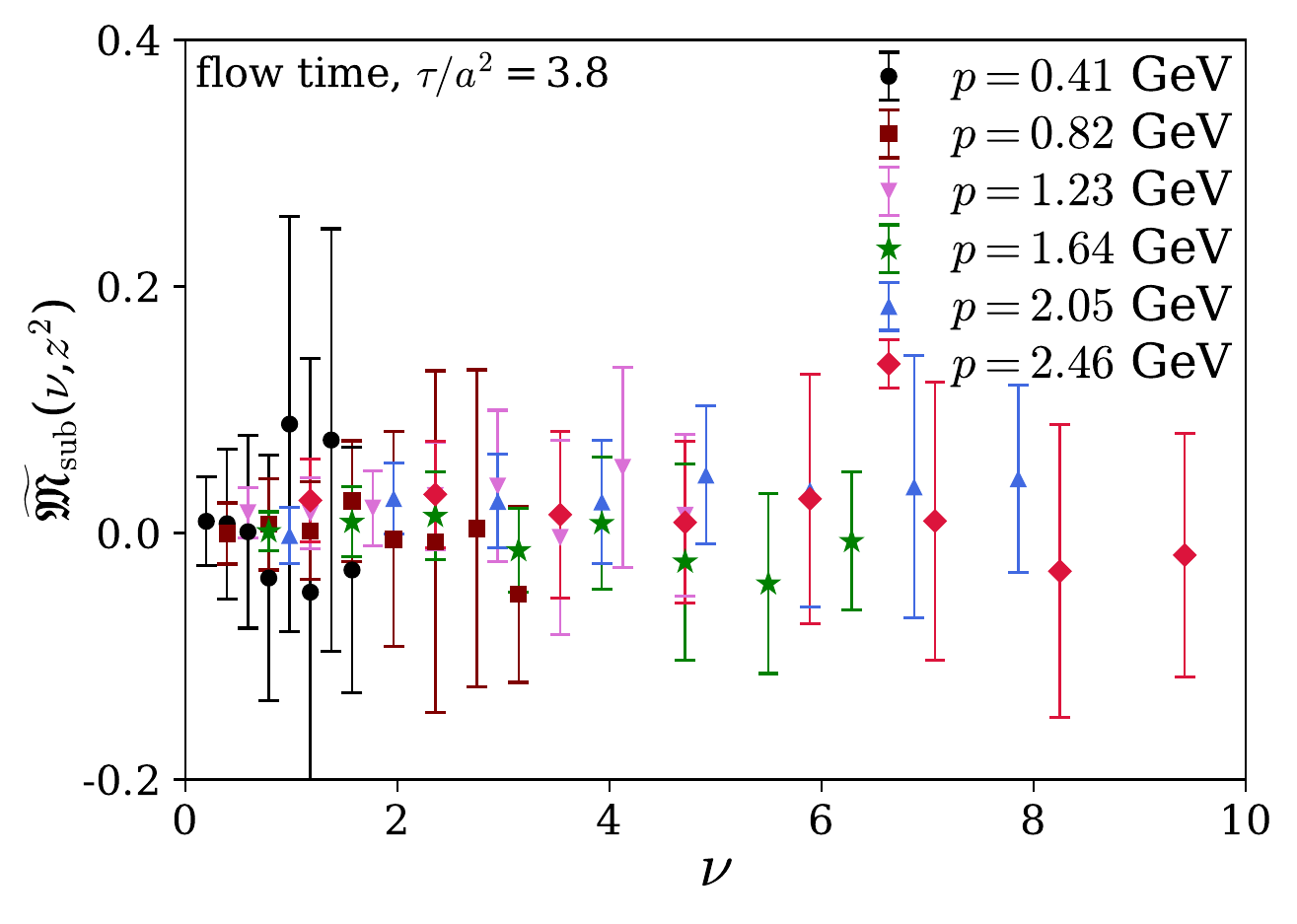}

\caption{\label{fig:ritdallsub} The reduced matrix elements, $\wt{\mathfrak{M}}_{\rm sub} (\nu, z^2)$ with respect to the Ioffe-time for different flow times. Starting from  the top-left to bottom-right panels,  the reduced matrix elements are shown for $\tau/a^2 =$ 1.0 to 3.8.}
\eefs{mockdemocn}

In a manner  similar to the procedure  described in Sec.~\ref{sec:zero-flowtime},  we determine $\wt{\mathfrak{M}}_{\rm sub} (\nu, z^2)$  in the zero flow-time limit and illustrate the results in the left and the
right panels of Fig.~\ref{fig:pseudo-rITDsub}. 
Fig.~\ref{fig:pseudo-rITDsub} shows that the data for
$\wt{\mathfrak{M}}_{\rm sub}(\nu,z^2)$ at the lowest momenta  have a factor of ten reduction compared to those for $\wt{\mathfrak{M}}(\nu,z^2)$. This means that the proposed subtraction has strongly decreased  the
magnitude of the contamination term. We note  the resulting 
statistical error at the smallest momentum is larger than at 
the higher momenta  due to the factor of $(p_z)^{-1}$ in 
the definition in Eq~\eqref{eq:subcalm}. In the left panel of
Fig.~\ref{fig:pseudo-rITDsub}, we compare the lattice data with 
the pseudo-ITD obtained by fits as we discussed previously in 
the bottom panel of Fig.~\ref{fig:mpcfit}. We can clearly 
see that the expectation for the ITD based on method-1 agrees
quite well with the data for ITD obtained via method-2, thus serving  as a cross-check of the two 
methods. In the right panel of Fig.~\ref{fig:pseudo-rITDsub}, we 
show the result of fits to $\wt{\mathfrak{M}}_{\rm sub}(\nu,z^2)$, using 
the functional form in 
Eq.~\eqref{eq:ITDfit} with the parameter $b_0=0$ and $a_0,a_1$ and $b_1$ left to be fit. Fig.~\ref{fig:pseudo-rITDsub} shows that
the effect of the residual correction after subtraction, as given by 
the $b_1\nu^3$ term, is rather minimal, thus explaining  why the 
subtracted data has an improved universal behavior with respect to
$\nu$.

For a comparison between the two different methods of treating the ${\cal O}(m_p^2/p_z^2)$ contamination term and determination of the gluon helicity  Ioffe-time pseudo-distribution, we plot the fit bands,  Fit-1 and Fit-2 from the analyses in Sec.~\ref{sec:methodI}, referred to as Method-I in Fig.~\ref{fig:pseudo-rITDsub}. The agreement between the two methods to correct for the ${\cal O}(m_p^2/p_z^2)$ contamination  demonstrates the consistency between  fitting using moments and the subtraction of the 
rest-frame 
$\wt{\mathcal{M}}_{pp}(z,\nu=0)$ contribution, bolstering confidence in  our determination of the gluon helicity pseudo Ioffe-time distribution.  Again,  we must normalize  the matrix element by the gluon momentum fraction from~\cite{Alexandrou:2020sml} and we apply this normalization to the contamination-corrected gluon helicity  pseudo-ITD  in Fig.~\ref{fig:pseudo-rITDsub}. 

\befs 
\centering

\includegraphics[scale=0.6]{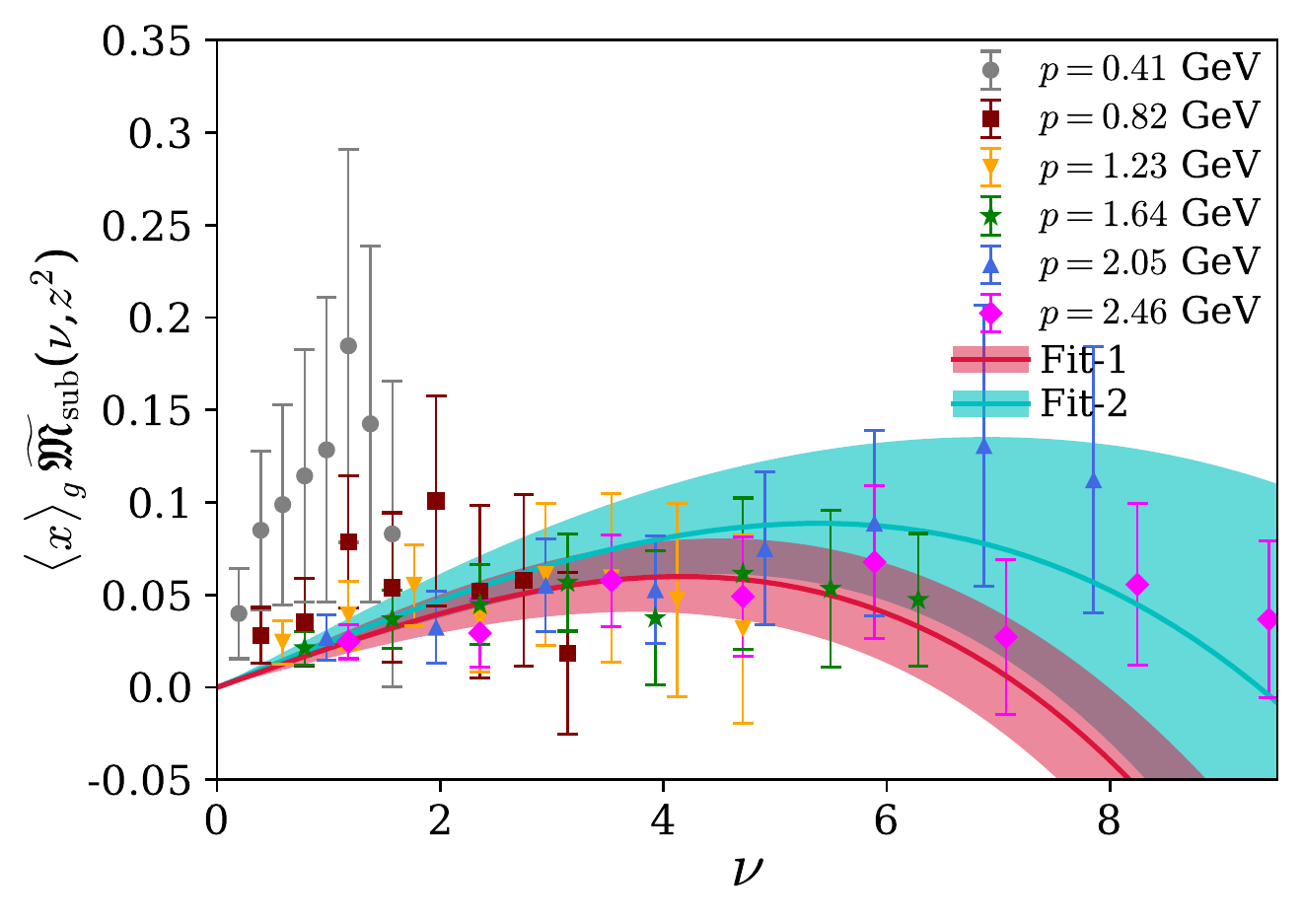}
\includegraphics[scale=0.6]{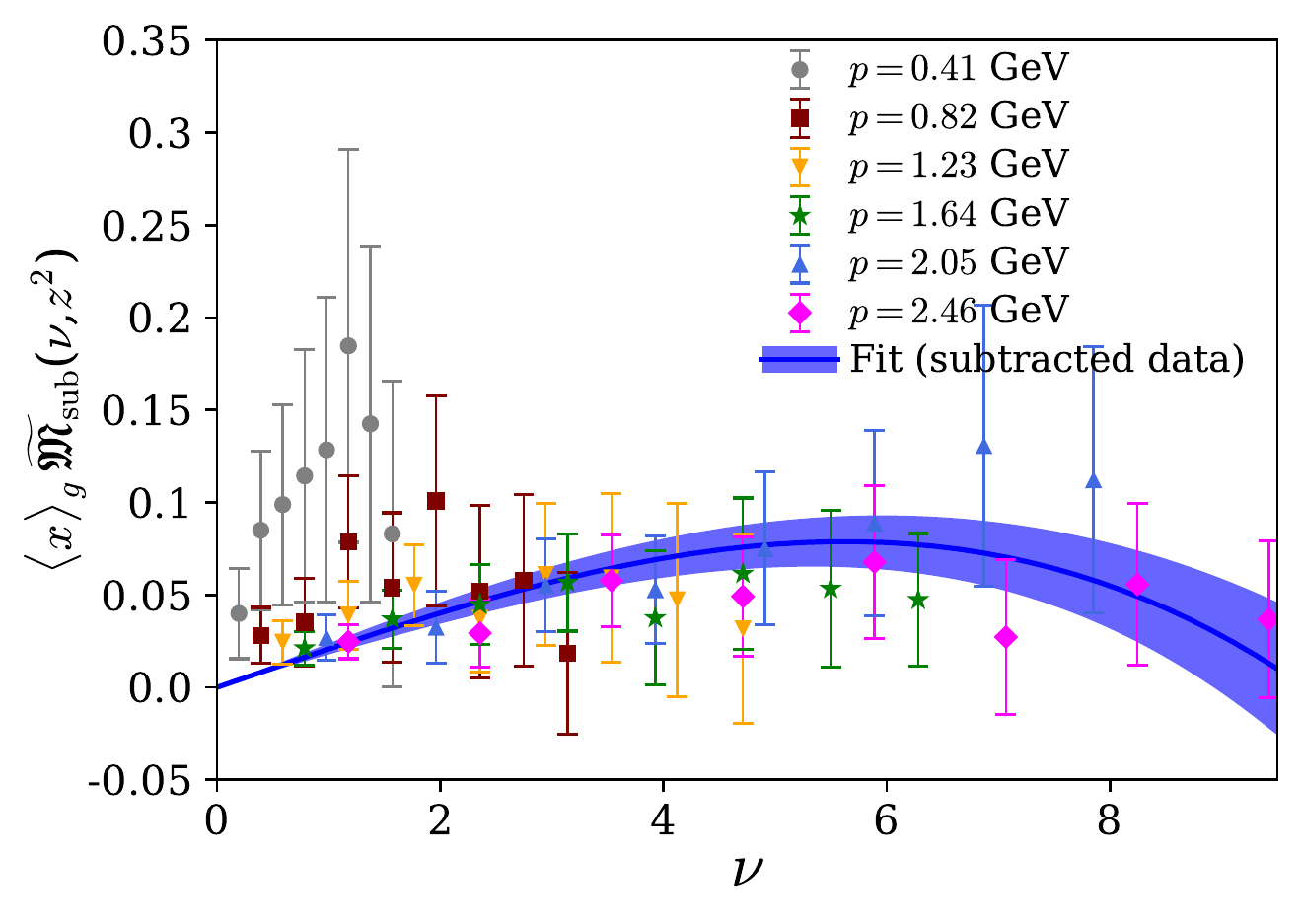}

\caption{\label{fig:pseudo-rITDsub} The lattice data points represent the reduced Ioffe-time pseudo-distribution, $\wt{\mathfrak{M}} (\nu, z^2)$ in the zero flow-time limit obtained through the subtraction method using $p=0$ matrix elements. The lattice data points and the fit bands are normalized  using the gluon momentum fraction, $\langle x\rangle_g$ from~\cite{Alexandrou:2020sml}. Left panel: the red and cyan bands represent the target mass corrected reduced Ioffe-time pseudo-distribution using the fit of moments in Sec.~\ref{sec:methodI}. 
Right panel: the blue band is a fit to the subtracted pseudo-ITD
using the  functional form in Eq.~\eqref{eq:ITDfit} with  $a_0,a_1,b_1$
as fit parameters and  $b_0=0$ fixed by construction. }
\eefs{mockdemocn}



\section{Comparison with phenomenological distribution and prospects for determining the gluon helicity PDF from lattice QCD}\label{sec:pdf_calc}

In general, determining PDFs from lattice calculations
requires the extraction of a continuous distribution  from discrete lattice data. This challenge is  compounded  by  the finite  range  of  Ioffe times accessible to lattice calculations, which limits the  number  of  data  points available. Several numerical techniques for the extraction of $x$-dependent distribution functions from lattice data have been studied, including  discrete Fourier transforms, the Backus-Gilbert method~\cite{Liang:2019frk,Karpie:2019eiq,Bhat:2020ktg,Alexandrou:2020qtt}, the Bayes-Gauss-Fourier transform~\cite{Alexandrou:2020qtt}, adapting phenomenologically-motivated functional forms for fitting lattice data~\cite{Sufian:2020vzb,Gao:2020ito,Joo:2020spy,Fan:2021bcr,Bhat:2022zrw}, parametrization  of the reduced pseudo-ITD using Jacobi polynomials~\cite{Karpie:2021pap, Egerer:2021ymv}, and finally the application of neural networks~\cite{Cichy:2019ebf,DelDebbio:2020rgv}. 

Our results for the corrected matrix elements, determined in Sec.~\ref{sec:sub}, have significant statistical uncertainties, because of the subtraction of lattice data of similar magnitudes required to remove the ${\cal O}(m_p^2/p_z^2)$ contamination. The current precision of the lattice data does not allow us to handle the inverse problem effectively using any of these methods to extract the gluon helicity PDF and gluon spin content  in the nucleon.   Most importantly, the current statistical precision prevents us from completely eliminating the contamination term from the lattice data in a model independent way. There are additional systematic uncertainties related to the truncation of fit parameters and model dependence in our extrapolated reduced ITD that we are unable to estimate with the current precision. Fig.~\ref{fig:pseudo-rITDsub} shows that we cannot observe any $z^2$-dependence in the lattice data, due to substantially larger uncertainties in the $\widetilde{\mathcal{M}}_{\rm sub}(\nu,z^2)$ data. 
Taking these considerations into account, we refrain from implementing the perturbative matching kernel in Eq.~\eqref{eq:matching} on the lattice reduced pseudo-ITD  data. Yet for demonstration purposes, we present the effect of such incomplete matching in the Appendix~\ref{appendix:matching}. At tree-level $\langle x\rangle_g\,\wt{\mathfrak{M}}(\nu)$ can be associated with the gluon helicity ITD $\wt{\mathcal{I}_p}(\nu)$ [see  Eq.~\eqref{eq:matching}]. We also neglect the contributions of the iso-scalar quark distribution.

In spite of these limitations, qualitative comparison of our results with those from global fits provide useful information and a sense of the future prospects for lattice QCD contributions to determinations  of the gluon helicity. Our results for the gluon helicity pseudo-ITD, shown in Fig.~\ref{fig:phenocomp}, are consistent with the gluon helicity ITDs extracted from global fits by the NNPDF and JAM collaborations in the $0\leq \nu \leq 9.42$ region. Moreover, our model-independent extraction of the gluon helicity pseudo-distribution over a range of Ioffe-time $\nu\lesssim 9$ hints at  a non-zero gluon spin contribution to the proton spin, as illustrated in Fig.~\ref{fig:phenocomp}.

Ref.~\cite{Alexandrou:2021mmi} demonstrated that it is possible to reconstruct PDFs using the lowest few  moments, but the two extracted moments determined in this calculation (listed in Table~\ref{tab:fitextrapol}) feature large  uncertainties. These uncertainties prohibit a proper reconstruction of the gluon helicity PDF. Nevertheless, this calculation provides the first lattice QCD estimates of the first two nonzero moments $a_0$ and $a_1$ of the gluon helicity PDF, shown in Table~\ref{tab:fitextrapol}. 


\befs 
\centering

\includegraphics[scale=0.6]{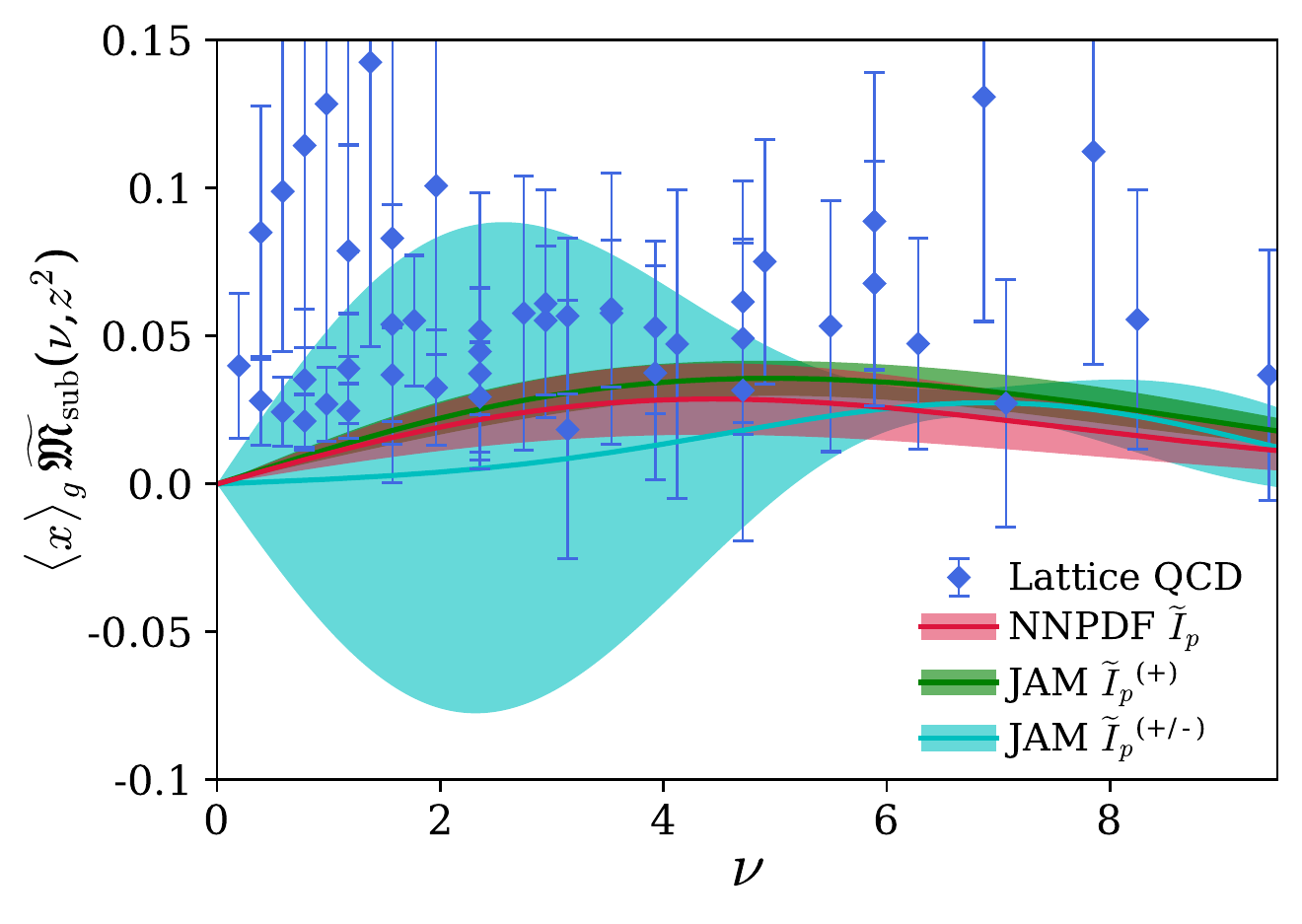}
\includegraphics[scale=0.6]{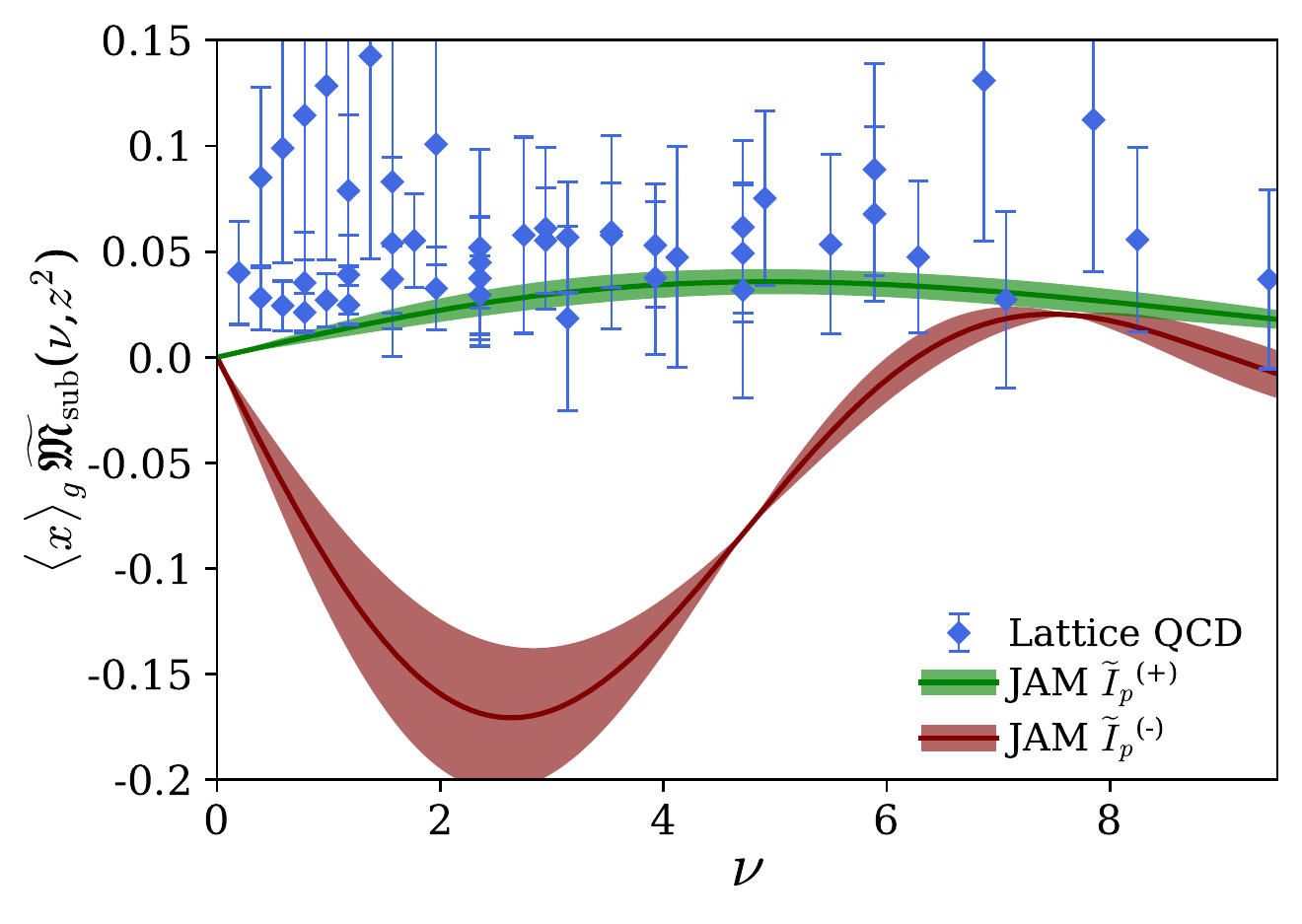}

\caption{\label{fig:phenocomp} A comparison between the lattice reduced Ioffe-time pseudo-distribution $\wt{\mathfrak{M}} (\nu, z^2)$ in the zero flow-time limit obtained through the subtraction method using the $p=0$ matrix elements, and the gluon helicity ITD constructed from global fits of PDFs. The lattice data points are the same as in Fig.~\ref{fig:pseudo-rITDsub}, plotted on a smaller vertical scale  for better comparison  with the phenomenological ITD bands. In the left plot, the red band denotes the ITD constructed from the gluon helicity distribution by the NNPDF collaboration. The green band labeled by $\wt{\mathcal{I}}_p^{(+)}$ and the cyan band labeled by $\wt{\mathcal{I}}_p^{(+/-)}$ represent the gluon helicity ITD determined by the JAM collaboration with and without the positivity constraint on the gluon helicity PDF, respectively. On the right plot, the gluon helicity ITDs for positive and negative helicity PDFs are compared with the lattice data. The green band labeled by $\wt{\mathcal{I}}_p^{(+)}$ and the maroon band  labeled by $\wt{\mathcal{I}}_p^{(-)}$ represent the gluon helicity ITD determined by the JAM collaboration associated with the positive and negative gluon helicity PDF solutions, respectively.}
\eefs{mockdemocn}


In their most recent global fit of the gluon helicity distribution~\cite{Zhou:2022wzm}, the JAM collaboration showed that without the assumption of positivity constraints on the quark and gluon helicity  PDFs, the magnitude or sign of the gluon polarization in the nucleon cannot be properly constrained. In other words, the ITD  extracted from the JAM global fit (labeled by JAM $\wt{\mathcal{I}}_p^{(+/-)}$ in Fig.~\ref{fig:phenocomp}) may have a similar or even larger magnitude of uncertainty than our lattice QCD calculation. We show a comparison of the polarized gluon ITDs obtained from global fits and our lattice calculation in Fig.~\ref{fig:phenocomp}. Most importantly, Fig.~\ref{fig:phenocomp} shows that the ITD data in the $\nu \lesssim 6$ region is primarily controlled by whether the gluon polarization in the nucleon is positive or negative, according to the JAM analysis.

The positivity constraint on the gluon distributions,  namely helicity-aligned and helicity-antialigned both being non-negative, in the analysis of experimental data in~\cite{Zhou:2022wzm} leads to a substantial reduction of the variance of $x\Delta g(x)$ in the large-$x$ region, as seen in Fig. 6 of~\cite{Zhou:2022wzm}. Specifically, the PDFs without the positivity assumption were organized into a band of solutions with a negative PDF and a band of solutions with a positive PDF. We compare the ITDs resulting from the two bands with positive and negative $x\Delta g(x)$ to our results in the right panel of Fig.~\ref{fig:phenocomp}. The current matrix elements, albeit with an unphysical pion mass and finite lattice spacing, are inconsistent within statistical uncertainties with the negative PDF branch. We note, however, that the large $\nu$ behavior of the lattice-calculated ITD, beyond the range currently accessible, could turn negative and compensate for the positive trend observed in Fig.~\ref{fig:phenocomp}.

By performing a simultaneous phenomenological fit to the NNPDF unpolarized and polarized gluon PDFs from~\cite{Ball:2017nwa}, it was  found in~\cite{Sufian:2020wcv} that the magnitude of the gluon helicity ITD in the small $\nu \leq 6$ region can benchmark the gluonic contribution to the nucleon spin budget to be  between roughly 40\% and  80\%. In conjunction with the findings of the JAM analysis, these observations indicate that the ITD data in the $\nu\lesssim6$-region are of tremendous importance in quantifying the gluonic contribution to the proton spin and the gluon helicity distribution as a whole. The small-$\nu$ region is most readily accessible with current leadership-class computing capabilities, so these observations signal the possibility of a timely  constraint on the gluon contribution to the proton spin and its $x$-dependent helicity distribution from lattice QCD. Future lattice data, when combined with experimental data and incorporated into a global analysis, as in Refs.~in~\cite{Lin:2017stx,Bringewatt:2020ixn,Barry:2022itu,Hou:2022sdf,Hou:2022ajg}, have the potential to provide a strong constraint on the gluon helicity distribution in the nucleon. 

\section{Conclusion and outlook}\label{conclusion}

This work demonstrates the possibility of determining the $x$-dependent gluon helicity distribution from lattice QCD, by calculating the polarized gluon Ioffe-time distribution using the pseudo-PDF approach.
We employed a combination of numerical techniques to facilitate this calculation, including: 
distillation for the nucleon interpolators, gradient flow for the gauge
fields, and the solution of a summed generalized eigenvalue problem for the
analysis of the gluonic matrix elements.  Our exploratory calculation demonstrates the significant challenge posed by the large statistical uncertainties in the gluonic
matrix elements, especially those evaluated at large nucleon momenta. These uncertainties limit our ability to properly estimate the systematic uncertainties associated with the determination of the relevant gluonic matrix elements.

The gluon helicity contribution to the proton spin can simply be written in a frame-independent
and gauge-invariant manner as an integral of the gluon-ITD over Ioffe-time. 
We have shown that
the Ioffe-time distribution in the low $\nu \lesssim 6$  region is
mostly controlled by the sign and magnitude of the gluon helicity
distribution determined from the existing experimental data in global analyses. Within the present statistical precision and through a qualitative
comparison with global analyses of the gluon helicity distribution, our result hints at a positive gluon polarization contribution to the nucleon spin budget.

In the pseudo-distribution approach, the leading order matrix element that defines the gluon helicity distribution cannot be calculated directly. To relate our numerical results to the gluon helicity distribution, a large ${\cal O}(m_p^2/p_z^2)$ contamination term must be removed. We 
developed two practical methods to remove this contamination. In the future, including more precise datasets at higher momenta 
to suppress such  
${\cal O}(m_p^2/p_z^2)$-contamination, in combination with more robust methods for removing higher order corrections, should lead to a reliable computation of the gluon helicity distribution in the proton. We note, however, that it was important for us to use our data at all momenta to 
estimate such ${\cal O}(m_p^2/p_z^2)$ corrections. The largest momentum dataset still contains a significant contribution ${\cal O}(m_p^2/p_z^2)$ that needs removal. Furthermore, it is unlikely to be practical to perform lattice QCD calculations at such large momentum that the contamination term is negligible for the entire range of accessible Ioffe time.  

We emphasize that the Ioffe-time distribution in the 
$\nu \lesssim 6$-domain, which is the most accessible region for a lattice QCD
calculation, has the potential to distinguish between positive and negative solutions of the gluon helicity PDFs determined in global fits of experimental data.  Our work demonstrates that, with improved statistical precision and precise data over a wider range of Ioffe times, a controlled determination of the gluon contribution to the 
proton spin from lattice QCD is possible.

\section{ACKNOWLEDGMENT}

We would like to thank all the members of the HadStruc collaboration for fruitful and stimulating exchanges. T.K. and R.S.S. acknowledge Luka Leskovec and Yiyu Zhou for offering their expertise, which greatly assisted this research. This work is supported by the U.S. Department of Energy, Office of Science, Office of Nuclear Physics under contract DE-AC05-06OR23177. J.K. is supported
by U.S. DOE grant \mbox{\#DE-SC0011941}. N.K., K.O., and R.S.S. are supported  by U.S. DOE Grant \mbox{\#DE-FG02-04ER41302.} A.R. and W.M. are also supported  by U.S. DOE Grant \mbox{\#DE-FG02-97ER41028.}   Computations for this work were carried out in part on facilities of the USQCD Collaboration, which are funded by the Office of Science of the U.S. Department of Energy. This work was performed in part using computing facilities at William and Mary which were provided by contributions from the National Science Foundation (MRI grant PHY-1626177), and the Commonwealth of Virginia Equipment Trust Fund. This work used the Extreme Science and Engineering Discovery Environment (XSEDE), which is supported by National Science Foundation grant number ACI-1548562. Specifically, it used the Bridges system, which is supported by NSF award number ACI-1445606, at the Pittsburgh Supercomputing Center (PSC) \cite{6866038, Nystrom:2015:BUF:2792745.2792775}. In addition, this work used resources at NERSC, a DOE Office of Science User Facility supported by the Office of Science of the U.S. Department of Energy under Contract \#DE-AC02-05CH11231, as well as resources of the Oak Ridge Leadership Computing Facility at the Oak Ridge National Laboratory, which is supported by the Office of Science of the U.S. Department of Energy under Contract No. \mbox{\#DE-AC05-00OR22725}. The software codes {\tt Chroma} \cite{Edwards:2004sx}, {\tt QUDA} \cite{Clark:2009wm, Babich:2010mu} and {\tt QPhiX} \cite{QPhiX2} were used in our work. The authors acknowledge support from the U.S. Department of Energy, Office of Science, Office of Advanced Scientific Computing Research and Office of Nuclear Physics, Scientific Discovery through Advanced Computing (SciDAC) program, and of the U.S. Department of Energy Exascale Computing Project. The authors also acknowledge the Texas Advanced Computing Center (TACC) at The University of Texas at Austin for providing HPC resources, like Frontera computing system~\cite{10.1145/3311790.3396656} that has contributed to the research results reported within this paper.
We acknowledge PRACE (Partnership for Advanced Computing in Europe) for awarding us access to the high performance computing system Marconi100 at CINECA (Consorzio Interuniversitario per il Calcolo Automatico dell’Italia Nord-orientale) under the grants Pra21$-$5389 and Pra23$-$0076. This work also benefited from access to the Jean Zay supercomputer at the Institute for Development and Resources in Intensive Scientific Computing (IDRIS) in Orsay, France under project A0080511504.

\bibliography{References.bib}

\begin{thebibliography}{120}%
\makeatletter
\providecommand \@ifxundefined [1]{%
 \@ifx{#1\undefined}
}%
\providecommand \@ifnum [1]{%
 \ifnum #1\expandafter \@firstoftwo
 \else \expandafter \@secondoftwo
 \fi
}%
\providecommand \@ifx [1]{%
 \ifx #1\expandafter \@firstoftwo
 \else \expandafter \@secondoftwo
 \fi
}%
\providecommand \natexlab [1]{#1}%
\providecommand \enquote  [1]{``#1''}%
\providecommand \bibnamefont  [1]{#1}%
\providecommand \bibfnamefont [1]{#1}%
\providecommand \citenamefont [1]{#1}%
\providecommand \href@noop [0]{\@secondoftwo}%
\providecommand \href [0]{\begingroup \@sanitize@url \@href}%
\providecommand \@href[1]{\@@startlink{#1}\@@href}%
\providecommand \@@href[1]{\endgroup#1\@@endlink}%
\providecommand \@sanitize@url [0]{\catcode `\\12\catcode `\$12\catcode
  `\&12\catcode `\#12\catcode `\^12\catcode `\_12\catcode `\%12\relax}%
\providecommand \@@startlink[1]{}%
\providecommand \@@endlink[0]{}%
\providecommand \url  [0]{\begingroup\@sanitize@url \@url }%
\providecommand \@url [1]{\endgroup\@href {#1}{\urlprefix }}%
\providecommand \urlprefix  [0]{URL }%
\providecommand \Eprint [0]{\href }%
\providecommand \doibase [0]{http://dx.doi.org/}%
\providecommand \selectlanguage [0]{\@gobble}%
\providecommand \bibinfo  [0]{\@secondoftwo}%
\providecommand \bibfield  [0]{\@secondoftwo}%
\providecommand \translation [1]{[#1]}%
\providecommand \BibitemOpen [0]{}%
\providecommand \bibitemStop [0]{}%
\providecommand \bibitemNoStop [0]{.\EOS\space}%
\providecommand \EOS [0]{\spacefactor3000\relax}%
\providecommand \BibitemShut  [1]{\csname bibitem#1\endcsname}%
\let\auto@bib@innerbib\@empty
\bibitem [{\citenamefont {Jaffe}\ and\ \citenamefont
  {Manohar}(1990)}]{Jaffe:1989jz}%
  \BibitemOpen
  \bibfield  {author} {\bibinfo {author} {\bibfnamefont {R.~L.}\ \bibnamefont
  {Jaffe}}\ and\ \bibinfo {author} {\bibfnamefont {A.}~\bibnamefont
  {Manohar}},\ }\href {\doibase 10.1016/0550-3213(90)90506-9} {\bibfield
  {journal} {\bibinfo  {journal} {Nucl. Phys. B}\ }\textbf {\bibinfo {volume}
  {337}},\ \bibinfo {pages} {509} (\bibinfo {year} {1990})}\BibitemShut
  {NoStop}%
\bibitem [{\citenamefont {Ji}(1997)}]{Ji:1996ek}%
  \BibitemOpen
  \bibfield  {author} {\bibinfo {author} {\bibfnamefont {X.-D.}\ \bibnamefont
  {Ji}},\ }\href {\doibase 10.1103/PhysRevLett.78.610} {\bibfield  {journal}
  {\bibinfo  {journal} {Phys. Rev. Lett.}\ }\textbf {\bibinfo {volume} {78}},\
  \bibinfo {pages} {610} (\bibinfo {year} {1997})},\ \Eprint
  {http://arxiv.org/abs/hep-ph/9603249} {arXiv:hep-ph/9603249} \BibitemShut
  {NoStop}%
\bibitem [{\citenamefont {Ashman}\ \emph {et~al.}(1988)\citenamefont {Ashman}
  \emph {et~al.}}]{EuropeanMuon:1987isl}%
  \BibitemOpen
  \bibfield  {author} {\bibinfo {author} {\bibfnamefont {J.}~\bibnamefont
  {Ashman}} \emph {et~al.} (\bibinfo {collaboration} {European Muon}),\ }\href
  {\doibase 10.1016/0370-2693(88)91523-7} {\bibfield  {journal} {\bibinfo
  {journal} {Phys. Lett. B}\ }\textbf {\bibinfo {volume} {206}},\ \bibinfo
  {pages} {364} (\bibinfo {year} {1988})}\BibitemShut {NoStop}%
\bibitem [{\citenamefont {Ashman}\ \emph {et~al.}(1989)\citenamefont {Ashman}
  \emph {et~al.}}]{EuropeanMuon:1989yki}%
  \BibitemOpen
  \bibfield  {author} {\bibinfo {author} {\bibfnamefont {J.}~\bibnamefont
  {Ashman}} \emph {et~al.} (\bibinfo {collaboration} {European Muon}),\ }\href
  {\doibase 10.1016/0550-3213(89)90089-8} {\bibfield  {journal} {\bibinfo
  {journal} {Nucl. Phys. B}\ }\textbf {\bibinfo {volume} {328}},\ \bibinfo
  {pages} {1} (\bibinfo {year} {1989})}\BibitemShut {NoStop}%
\bibitem [{\citenamefont {de~Florian}\ \emph {et~al.}(2009)\citenamefont
  {de~Florian}, \citenamefont {Sassot}, \citenamefont {Stratmann},\ and\
  \citenamefont {Vogelsang}}]{deFlorian:2009vb}%
  \BibitemOpen
  \bibfield  {author} {\bibinfo {author} {\bibfnamefont {D.}~\bibnamefont
  {de~Florian}}, \bibinfo {author} {\bibfnamefont {R.}~\bibnamefont {Sassot}},
  \bibinfo {author} {\bibfnamefont {M.}~\bibnamefont {Stratmann}}, \ and\
  \bibinfo {author} {\bibfnamefont {W.}~\bibnamefont {Vogelsang}},\ }\href
  {\doibase 10.1103/PhysRevD.80.034030} {\bibfield  {journal} {\bibinfo
  {journal} {Phys. Rev. D}\ }\textbf {\bibinfo {volume} {80}},\ \bibinfo
  {pages} {034030} (\bibinfo {year} {2009})},\ \Eprint
  {http://arxiv.org/abs/0904.3821} {arXiv:0904.3821 [hep-ph]} \BibitemShut
  {NoStop}%
\bibitem [{\citenamefont {Nocera}\ \emph {et~al.}(2014)\citenamefont {Nocera},
  \citenamefont {Ball}, \citenamefont {Forte}, \citenamefont {Ridolfi},\ and\
  \citenamefont {Rojo}}]{Nocera:2014gqa}%
  \BibitemOpen
  \bibfield  {author} {\bibinfo {author} {\bibfnamefont {E.~R.}\ \bibnamefont
  {Nocera}}, \bibinfo {author} {\bibfnamefont {R.~D.}\ \bibnamefont {Ball}},
  \bibinfo {author} {\bibfnamefont {S.}~\bibnamefont {Forte}}, \bibinfo
  {author} {\bibfnamefont {G.}~\bibnamefont {Ridolfi}}, \ and\ \bibinfo
  {author} {\bibfnamefont {J.}~\bibnamefont {Rojo}} (\bibinfo {collaboration}
  {NNPDF}),\ }\href {\doibase 10.1016/j.nuclphysb.2014.08.008} {\bibfield
  {journal} {\bibinfo  {journal} {Nucl. Phys. B}\ }\textbf {\bibinfo {volume}
  {887}},\ \bibinfo {pages} {276} (\bibinfo {year} {2014})},\ \Eprint
  {http://arxiv.org/abs/1406.5539} {arXiv:1406.5539 [hep-ph]} \BibitemShut
  {NoStop}%
\bibitem [{\citenamefont {Ethier}\ \emph {et~al.}(2017)\citenamefont {Ethier},
  \citenamefont {Sato},\ and\ \citenamefont {Melnitchouk}}]{Ethier:2017zbq}%
  \BibitemOpen
  \bibfield  {author} {\bibinfo {author} {\bibfnamefont {J.~J.}\ \bibnamefont
  {Ethier}}, \bibinfo {author} {\bibfnamefont {N.}~\bibnamefont {Sato}}, \ and\
  \bibinfo {author} {\bibfnamefont {W.}~\bibnamefont {Melnitchouk}},\ }\href
  {\doibase 10.1103/PhysRevLett.119.132001} {\bibfield  {journal} {\bibinfo
  {journal} {Phys. Rev. Lett.}\ }\textbf {\bibinfo {volume} {119}},\ \bibinfo
  {pages} {132001} (\bibinfo {year} {2017})},\ \Eprint
  {http://arxiv.org/abs/1705.05889} {arXiv:1705.05889 [hep-ph]} \BibitemShut
  {NoStop}%
\bibitem [{\citenamefont {de~Florian}\ \emph {et~al.}(2014)\citenamefont
  {de~Florian}, \citenamefont {Sassot}, \citenamefont {Stratmann},\ and\
  \citenamefont {Vogelsang}}]{deFlorian:2014yva}%
  \BibitemOpen
  \bibfield  {author} {\bibinfo {author} {\bibfnamefont {D.}~\bibnamefont
  {de~Florian}}, \bibinfo {author} {\bibfnamefont {R.}~\bibnamefont {Sassot}},
  \bibinfo {author} {\bibfnamefont {M.}~\bibnamefont {Stratmann}}, \ and\
  \bibinfo {author} {\bibfnamefont {W.}~\bibnamefont {Vogelsang}},\ }\href
  {\doibase 10.1103/PhysRevLett.113.012001} {\bibfield  {journal} {\bibinfo
  {journal} {Phys. Rev. Lett.}\ }\textbf {\bibinfo {volume} {113}},\ \bibinfo
  {pages} {012001} (\bibinfo {year} {2014})},\ \Eprint
  {http://arxiv.org/abs/1404.4293} {arXiv:1404.4293 [hep-ph]} \BibitemShut
  {NoStop}%
\bibitem [{\citenamefont {Djawotho}(2013)}]{Djawotho:2013pga}%
  \BibitemOpen
  \bibfield  {author} {\bibinfo {author} {\bibfnamefont {P.}~\bibnamefont
  {Djawotho}} (\bibinfo {collaboration} {STAR}),\ }\href {\doibase
  10.1393/ncc/i2013-11569-3} {\bibfield  {journal} {\bibinfo  {journal} {Nuovo
  Cim. C}\ }\textbf {\bibinfo {volume} {036}},\ \bibinfo {pages} {35} (\bibinfo
  {year} {2013})},\ \Eprint {http://arxiv.org/abs/1303.0543} {arXiv:1303.0543
  [nucl-ex]} \BibitemShut {NoStop}%
\bibitem [{\citenamefont {Adare}\ \emph {et~al.}(2014)\citenamefont {Adare}
  \emph {et~al.}}]{PHENIX:2014gbf}%
  \BibitemOpen
  \bibfield  {author} {\bibinfo {author} {\bibfnamefont {A.}~\bibnamefont
  {Adare}} \emph {et~al.} (\bibinfo {collaboration} {PHENIX}),\ }\href
  {\doibase 10.1103/PhysRevD.90.012007} {\bibfield  {journal} {\bibinfo
  {journal} {Phys. Rev. D}\ }\textbf {\bibinfo {volume} {90}},\ \bibinfo
  {pages} {012007} (\bibinfo {year} {2014})},\ \Eprint
  {http://arxiv.org/abs/1402.6296} {arXiv:1402.6296 [hep-ex]} \BibitemShut
  {NoStop}%
\bibitem [{\citenamefont {Zhou}\ \emph {et~al.}(2022)\citenamefont {Zhou},
  \citenamefont {Sato},\ and\ \citenamefont {Melnitchouk}}]{Zhou:2022wzm}%
  \BibitemOpen
  \bibfield  {author} {\bibinfo {author} {\bibfnamefont {Y.}~\bibnamefont
  {Zhou}}, \bibinfo {author} {\bibfnamefont {N.}~\bibnamefont {Sato}}, \ and\
  \bibinfo {author} {\bibfnamefont {W.}~\bibnamefont {Melnitchouk}} (\bibinfo
  {collaboration} {Jefferson Lab Angular Momentum (JAM)}),\ }\href {\doibase
  10.1103/PhysRevD.105.074022} {\bibfield  {journal} {\bibinfo  {journal}
  {Phys. Rev. D}\ }\textbf {\bibinfo {volume} {105}},\ \bibinfo {pages}
  {074022} (\bibinfo {year} {2022})},\ \Eprint
  {http://arxiv.org/abs/2201.02075} {arXiv:2201.02075 [hep-ph]} \BibitemShut
  {NoStop}%
\bibitem [{\citenamefont {Bunce}\ \emph {et~al.}(2000)\citenamefont {Bunce},
  \citenamefont {Saito}, \citenamefont {Soffer},\ and\ \citenamefont
  {Vogelsang}}]{Bunce:2000uv}%
  \BibitemOpen
  \bibfield  {author} {\bibinfo {author} {\bibfnamefont {G.}~\bibnamefont
  {Bunce}}, \bibinfo {author} {\bibfnamefont {N.}~\bibnamefont {Saito}},
  \bibinfo {author} {\bibfnamefont {J.}~\bibnamefont {Soffer}}, \ and\ \bibinfo
  {author} {\bibfnamefont {W.}~\bibnamefont {Vogelsang}},\ }\href {\doibase
  10.1146/annurev.nucl.50.1.525} {\bibfield  {journal} {\bibinfo  {journal}
  {Ann. Rev. Nucl. Part. Sci.}\ }\textbf {\bibinfo {volume} {50}},\ \bibinfo
  {pages} {525} (\bibinfo {year} {2000})},\ \Eprint
  {http://arxiv.org/abs/hep-ph/0007218} {arXiv:hep-ph/0007218} \BibitemShut
  {NoStop}%
\bibitem [{\citenamefont {Airapetian}\ \emph {et~al.}(2008)\citenamefont
  {Airapetian} \emph {et~al.}}]{HERMES:2008abz}%
  \BibitemOpen
  \bibfield  {author} {\bibinfo {author} {\bibfnamefont {A.}~\bibnamefont
  {Airapetian}} \emph {et~al.} (\bibinfo {collaboration} {HERMES}),\ }\href
  {\doibase 10.1088/1126-6708/2008/06/066} {\bibfield  {journal} {\bibinfo
  {journal} {JHEP}\ }\textbf {\bibinfo {volume} {06}},\ \bibinfo {pages} {066}
  (\bibinfo {year} {2008})},\ \Eprint {http://arxiv.org/abs/0802.2499}
  {arXiv:0802.2499 [hep-ex]} \BibitemShut {NoStop}%
\bibitem [{\citenamefont {Dudek}\ \emph
  {et~al.}(2012{\natexlab{a}})\citenamefont {Dudek} \emph
  {et~al.}}]{Dudek:2012vr}%
  \BibitemOpen
  \bibfield  {author} {\bibinfo {author} {\bibfnamefont {J.}~\bibnamefont
  {Dudek}} \emph {et~al.},\ }\href {\doibase 10.1140/epja/i2012-12187-1}
  {\bibfield  {journal} {\bibinfo  {journal} {Eur. Phys. J. A}\ }\textbf
  {\bibinfo {volume} {48}},\ \bibinfo {pages} {187} (\bibinfo {year}
  {2012}{\natexlab{a}})},\ \Eprint {http://arxiv.org/abs/1208.1244}
  {arXiv:1208.1244 [hep-ex]} \BibitemShut {NoStop}%
\bibitem [{\citenamefont {Akhunzyanov}\ \emph {et~al.}(2019)\citenamefont
  {Akhunzyanov} \emph {et~al.}}]{COMPASS:2018pup}%
  \BibitemOpen
  \bibfield  {author} {\bibinfo {author} {\bibfnamefont {R.}~\bibnamefont
  {Akhunzyanov}} \emph {et~al.} (\bibinfo {collaboration} {COMPASS}),\ }\href
  {\doibase 10.1016/j.physletb.2019.04.038} {\bibfield  {journal} {\bibinfo
  {journal} {Phys. Lett. B}\ }\textbf {\bibinfo {volume} {793}},\ \bibinfo
  {pages} {188} (\bibinfo {year} {2019})},\ \bibinfo {note} {[Erratum:
  Phys.Lett.B 800, 135129 (2020)]},\ \Eprint {http://arxiv.org/abs/1802.02739}
  {arXiv:1802.02739 [hep-ex]} \BibitemShut {NoStop}%
\bibitem [{\citenamefont {{Accardi, A. \textit{et
  al.}}}(2016)}]{Accardi:2012qut}%
  \BibitemOpen
  \bibfield  {author} {\bibinfo {author} {\bibnamefont {{Accardi, A. \textit{et
  al.}}}},\ }\href {\doibase 10.1140/epja/i2016-16268-9} {\bibfield  {journal}
  {\bibinfo  {journal} {Eur. Phys. J. A}\ }\textbf {\bibinfo {volume} {52}},\
  \bibinfo {pages} {268} (\bibinfo {year} {2016})},\ \Eprint
  {http://arxiv.org/abs/1212.1701} {arXiv:1212.1701 [nucl-ex]} \BibitemShut
  {NoStop}%
\bibitem [{\citenamefont {Alexandrou}\ \emph {et~al.}(2020)\citenamefont
  {Alexandrou}, \citenamefont {Bacchio}, \citenamefont {Constantinou},
  \citenamefont {Finkenrath}, \citenamefont {Hadjiyiannakou}, \citenamefont
  {Jansen}, \citenamefont {Koutsou}, \citenamefont {Panagopoulos},\ and\
  \citenamefont {Spanoudes}}]{Alexandrou:2020sml}%
  \BibitemOpen
  \bibfield  {author} {\bibinfo {author} {\bibfnamefont {C.}~\bibnamefont
  {Alexandrou}}, \bibinfo {author} {\bibfnamefont {S.}~\bibnamefont {Bacchio}},
  \bibinfo {author} {\bibfnamefont {M.}~\bibnamefont {Constantinou}}, \bibinfo
  {author} {\bibfnamefont {J.}~\bibnamefont {Finkenrath}}, \bibinfo {author}
  {\bibfnamefont {K.}~\bibnamefont {Hadjiyiannakou}}, \bibinfo {author}
  {\bibfnamefont {K.}~\bibnamefont {Jansen}}, \bibinfo {author} {\bibfnamefont
  {G.}~\bibnamefont {Koutsou}}, \bibinfo {author} {\bibfnamefont
  {H.}~\bibnamefont {Panagopoulos}}, \ and\ \bibinfo {author} {\bibfnamefont
  {G.}~\bibnamefont {Spanoudes}},\ }\href {\doibase
  10.1103/PhysRevD.101.094513} {\bibfield  {journal} {\bibinfo  {journal}
  {Phys. Rev. D}\ }\textbf {\bibinfo {volume} {101}},\ \bibinfo {pages}
  {094513} (\bibinfo {year} {2020})},\ \Eprint
  {http://arxiv.org/abs/2003.08486} {arXiv:2003.08486 [hep-lat]} \BibitemShut
  {NoStop}%
\bibitem [{\citenamefont {Wang}\ \emph {et~al.}(2021)\citenamefont {Wang},
  \citenamefont {Yang}, \citenamefont {Liang}, \citenamefont {Draper},\ and\
  \citenamefont {Liu}}]{Wang:2021vqy}%
  \BibitemOpen
  \bibfield  {author} {\bibinfo {author} {\bibfnamefont {G.}~\bibnamefont
  {Wang}}, \bibinfo {author} {\bibfnamefont {Y.-B.}\ \bibnamefont {Yang}},
  \bibinfo {author} {\bibfnamefont {J.}~\bibnamefont {Liang}}, \bibinfo
  {author} {\bibfnamefont {T.}~\bibnamefont {Draper}}, \ and\ \bibinfo {author}
  {\bibfnamefont {K.-F.}\ \bibnamefont {Liu}} (\bibinfo {collaboration}
  {chiQCD}),\ }\href@noop {} {\  (\bibinfo {year} {2021})},\ \Eprint
  {http://arxiv.org/abs/2111.09329} {arXiv:2111.09329 [hep-lat]} \BibitemShut
  {NoStop}%
\bibitem [{\citenamefont {Engelhardt}\ \emph {et~al.}(2020)\citenamefont
  {Engelhardt}, \citenamefont {Green}, \citenamefont {Hasan}, \citenamefont
  {Krieg}, \citenamefont {Meinel}, \citenamefont {Negele}, \citenamefont
  {Pochinsky},\ and\ \citenamefont {Syritsyn}}]{Engelhardt:2020qtg}%
  \BibitemOpen
  \bibfield  {author} {\bibinfo {author} {\bibfnamefont {M.}~\bibnamefont
  {Engelhardt}}, \bibinfo {author} {\bibfnamefont {J.~R.}\ \bibnamefont
  {Green}}, \bibinfo {author} {\bibfnamefont {N.}~\bibnamefont {Hasan}},
  \bibinfo {author} {\bibfnamefont {S.}~\bibnamefont {Krieg}}, \bibinfo
  {author} {\bibfnamefont {S.}~\bibnamefont {Meinel}}, \bibinfo {author}
  {\bibfnamefont {J.}~\bibnamefont {Negele}}, \bibinfo {author} {\bibfnamefont
  {A.}~\bibnamefont {Pochinsky}}, \ and\ \bibinfo {author} {\bibfnamefont
  {S.}~\bibnamefont {Syritsyn}},\ }\href {\doibase 10.1103/PhysRevD.102.074505}
  {\bibfield  {journal} {\bibinfo  {journal} {Phys. Rev. D}\ }\textbf {\bibinfo
  {volume} {102}},\ \bibinfo {pages} {074505} (\bibinfo {year} {2020})},\
  \Eprint {http://arxiv.org/abs/2008.03660} {arXiv:2008.03660 [hep-lat]}
  \BibitemShut {NoStop}%
\bibitem [{\citenamefont {Ji}\ \emph {et~al.}(2013)\citenamefont {Ji},
  \citenamefont {Zhang},\ and\ \citenamefont {Zhao}}]{Ji:2013fga}%
  \BibitemOpen
  \bibfield  {author} {\bibinfo {author} {\bibfnamefont {X.}~\bibnamefont
  {Ji}}, \bibinfo {author} {\bibfnamefont {J.-H.}\ \bibnamefont {Zhang}}, \
  and\ \bibinfo {author} {\bibfnamefont {Y.}~\bibnamefont {Zhao}},\ }\href
  {\doibase 10.1103/PhysRevLett.111.112002} {\bibfield  {journal} {\bibinfo
  {journal} {Phys. Rev. Lett.}\ }\textbf {\bibinfo {volume} {111}},\ \bibinfo
  {pages} {112002} (\bibinfo {year} {2013})},\ \Eprint
  {http://arxiv.org/abs/1304.6708} {arXiv:1304.6708 [hep-ph]} \BibitemShut
  {NoStop}%
\bibitem [{\citenamefont {Manohar}(1991)}]{Manohar:1990jx}%
  \BibitemOpen
  \bibfield  {author} {\bibinfo {author} {\bibfnamefont {A.~V.}\ \bibnamefont
  {Manohar}},\ }\href {\doibase 10.1103/PhysRevLett.66.289} {\bibfield
  {journal} {\bibinfo  {journal} {Phys. Rev. Lett.}\ }\textbf {\bibinfo
  {volume} {66}},\ \bibinfo {pages} {289} (\bibinfo {year} {1991})}\BibitemShut
  {NoStop}%
\bibitem [{\citenamefont {Yang}\ \emph {et~al.}(2017)\citenamefont {Yang},
  \citenamefont {Sufian}, \citenamefont {Alexandru}, \citenamefont {Draper},
  \citenamefont {Glatzmaier}, \citenamefont {Liu},\ and\ \citenamefont
  {Zhao}}]{Yang:2016plb}%
  \BibitemOpen
  \bibfield  {author} {\bibinfo {author} {\bibfnamefont {Y.-B.}\ \bibnamefont
  {Yang}}, \bibinfo {author} {\bibfnamefont {R.~S.}\ \bibnamefont {Sufian}},
  \bibinfo {author} {\bibfnamefont {A.}~\bibnamefont {Alexandru}}, \bibinfo
  {author} {\bibfnamefont {T.}~\bibnamefont {Draper}}, \bibinfo {author}
  {\bibfnamefont {M.~J.}\ \bibnamefont {Glatzmaier}}, \bibinfo {author}
  {\bibfnamefont {K.-F.}\ \bibnamefont {Liu}}, \ and\ \bibinfo {author}
  {\bibfnamefont {Y.}~\bibnamefont {Zhao}},\ }\href {\doibase
  10.1103/PhysRevLett.118.102001} {\bibfield  {journal} {\bibinfo  {journal}
  {Phys. Rev. Lett.}\ }\textbf {\bibinfo {volume} {118}},\ \bibinfo {pages}
  {102001} (\bibinfo {year} {2017})},\ \Eprint
  {http://arxiv.org/abs/1609.05937} {arXiv:1609.05937 [hep-ph]} \BibitemShut
  {NoStop}%
\bibitem [{\citenamefont {Ji}(2014)}]{Ji:2014gla}%
  \BibitemOpen
  \bibfield  {author} {\bibinfo {author} {\bibfnamefont {X.}~\bibnamefont
  {Ji}},\ }\href {\doibase 10.1007/s11433-014-5492-3} {\bibfield  {journal}
  {\bibinfo  {journal} {Sci. China Phys. Mech. Astron.}\ }\textbf {\bibinfo
  {volume} {57}},\ \bibinfo {pages} {1407} (\bibinfo {year} {2014})},\ \Eprint
  {http://arxiv.org/abs/1404.6680} {arXiv:1404.6680 [hep-ph]} \BibitemShut
  {NoStop}%
\bibitem [{\citenamefont {Liu}(2022)}]{Liu:2021lke}%
  \BibitemOpen
  \bibfield  {author} {\bibinfo {author} {\bibfnamefont {K.-F.}\ \bibnamefont
  {Liu}},\ }\href {\doibase 10.1007/s43673-022-00037-4} {\bibfield  {journal}
  {\bibinfo  {journal} {AAPPS Bull.}\ }\textbf {\bibinfo {volume} {32}},\
  \bibinfo {pages} {8} (\bibinfo {year} {2022})},\ \Eprint
  {http://arxiv.org/abs/2112.08416} {arXiv:2112.08416 [hep-lat]} \BibitemShut
  {NoStop}%
\bibitem [{\citenamefont {Radyushkin}(2017)}]{Radyushkin:2017cyf}%
  \BibitemOpen
  \bibfield  {author} {\bibinfo {author} {\bibfnamefont {A.~V.}\ \bibnamefont
  {Radyushkin}},\ }\href {\doibase 10.1103/PhysRevD.96.034025} {\bibfield
  {journal} {\bibinfo  {journal} {Phys. Rev. D}\ }\textbf {\bibinfo {volume}
  {96}},\ \bibinfo {pages} {034025} (\bibinfo {year} {2017})},\ \Eprint
  {http://arxiv.org/abs/1705.01488} {arXiv:1705.01488 [hep-ph]} \BibitemShut
  {NoStop}%
\bibitem [{\citenamefont {Braun}\ \emph {et~al.}(1995)\citenamefont {Braun},
  \citenamefont {Gornicki},\ and\ \citenamefont {Mankiewicz}}]{Braun:1994jq}%
  \BibitemOpen
  \bibfield  {author} {\bibinfo {author} {\bibfnamefont {V.}~\bibnamefont
  {Braun}}, \bibinfo {author} {\bibfnamefont {P.}~\bibnamefont {Gornicki}}, \
  and\ \bibinfo {author} {\bibfnamefont {L.}~\bibnamefont {Mankiewicz}},\
  }\href {\doibase 10.1103/PhysRevD.51.6036} {\bibfield  {journal} {\bibinfo
  {journal} {Phys. Rev. D}\ }\textbf {\bibinfo {volume} {51}},\ \bibinfo
  {pages} {6036} (\bibinfo {year} {1995})},\ \Eprint
  {http://arxiv.org/abs/hep-ph/9410318} {arXiv:hep-ph/9410318} \BibitemShut
  {NoStop}%
\bibitem [{\citenamefont {Balitsky}\ \emph {et~al.}(2022)\citenamefont
  {Balitsky}, \citenamefont {Morris},\ and\ \citenamefont
  {Radyushkin}}]{Balitsky:2021cwr}%
  \BibitemOpen
  \bibfield  {author} {\bibinfo {author} {\bibfnamefont {I.}~\bibnamefont
  {Balitsky}}, \bibinfo {author} {\bibfnamefont {W.}~\bibnamefont {Morris}}, \
  and\ \bibinfo {author} {\bibfnamefont {A.}~\bibnamefont {Radyushkin}},\
  }\href {\doibase 10.1007/JHEP02(2022)193} {\bibfield  {journal} {\bibinfo
  {journal} {JHEP}\ }\textbf {\bibinfo {volume} {02}},\ \bibinfo {pages} {193}
  (\bibinfo {year} {2022})},\ \Eprint {http://arxiv.org/abs/2112.02011}
  {arXiv:2112.02011 [hep-ph]} \BibitemShut {NoStop}%
\bibitem [{\citenamefont
  {Radyushkin}(2018{\natexlab{a}})}]{Radyushkin:2017lvu}%
  \BibitemOpen
  \bibfield  {author} {\bibinfo {author} {\bibfnamefont {A.~V.}\ \bibnamefont
  {Radyushkin}},\ }\href {\doibase 10.1016/j.physletb.2018.04.023} {\bibfield
  {journal} {\bibinfo  {journal} {Phys. Lett. B}\ }\textbf {\bibinfo {volume}
  {781}},\ \bibinfo {pages} {433} (\bibinfo {year} {2018}{\natexlab{a}})},\
  \Eprint {http://arxiv.org/abs/1710.08813} {arXiv:1710.08813 [hep-ph]}
  \BibitemShut {NoStop}%
\bibitem [{\citenamefont
  {Radyushkin}(2018{\natexlab{b}})}]{Radyushkin:2018cvn}%
  \BibitemOpen
  \bibfield  {author} {\bibinfo {author} {\bibfnamefont {A.}~\bibnamefont
  {Radyushkin}},\ }\href {\doibase 10.1103/PhysRevD.98.014019} {\bibfield
  {journal} {\bibinfo  {journal} {Phys. Rev. D}\ }\textbf {\bibinfo {volume}
  {98}},\ \bibinfo {pages} {014019} (\bibinfo {year} {2018}{\natexlab{b}})},\
  \Eprint {http://arxiv.org/abs/1801.02427} {arXiv:1801.02427 [hep-ph]}
  \BibitemShut {NoStop}%
\bibitem [{\citenamefont {Ioffe}(1969)}]{Ioffe:1969kf}%
  \BibitemOpen
  \bibfield  {author} {\bibinfo {author} {\bibfnamefont {B.~L.}\ \bibnamefont
  {Ioffe}},\ }\href {\doibase 10.1016/0370-2693(69)90415-8} {\bibfield
  {journal} {\bibinfo  {journal} {Phys. Lett. B}\ }\textbf {\bibinfo {volume}
  {30}},\ \bibinfo {pages} {123} (\bibinfo {year} {1969})}\BibitemShut
  {NoStop}%
\bibitem [{\citenamefont {Orginos}\ \emph {et~al.}(2017)\citenamefont
  {Orginos}, \citenamefont {Radyushkin}, \citenamefont {Karpie},\ and\
  \citenamefont {Zafeiropoulos}}]{Orginos:2017kos}%
  \BibitemOpen
  \bibfield  {author} {\bibinfo {author} {\bibfnamefont {K.}~\bibnamefont
  {Orginos}}, \bibinfo {author} {\bibfnamefont {A.}~\bibnamefont {Radyushkin}},
  \bibinfo {author} {\bibfnamefont {J.}~\bibnamefont {Karpie}}, \ and\ \bibinfo
  {author} {\bibfnamefont {S.}~\bibnamefont {Zafeiropoulos}},\ }\href {\doibase
  10.1103/PhysRevD.96.094503} {\bibfield  {journal} {\bibinfo  {journal} {Phys.
  Rev. D}\ }\textbf {\bibinfo {volume} {96}},\ \bibinfo {pages} {094503}
  (\bibinfo {year} {2017})},\ \Eprint {http://arxiv.org/abs/1706.05373}
  {arXiv:1706.05373 [hep-ph]} \BibitemShut {NoStop}%
\bibitem [{\citenamefont {Lin}\ \emph {et~al.}(2018{\natexlab{a}})\citenamefont
  {Lin}, \citenamefont {Chen}, \citenamefont {Ji}, \citenamefont {Jin},
  \citenamefont {Li}, \citenamefont {Liu}, \citenamefont {Yang}, \citenamefont
  {Zhang},\ and\ \citenamefont {Zhao}}]{Lin:2018pvv}%
  \BibitemOpen
  \bibfield  {author} {\bibinfo {author} {\bibfnamefont {H.-W.}\ \bibnamefont
  {Lin}}, \bibinfo {author} {\bibfnamefont {J.-W.}\ \bibnamefont {Chen}},
  \bibinfo {author} {\bibfnamefont {X.}~\bibnamefont {Ji}}, \bibinfo {author}
  {\bibfnamefont {L.}~\bibnamefont {Jin}}, \bibinfo {author} {\bibfnamefont
  {R.}~\bibnamefont {Li}}, \bibinfo {author} {\bibfnamefont {Y.-S.}\
  \bibnamefont {Liu}}, \bibinfo {author} {\bibfnamefont {Y.-B.}\ \bibnamefont
  {Yang}}, \bibinfo {author} {\bibfnamefont {J.-H.}\ \bibnamefont {Zhang}}, \
  and\ \bibinfo {author} {\bibfnamefont {Y.}~\bibnamefont {Zhao}},\ }\href
  {\doibase 10.1103/PhysRevLett.121.242003} {\bibfield  {journal} {\bibinfo
  {journal} {Phys. Rev. Lett.}\ }\textbf {\bibinfo {volume} {121}},\ \bibinfo
  {pages} {242003} (\bibinfo {year} {2018}{\natexlab{a}})},\ \Eprint
  {http://arxiv.org/abs/1807.07431} {arXiv:1807.07431 [hep-lat]} \BibitemShut
  {NoStop}%
\bibitem [{\citenamefont {Alexandrou}\ \emph {et~al.}(2018)\citenamefont
  {Alexandrou}, \citenamefont {Cichy}, \citenamefont {Constantinou},
  \citenamefont {Jansen}, \citenamefont {Scapellato},\ and\ \citenamefont
  {Steffens}}]{Alexandrou:2018pbm}%
  \BibitemOpen
  \bibfield  {author} {\bibinfo {author} {\bibfnamefont {C.}~\bibnamefont
  {Alexandrou}}, \bibinfo {author} {\bibfnamefont {K.}~\bibnamefont {Cichy}},
  \bibinfo {author} {\bibfnamefont {M.}~\bibnamefont {Constantinou}}, \bibinfo
  {author} {\bibfnamefont {K.}~\bibnamefont {Jansen}}, \bibinfo {author}
  {\bibfnamefont {A.}~\bibnamefont {Scapellato}}, \ and\ \bibinfo {author}
  {\bibfnamefont {F.}~\bibnamefont {Steffens}},\ }\href {\doibase
  10.1103/PhysRevLett.121.112001} {\bibfield  {journal} {\bibinfo  {journal}
  {Phys. Rev. Lett.}\ }\textbf {\bibinfo {volume} {121}},\ \bibinfo {pages}
  {112001} (\bibinfo {year} {2018})},\ \Eprint
  {http://arxiv.org/abs/1803.02685} {arXiv:1803.02685 [hep-lat]} \BibitemShut
  {NoStop}%
\bibitem [{\citenamefont {Liang}\ \emph {et~al.}(2020)\citenamefont {Liang},
  \citenamefont {Draper}, \citenamefont {Liu}, \citenamefont {Rothkopf},\ and\
  \citenamefont {Yang}}]{Liang:2019frk}%
  \BibitemOpen
  \bibfield  {author} {\bibinfo {author} {\bibfnamefont {J.}~\bibnamefont
  {Liang}}, \bibinfo {author} {\bibfnamefont {T.}~\bibnamefont {Draper}},
  \bibinfo {author} {\bibfnamefont {K.-F.}\ \bibnamefont {Liu}}, \bibinfo
  {author} {\bibfnamefont {A.}~\bibnamefont {Rothkopf}}, \ and\ \bibinfo
  {author} {\bibfnamefont {Y.-B.}\ \bibnamefont {Yang}} (\bibinfo
  {collaboration} {XQCD}),\ }\href {\doibase 10.1103/PhysRevD.101.114503}
  {\bibfield  {journal} {\bibinfo  {journal} {Phys. Rev. D}\ }\textbf {\bibinfo
  {volume} {101}},\ \bibinfo {pages} {114503} (\bibinfo {year} {2020})},\
  \Eprint {http://arxiv.org/abs/1906.05312} {arXiv:1906.05312 [hep-ph]}
  \BibitemShut {NoStop}%
\bibitem [{\citenamefont {Karpie}\ \emph {et~al.}(2021)\citenamefont {Karpie},
  \citenamefont {Orginos}, \citenamefont {Radyushkin},\ and\ \citenamefont
  {Zafeiropoulos}}]{Karpie:2021pap}%
  \BibitemOpen
  \bibfield  {author} {\bibinfo {author} {\bibfnamefont {J.}~\bibnamefont
  {Karpie}}, \bibinfo {author} {\bibfnamefont {K.}~\bibnamefont {Orginos}},
  \bibinfo {author} {\bibfnamefont {A.}~\bibnamefont {Radyushkin}}, \ and\
  \bibinfo {author} {\bibfnamefont {S.}~\bibnamefont {Zafeiropoulos}} (\bibinfo
  {collaboration} {HadStruc}),\ }\href {\doibase 10.1007/JHEP11(2021)024}
  {\bibfield  {journal} {\bibinfo  {journal} {JHEP}\ }\textbf {\bibinfo
  {volume} {11}},\ \bibinfo {pages} {024} (\bibinfo {year} {2021})},\ \Eprint
  {http://arxiv.org/abs/2105.13313} {arXiv:2105.13313 [hep-lat]} \BibitemShut
  {NoStop}%
\bibitem [{\citenamefont {Sufian}\ \emph {et~al.}(2019)\citenamefont {Sufian},
  \citenamefont {Karpie}, \citenamefont {Egerer}, \citenamefont {Orginos},
  \citenamefont {Qiu},\ and\ \citenamefont {Richards}}]{Sufian:2019bol}%
  \BibitemOpen
  \bibfield  {author} {\bibinfo {author} {\bibfnamefont {R.~S.}\ \bibnamefont
  {Sufian}}, \bibinfo {author} {\bibfnamefont {J.}~\bibnamefont {Karpie}},
  \bibinfo {author} {\bibfnamefont {C.}~\bibnamefont {Egerer}}, \bibinfo
  {author} {\bibfnamefont {K.}~\bibnamefont {Orginos}}, \bibinfo {author}
  {\bibfnamefont {J.-W.}\ \bibnamefont {Qiu}}, \ and\ \bibinfo {author}
  {\bibfnamefont {D.~G.}\ \bibnamefont {Richards}},\ }\href {\doibase
  10.1103/PhysRevD.99.074507} {\bibfield  {journal} {\bibinfo  {journal} {Phys.
  Rev. D}\ }\textbf {\bibinfo {volume} {99}},\ \bibinfo {pages} {074507}
  (\bibinfo {year} {2019})},\ \Eprint {http://arxiv.org/abs/1901.03921}
  {arXiv:1901.03921 [hep-lat]} \BibitemShut {NoStop}%
\bibitem [{\citenamefont {Sufian}\ \emph {et~al.}(2020)\citenamefont {Sufian},
  \citenamefont {Egerer}, \citenamefont {Karpie}, \citenamefont {Edwards},
  \citenamefont {Jo\'o}, \citenamefont {Ma}, \citenamefont {Orginos},
  \citenamefont {Qiu},\ and\ \citenamefont {Richards}}]{Sufian:2020vzb}%
  \BibitemOpen
  \bibfield  {author} {\bibinfo {author} {\bibfnamefont {R.~S.}\ \bibnamefont
  {Sufian}}, \bibinfo {author} {\bibfnamefont {C.}~\bibnamefont {Egerer}},
  \bibinfo {author} {\bibfnamefont {J.}~\bibnamefont {Karpie}}, \bibinfo
  {author} {\bibfnamefont {R.~G.}\ \bibnamefont {Edwards}}, \bibinfo {author}
  {\bibfnamefont {B.}~\bibnamefont {Jo\'o}}, \bibinfo {author} {\bibfnamefont
  {Y.-Q.}\ \bibnamefont {Ma}}, \bibinfo {author} {\bibfnamefont
  {K.}~\bibnamefont {Orginos}}, \bibinfo {author} {\bibfnamefont {J.-W.}\
  \bibnamefont {Qiu}}, \ and\ \bibinfo {author} {\bibfnamefont {D.~G.}\
  \bibnamefont {Richards}},\ }\href {\doibase 10.1103/PhysRevD.102.054508}
  {\bibfield  {journal} {\bibinfo  {journal} {Phys. Rev. D}\ }\textbf {\bibinfo
  {volume} {102}},\ \bibinfo {pages} {054508} (\bibinfo {year} {2020})},\
  \Eprint {http://arxiv.org/abs/2001.04960} {arXiv:2001.04960 [hep-lat]}
  \BibitemShut {NoStop}%
\bibitem [{\citenamefont {Zhang}\ \emph {et~al.}(2020)\citenamefont {Zhang},
  \citenamefont {Li}, \citenamefont {Huo}, \citenamefont {Sun},\ and\
  \citenamefont {Yang}}]{Zhang:2020rsx}%
  \BibitemOpen
  \bibfield  {author} {\bibinfo {author} {\bibfnamefont {K.}~\bibnamefont
  {Zhang}}, \bibinfo {author} {\bibfnamefont {Y.-Y.}\ \bibnamefont {Li}},
  \bibinfo {author} {\bibfnamefont {Y.-K.}\ \bibnamefont {Huo}}, \bibinfo
  {author} {\bibfnamefont {P.}~\bibnamefont {Sun}}, \ and\ \bibinfo {author}
  {\bibfnamefont {Y.-B.}\ \bibnamefont {Yang}},\ }\href@noop {} {\  (\bibinfo
  {year} {2020})},\ \Eprint {http://arxiv.org/abs/2012.05448} {arXiv:2012.05448
  [hep-lat]} \BibitemShut {NoStop}%
\bibitem [{\citenamefont {Izubuchi}\ \emph {et~al.}(2019)\citenamefont
  {Izubuchi}, \citenamefont {Jin}, \citenamefont {Kallidonis}, \citenamefont
  {Karthik}, \citenamefont {Mukherjee}, \citenamefont {Petreczky},
  \citenamefont {Shugert},\ and\ \citenamefont {Syritsyn}}]{Izubuchi:2019lyk}%
  \BibitemOpen
  \bibfield  {author} {\bibinfo {author} {\bibfnamefont {T.}~\bibnamefont
  {Izubuchi}}, \bibinfo {author} {\bibfnamefont {L.}~\bibnamefont {Jin}},
  \bibinfo {author} {\bibfnamefont {C.}~\bibnamefont {Kallidonis}}, \bibinfo
  {author} {\bibfnamefont {N.}~\bibnamefont {Karthik}}, \bibinfo {author}
  {\bibfnamefont {S.}~\bibnamefont {Mukherjee}}, \bibinfo {author}
  {\bibfnamefont {P.}~\bibnamefont {Petreczky}}, \bibinfo {author}
  {\bibfnamefont {C.}~\bibnamefont {Shugert}}, \ and\ \bibinfo {author}
  {\bibfnamefont {S.}~\bibnamefont {Syritsyn}},\ }\href {\doibase
  10.1103/PhysRevD.100.034516} {\bibfield  {journal} {\bibinfo  {journal}
  {Phys. Rev. D}\ }\textbf {\bibinfo {volume} {100}},\ \bibinfo {pages}
  {034516} (\bibinfo {year} {2019})},\ \Eprint
  {http://arxiv.org/abs/1905.06349} {arXiv:1905.06349 [hep-lat]} \BibitemShut
  {NoStop}%
\bibitem [{\citenamefont {Gao}\ \emph {et~al.}(2020)\citenamefont {Gao},
  \citenamefont {Jin}, \citenamefont {Kallidonis}, \citenamefont {Karthik},
  \citenamefont {Mukherjee}, \citenamefont {Petreczky}, \citenamefont
  {Shugert}, \citenamefont {Syritsyn},\ and\ \citenamefont
  {Zhao}}]{Gao:2020ito}%
  \BibitemOpen
  \bibfield  {author} {\bibinfo {author} {\bibfnamefont {X.}~\bibnamefont
  {Gao}}, \bibinfo {author} {\bibfnamefont {L.}~\bibnamefont {Jin}}, \bibinfo
  {author} {\bibfnamefont {C.}~\bibnamefont {Kallidonis}}, \bibinfo {author}
  {\bibfnamefont {N.}~\bibnamefont {Karthik}}, \bibinfo {author} {\bibfnamefont
  {S.}~\bibnamefont {Mukherjee}}, \bibinfo {author} {\bibfnamefont
  {P.}~\bibnamefont {Petreczky}}, \bibinfo {author} {\bibfnamefont
  {C.}~\bibnamefont {Shugert}}, \bibinfo {author} {\bibfnamefont
  {S.}~\bibnamefont {Syritsyn}}, \ and\ \bibinfo {author} {\bibfnamefont
  {Y.}~\bibnamefont {Zhao}},\ }\href {\doibase 10.1103/PhysRevD.102.094513}
  {\bibfield  {journal} {\bibinfo  {journal} {Phys. Rev. D}\ }\textbf {\bibinfo
  {volume} {102}},\ \bibinfo {pages} {094513} (\bibinfo {year} {2020})},\
  \Eprint {http://arxiv.org/abs/2007.06590} {arXiv:2007.06590 [hep-lat]}
  \BibitemShut {NoStop}%
\bibitem [{\citenamefont {Alexandrou}\ \emph
  {et~al.}(2021{\natexlab{a}})\citenamefont {Alexandrou}, \citenamefont
  {Cichy}, \citenamefont {Constantinou}, \citenamefont {Green}, \citenamefont
  {Hadjiyiannakou}, \citenamefont {Jansen}, \citenamefont {Manigrasso},
  \citenamefont {Scapellato},\ and\ \citenamefont
  {Steffens}}]{Alexandrou:2020qtt}%
  \BibitemOpen
  \bibfield  {author} {\bibinfo {author} {\bibfnamefont {C.}~\bibnamefont
  {Alexandrou}}, \bibinfo {author} {\bibfnamefont {K.}~\bibnamefont {Cichy}},
  \bibinfo {author} {\bibfnamefont {M.}~\bibnamefont {Constantinou}}, \bibinfo
  {author} {\bibfnamefont {J.~R.}\ \bibnamefont {Green}}, \bibinfo {author}
  {\bibfnamefont {K.}~\bibnamefont {Hadjiyiannakou}}, \bibinfo {author}
  {\bibfnamefont {K.}~\bibnamefont {Jansen}}, \bibinfo {author} {\bibfnamefont
  {F.}~\bibnamefont {Manigrasso}}, \bibinfo {author} {\bibfnamefont
  {A.}~\bibnamefont {Scapellato}}, \ and\ \bibinfo {author} {\bibfnamefont
  {F.}~\bibnamefont {Steffens}},\ }\href {\doibase 10.1103/PhysRevD.103.094512}
  {\bibfield  {journal} {\bibinfo  {journal} {Phys. Rev. D}\ }\textbf {\bibinfo
  {volume} {103}},\ \bibinfo {pages} {094512} (\bibinfo {year}
  {2021}{\natexlab{a}})},\ \Eprint {http://arxiv.org/abs/2011.00964}
  {arXiv:2011.00964 [hep-lat]} \BibitemShut {NoStop}%
\bibitem [{\citenamefont {Alexandrou}\ \emph
  {et~al.}(2021{\natexlab{b}})\citenamefont {Alexandrou}, \citenamefont
  {Constantinou}, \citenamefont {Hadjiyiannakou}, \citenamefont {Jansen},\ and\
  \citenamefont {Manigrasso}}]{Alexandrou:2020uyt}%
  \BibitemOpen
  \bibfield  {author} {\bibinfo {author} {\bibfnamefont {C.}~\bibnamefont
  {Alexandrou}}, \bibinfo {author} {\bibfnamefont {M.}~\bibnamefont
  {Constantinou}}, \bibinfo {author} {\bibfnamefont {K.}~\bibnamefont
  {Hadjiyiannakou}}, \bibinfo {author} {\bibfnamefont {K.}~\bibnamefont
  {Jansen}}, \ and\ \bibinfo {author} {\bibfnamefont {F.}~\bibnamefont
  {Manigrasso}},\ }\href {\doibase 10.1103/PhysRevLett.126.102003} {\bibfield
  {journal} {\bibinfo  {journal} {Phys. Rev. Lett.}\ }\textbf {\bibinfo
  {volume} {126}},\ \bibinfo {pages} {102003} (\bibinfo {year}
  {2021}{\natexlab{b}})},\ \Eprint {http://arxiv.org/abs/2009.13061}
  {arXiv:2009.13061 [hep-lat]} \BibitemShut {NoStop}%
\bibitem [{\citenamefont {Fan}\ \emph {et~al.}(2020)\citenamefont {Fan},
  \citenamefont {Gao}, \citenamefont {Li}, \citenamefont {Lin}, \citenamefont
  {Karthik}, \citenamefont {Mukherjee}, \citenamefont {Petreczky},
  \citenamefont {Syritsyn}, \citenamefont {Yang},\ and\ \citenamefont
  {Zhang}}]{Fan:2020nzz}%
  \BibitemOpen
  \bibfield  {author} {\bibinfo {author} {\bibfnamefont {Z.}~\bibnamefont
  {Fan}}, \bibinfo {author} {\bibfnamefont {X.}~\bibnamefont {Gao}}, \bibinfo
  {author} {\bibfnamefont {R.}~\bibnamefont {Li}}, \bibinfo {author}
  {\bibfnamefont {H.-W.}\ \bibnamefont {Lin}}, \bibinfo {author} {\bibfnamefont
  {N.}~\bibnamefont {Karthik}}, \bibinfo {author} {\bibfnamefont
  {S.}~\bibnamefont {Mukherjee}}, \bibinfo {author} {\bibfnamefont
  {P.}~\bibnamefont {Petreczky}}, \bibinfo {author} {\bibfnamefont
  {S.}~\bibnamefont {Syritsyn}}, \bibinfo {author} {\bibfnamefont {Y.-B.}\
  \bibnamefont {Yang}}, \ and\ \bibinfo {author} {\bibfnamefont
  {R.}~\bibnamefont {Zhang}},\ }\href {\doibase 10.1103/PhysRevD.102.074504}
  {\bibfield  {journal} {\bibinfo  {journal} {Phys. Rev. D}\ }\textbf {\bibinfo
  {volume} {102}},\ \bibinfo {pages} {074504} (\bibinfo {year} {2020})},\
  \Eprint {http://arxiv.org/abs/2005.12015} {arXiv:2005.12015 [hep-lat]}
  \BibitemShut {NoStop}%
\bibitem [{\citenamefont {Lin}\ \emph {et~al.}(2020)\citenamefont {Lin},
  \citenamefont {Chen},\ and\ \citenamefont {Zhang}}]{Lin:2020fsj}%
  \BibitemOpen
  \bibfield  {author} {\bibinfo {author} {\bibfnamefont {H.-W.}\ \bibnamefont
  {Lin}}, \bibinfo {author} {\bibfnamefont {J.-W.}\ \bibnamefont {Chen}}, \
  and\ \bibinfo {author} {\bibfnamefont {R.}~\bibnamefont {Zhang}},\
  }\href@noop {} {\  (\bibinfo {year} {2020})},\ \Eprint
  {http://arxiv.org/abs/2011.14971} {arXiv:2011.14971 [hep-lat]} \BibitemShut
  {NoStop}%
\bibitem [{\citenamefont {Zhang}\ \emph {et~al.}(2021)\citenamefont {Zhang},
  \citenamefont {Lin},\ and\ \citenamefont {Yoon}}]{Zhang:2020dkn}%
  \BibitemOpen
  \bibfield  {author} {\bibinfo {author} {\bibfnamefont {R.}~\bibnamefont
  {Zhang}}, \bibinfo {author} {\bibfnamefont {H.-W.}\ \bibnamefont {Lin}}, \
  and\ \bibinfo {author} {\bibfnamefont {B.}~\bibnamefont {Yoon}},\ }\href
  {\doibase 10.1103/PhysRevD.104.094511} {\bibfield  {journal} {\bibinfo
  {journal} {Phys. Rev. D}\ }\textbf {\bibinfo {volume} {104}},\ \bibinfo
  {pages} {094511} (\bibinfo {year} {2021})},\ \Eprint
  {http://arxiv.org/abs/2005.01124} {arXiv:2005.01124 [hep-lat]} \BibitemShut
  {NoStop}%
\bibitem [{\citenamefont {Gao}\ \emph {et~al.}(2022{\natexlab{a}})\citenamefont
  {Gao}, \citenamefont {Hanlon}, \citenamefont {Mukherjee}, \citenamefont
  {Petreczky}, \citenamefont {Scior}, \citenamefont {Syritsyn},\ and\
  \citenamefont {Zhao}}]{Gao:2021dbh}%
  \BibitemOpen
  \bibfield  {author} {\bibinfo {author} {\bibfnamefont {X.}~\bibnamefont
  {Gao}}, \bibinfo {author} {\bibfnamefont {A.~D.}\ \bibnamefont {Hanlon}},
  \bibinfo {author} {\bibfnamefont {S.}~\bibnamefont {Mukherjee}}, \bibinfo
  {author} {\bibfnamefont {P.}~\bibnamefont {Petreczky}}, \bibinfo {author}
  {\bibfnamefont {P.}~\bibnamefont {Scior}}, \bibinfo {author} {\bibfnamefont
  {S.}~\bibnamefont {Syritsyn}}, \ and\ \bibinfo {author} {\bibfnamefont
  {Y.}~\bibnamefont {Zhao}},\ }\href {\doibase 10.1103/PhysRevLett.128.142003}
  {\bibfield  {journal} {\bibinfo  {journal} {Phys. Rev. Lett.}\ }\textbf
  {\bibinfo {volume} {128}},\ \bibinfo {pages} {142003} (\bibinfo {year}
  {2022}{\natexlab{a}})},\ \Eprint {http://arxiv.org/abs/2112.02208}
  {arXiv:2112.02208 [hep-lat]} \BibitemShut {NoStop}%
\bibitem [{\citenamefont {Hua}\ \emph {et~al.}(2021)\citenamefont {Hua},
  \citenamefont {Chu}, \citenamefont {Sun}, \citenamefont {Wang}, \citenamefont
  {Xu}, \citenamefont {Yang}, \citenamefont {Zhang},\ and\ \citenamefont
  {Zhang}}]{Hua:2020gnw}%
  \BibitemOpen
  \bibfield  {author} {\bibinfo {author} {\bibfnamefont {J.}~\bibnamefont
  {Hua}}, \bibinfo {author} {\bibfnamefont {M.-H.}\ \bibnamefont {Chu}},
  \bibinfo {author} {\bibfnamefont {P.}~\bibnamefont {Sun}}, \bibinfo {author}
  {\bibfnamefont {W.}~\bibnamefont {Wang}}, \bibinfo {author} {\bibfnamefont
  {J.}~\bibnamefont {Xu}}, \bibinfo {author} {\bibfnamefont {Y.-B.}\
  \bibnamefont {Yang}}, \bibinfo {author} {\bibfnamefont {J.-H.}\ \bibnamefont
  {Zhang}}, \ and\ \bibinfo {author} {\bibfnamefont {Q.-A.}\ \bibnamefont
  {Zhang}} (\bibinfo {collaboration} {Lattice Parton}),\ }\href {\doibase
  10.1103/PhysRevLett.127.062002} {\bibfield  {journal} {\bibinfo  {journal}
  {Phys. Rev. Lett.}\ }\textbf {\bibinfo {volume} {127}},\ \bibinfo {pages}
  {062002} (\bibinfo {year} {2021})},\ \Eprint
  {http://arxiv.org/abs/2011.09788} {arXiv:2011.09788 [hep-lat]} \BibitemShut
  {NoStop}%
\bibitem [{\citenamefont {Jo\'o}\ \emph
  {et~al.}(2019{\natexlab{a}})\citenamefont {Jo\'o}, \citenamefont {Karpie},
  \citenamefont {Orginos}, \citenamefont {Radyushkin}, \citenamefont
  {Richards},\ and\ \citenamefont {Zafeiropoulos}}]{Joo:2019jct}%
  \BibitemOpen
  \bibfield  {author} {\bibinfo {author} {\bibfnamefont {B.}~\bibnamefont
  {Jo\'o}}, \bibinfo {author} {\bibfnamefont {J.}~\bibnamefont {Karpie}},
  \bibinfo {author} {\bibfnamefont {K.}~\bibnamefont {Orginos}}, \bibinfo
  {author} {\bibfnamefont {A.}~\bibnamefont {Radyushkin}}, \bibinfo {author}
  {\bibfnamefont {D.}~\bibnamefont {Richards}}, \ and\ \bibinfo {author}
  {\bibfnamefont {S.}~\bibnamefont {Zafeiropoulos}},\ }\href {\doibase
  10.1007/JHEP12(2019)081} {\bibfield  {journal} {\bibinfo  {journal} {JHEP}\
  }\textbf {\bibinfo {volume} {12}},\ \bibinfo {pages} {081} (\bibinfo {year}
  {2019}{\natexlab{a}})},\ \Eprint {http://arxiv.org/abs/1908.09771}
  {arXiv:1908.09771 [hep-lat]} \BibitemShut {NoStop}%
\bibitem [{\citenamefont {Jo\'o}\ \emph
  {et~al.}(2019{\natexlab{b}})\citenamefont {Jo\'o}, \citenamefont {Karpie},
  \citenamefont {Orginos}, \citenamefont {Radyushkin}, \citenamefont
  {Richards}, \citenamefont {Sufian},\ and\ \citenamefont
  {Zafeiropoulos}}]{Joo:2019bzr}%
  \BibitemOpen
  \bibfield  {author} {\bibinfo {author} {\bibfnamefont {B.}~\bibnamefont
  {Jo\'o}}, \bibinfo {author} {\bibfnamefont {J.}~\bibnamefont {Karpie}},
  \bibinfo {author} {\bibfnamefont {K.}~\bibnamefont {Orginos}}, \bibinfo
  {author} {\bibfnamefont {A.~V.}\ \bibnamefont {Radyushkin}}, \bibinfo
  {author} {\bibfnamefont {D.~G.}\ \bibnamefont {Richards}}, \bibinfo {author}
  {\bibfnamefont {R.~S.}\ \bibnamefont {Sufian}}, \ and\ \bibinfo {author}
  {\bibfnamefont {S.}~\bibnamefont {Zafeiropoulos}},\ }\href {\doibase
  10.1103/PhysRevD.100.114512} {\bibfield  {journal} {\bibinfo  {journal}
  {Phys. Rev. D}\ }\textbf {\bibinfo {volume} {100}},\ \bibinfo {pages}
  {114512} (\bibinfo {year} {2019}{\natexlab{b}})},\ \Eprint
  {http://arxiv.org/abs/1909.08517} {arXiv:1909.08517 [hep-lat]} \BibitemShut
  {NoStop}%
\bibitem [{\citenamefont {Egerer}\ \emph {et~al.}(2022)\citenamefont {Egerer}
  \emph {et~al.}}]{HadStruc:2021qdf}%
  \BibitemOpen
  \bibfield  {author} {\bibinfo {author} {\bibfnamefont {C.}~\bibnamefont
  {Egerer}} \emph {et~al.} (\bibinfo {collaboration} {HadStruc}),\ }\href
  {\doibase 10.1103/PhysRevD.105.034507} {\bibfield  {journal} {\bibinfo
  {journal} {Phys. Rev. D}\ }\textbf {\bibinfo {volume} {105}},\ \bibinfo
  {pages} {034507} (\bibinfo {year} {2022})},\ \Eprint
  {http://arxiv.org/abs/2111.01808} {arXiv:2111.01808 [hep-lat]} \BibitemShut
  {NoStop}%
\bibitem [{\citenamefont {Gao}\ \emph {et~al.}(2022{\natexlab{b}})\citenamefont
  {Gao}, \citenamefont {Hanlon}, \citenamefont {Karthik}, \citenamefont
  {Mukherjee}, \citenamefont {Petreczky}, \citenamefont {Scior}, \citenamefont
  {Syritsyn},\ and\ \citenamefont {Zhao}}]{Gao:2022vyh}%
  \BibitemOpen
  \bibfield  {author} {\bibinfo {author} {\bibfnamefont {X.}~\bibnamefont
  {Gao}}, \bibinfo {author} {\bibfnamefont {A.~D.}\ \bibnamefont {Hanlon}},
  \bibinfo {author} {\bibfnamefont {N.}~\bibnamefont {Karthik}}, \bibinfo
  {author} {\bibfnamefont {S.}~\bibnamefont {Mukherjee}}, \bibinfo {author}
  {\bibfnamefont {P.}~\bibnamefont {Petreczky}}, \bibinfo {author}
  {\bibfnamefont {P.}~\bibnamefont {Scior}}, \bibinfo {author} {\bibfnamefont
  {S.}~\bibnamefont {Syritsyn}}, \ and\ \bibinfo {author} {\bibfnamefont
  {Y.}~\bibnamefont {Zhao}},\ }\href@noop {} {\  (\bibinfo {year}
  {2022}{\natexlab{b}})},\ \Eprint {http://arxiv.org/abs/2206.04084}
  {arXiv:2206.04084 [hep-lat]} \BibitemShut {NoStop}%
\bibitem [{\citenamefont {Bhattacharya}\ \emph {et~al.}(2021)\citenamefont
  {Bhattacharya}, \citenamefont {Cichy}, \citenamefont {Constantinou},
  \citenamefont {Metz}, \citenamefont {Scapellato},\ and\ \citenamefont
  {Steffens}}]{Bhattacharya:2021moj}%
  \BibitemOpen
  \bibfield  {author} {\bibinfo {author} {\bibfnamefont {S.}~\bibnamefont
  {Bhattacharya}}, \bibinfo {author} {\bibfnamefont {K.}~\bibnamefont {Cichy}},
  \bibinfo {author} {\bibfnamefont {M.}~\bibnamefont {Constantinou}}, \bibinfo
  {author} {\bibfnamefont {A.}~\bibnamefont {Metz}}, \bibinfo {author}
  {\bibfnamefont {A.}~\bibnamefont {Scapellato}}, \ and\ \bibinfo {author}
  {\bibfnamefont {F.}~\bibnamefont {Steffens}},\ }\href {\doibase
  10.1103/PhysRevD.104.114510} {\bibfield  {journal} {\bibinfo  {journal}
  {Phys. Rev. D}\ }\textbf {\bibinfo {volume} {104}},\ \bibinfo {pages}
  {114510} (\bibinfo {year} {2021})},\ \Eprint
  {http://arxiv.org/abs/2107.02574} {arXiv:2107.02574 [hep-lat]} \BibitemShut
  {NoStop}%
\bibitem [{\citenamefont {Alexandrou}\ \emph
  {et~al.}(2021{\natexlab{c}})\citenamefont {Alexandrou}, \citenamefont
  {Constantinou}, \citenamefont {Hadjiyiannakou}, \citenamefont {Jansen},\ and\
  \citenamefont {Manigrasso}}]{Alexandrou:2021oih}%
  \BibitemOpen
  \bibfield  {author} {\bibinfo {author} {\bibfnamefont {C.}~\bibnamefont
  {Alexandrou}}, \bibinfo {author} {\bibfnamefont {M.}~\bibnamefont
  {Constantinou}}, \bibinfo {author} {\bibfnamefont {K.}~\bibnamefont
  {Hadjiyiannakou}}, \bibinfo {author} {\bibfnamefont {K.}~\bibnamefont
  {Jansen}}, \ and\ \bibinfo {author} {\bibfnamefont {F.}~\bibnamefont
  {Manigrasso}},\ }\href {\doibase 10.1103/PhysRevD.104.054503} {\bibfield
  {journal} {\bibinfo  {journal} {Phys. Rev. D}\ }\textbf {\bibinfo {volume}
  {104}},\ \bibinfo {pages} {054503} (\bibinfo {year} {2021}{\natexlab{c}})},\
  \Eprint {http://arxiv.org/abs/2106.16065} {arXiv:2106.16065 [hep-lat]}
  \BibitemShut {NoStop}%
\bibitem [{\citenamefont {Detmold}\ \emph {et~al.}(2022)\citenamefont
  {Detmold}, \citenamefont {Grebe}, \citenamefont {Kanamori}, \citenamefont
  {Lin}, \citenamefont {Mondal}, \citenamefont {Perry},\ and\ \citenamefont
  {Zhao}}]{Detmold:2021qln}%
  \BibitemOpen
  \bibfield  {author} {\bibinfo {author} {\bibfnamefont {W.}~\bibnamefont
  {Detmold}}, \bibinfo {author} {\bibfnamefont {A.~V.}\ \bibnamefont {Grebe}},
  \bibinfo {author} {\bibfnamefont {I.}~\bibnamefont {Kanamori}}, \bibinfo
  {author} {\bibfnamefont {C.~J.~D.}\ \bibnamefont {Lin}}, \bibinfo {author}
  {\bibfnamefont {S.}~\bibnamefont {Mondal}}, \bibinfo {author} {\bibfnamefont
  {R.~J.}\ \bibnamefont {Perry}}, \ and\ \bibinfo {author} {\bibfnamefont
  {Y.}~\bibnamefont {Zhao}} (\bibinfo {collaboration} {HOPE}),\ }\href
  {\doibase 10.1103/PhysRevD.105.034506} {\bibfield  {journal} {\bibinfo
  {journal} {Phys. Rev. D}\ }\textbf {\bibinfo {volume} {105}},\ \bibinfo
  {pages} {034506} (\bibinfo {year} {2022})},\ \Eprint
  {http://arxiv.org/abs/2109.15241} {arXiv:2109.15241 [hep-lat]} \BibitemShut
  {NoStop}%
\bibitem [{\citenamefont {Bali}\ \emph {et~al.}(2018)\citenamefont {Bali},
  \citenamefont {Braun}, \citenamefont {Gl\"a\ss{}le}, \citenamefont
  {G\"ockeler}, \citenamefont {Gruber}, \citenamefont {Hutzler}, \citenamefont
  {Korcyl}, \citenamefont {Sch\"afer}, \citenamefont {Wein},\ and\
  \citenamefont {Zhang}}]{Bali:2018spj}%
  \BibitemOpen
  \bibfield  {author} {\bibinfo {author} {\bibfnamefont {G.~S.}\ \bibnamefont
  {Bali}}, \bibinfo {author} {\bibfnamefont {V.~M.}\ \bibnamefont {Braun}},
  \bibinfo {author} {\bibfnamefont {B.}~\bibnamefont {Gl\"a\ss{}le}}, \bibinfo
  {author} {\bibfnamefont {M.}~\bibnamefont {G\"ockeler}}, \bibinfo {author}
  {\bibfnamefont {M.}~\bibnamefont {Gruber}}, \bibinfo {author} {\bibfnamefont
  {F.}~\bibnamefont {Hutzler}}, \bibinfo {author} {\bibfnamefont
  {P.}~\bibnamefont {Korcyl}}, \bibinfo {author} {\bibfnamefont
  {A.}~\bibnamefont {Sch\"afer}}, \bibinfo {author} {\bibfnamefont
  {P.}~\bibnamefont {Wein}}, \ and\ \bibinfo {author} {\bibfnamefont {J.-H.}\
  \bibnamefont {Zhang}},\ }\href {\doibase 10.1103/PhysRevD.98.094507}
  {\bibfield  {journal} {\bibinfo  {journal} {Phys. Rev. D}\ }\textbf {\bibinfo
  {volume} {98}},\ \bibinfo {pages} {094507} (\bibinfo {year} {2018})},\
  \Eprint {http://arxiv.org/abs/1807.06671} {arXiv:1807.06671 [hep-lat]}
  \BibitemShut {NoStop}%
\bibitem [{\citenamefont {Bhat}\ \emph {et~al.}(2022)\citenamefont {Bhat},
  \citenamefont {Chomicki}, \citenamefont {Cichy}, \citenamefont
  {Constantinou}, \citenamefont {Green},\ and\ \citenamefont
  {Scapellato}}]{Bhat:2022zrw}%
  \BibitemOpen
  \bibfield  {author} {\bibinfo {author} {\bibfnamefont {M.}~\bibnamefont
  {Bhat}}, \bibinfo {author} {\bibfnamefont {W.}~\bibnamefont {Chomicki}},
  \bibinfo {author} {\bibfnamefont {K.}~\bibnamefont {Cichy}}, \bibinfo
  {author} {\bibfnamefont {M.}~\bibnamefont {Constantinou}}, \bibinfo {author}
  {\bibfnamefont {J.~R.}\ \bibnamefont {Green}}, \ and\ \bibinfo {author}
  {\bibfnamefont {A.}~\bibnamefont {Scapellato}},\ }\href@noop {} {\  (\bibinfo
  {year} {2022})},\ \Eprint {http://arxiv.org/abs/2205.07585} {arXiv:2205.07585
  [hep-lat]} \BibitemShut {NoStop}%
\bibitem [{\citenamefont {Liu}\ and\ \citenamefont {Dong}(1994)}]{Liu:1993cv}%
  \BibitemOpen
  \bibfield  {author} {\bibinfo {author} {\bibfnamefont {K.-F.}\ \bibnamefont
  {Liu}}\ and\ \bibinfo {author} {\bibfnamefont {S.-J.}\ \bibnamefont {Dong}},\
  }\href {\doibase 10.1103/PhysRevLett.72.1790} {\bibfield  {journal} {\bibinfo
   {journal} {Phys. Rev. Lett.}\ }\textbf {\bibinfo {volume} {72}},\ \bibinfo
  {pages} {1790} (\bibinfo {year} {1994})},\ \Eprint
  {http://arxiv.org/abs/hep-ph/9306299} {arXiv:hep-ph/9306299} \BibitemShut
  {NoStop}%
\bibitem [{\citenamefont {Detmold}\ and\ \citenamefont
  {Lin}(2006)}]{Detmold:2005gg}%
  \BibitemOpen
  \bibfield  {author} {\bibinfo {author} {\bibfnamefont {W.}~\bibnamefont
  {Detmold}}\ and\ \bibinfo {author} {\bibfnamefont {C.~J.~D.}\ \bibnamefont
  {Lin}},\ }\href {\doibase 10.1103/PhysRevD.73.014501} {\bibfield  {journal}
  {\bibinfo  {journal} {Phys. Rev. D}\ }\textbf {\bibinfo {volume} {73}},\
  \bibinfo {pages} {014501} (\bibinfo {year} {2006})},\ \Eprint
  {http://arxiv.org/abs/hep-lat/0507007} {arXiv:hep-lat/0507007} \BibitemShut
  {NoStop}%
\bibitem [{\citenamefont {Braun}\ and\ \citenamefont
  {M\"uller}(2008)}]{Braun:2007wv}%
  \BibitemOpen
  \bibfield  {author} {\bibinfo {author} {\bibfnamefont {V.}~\bibnamefont
  {Braun}}\ and\ \bibinfo {author} {\bibfnamefont {D.}~\bibnamefont
  {M\"uller}},\ }\href {\doibase 10.1140/epjc/s10052-008-0608-4} {\bibfield
  {journal} {\bibinfo  {journal} {Eur. Phys. J. C}\ }\textbf {\bibinfo {volume}
  {55}},\ \bibinfo {pages} {349} (\bibinfo {year} {2008})},\ \Eprint
  {http://arxiv.org/abs/0709.1348} {arXiv:0709.1348 [hep-ph]} \BibitemShut
  {NoStop}%
\bibitem [{\citenamefont {Ji}(2013)}]{Ji:2013dva}%
  \BibitemOpen
  \bibfield  {author} {\bibinfo {author} {\bibfnamefont {X.}~\bibnamefont
  {Ji}},\ }\href {\doibase 10.1103/PhysRevLett.110.262002} {\bibfield
  {journal} {\bibinfo  {journal} {Phys. Rev. Lett.}\ }\textbf {\bibinfo
  {volume} {110}},\ \bibinfo {pages} {262002} (\bibinfo {year} {2013})},\
  \Eprint {http://arxiv.org/abs/1305.1539} {arXiv:1305.1539 [hep-ph]}
  \BibitemShut {NoStop}%
\bibitem [{\citenamefont {Ma}\ and\ \citenamefont
  {Qiu}(2018{\natexlab{a}})}]{Ma:2014jla}%
  \BibitemOpen
  \bibfield  {author} {\bibinfo {author} {\bibfnamefont {Y.-Q.}\ \bibnamefont
  {Ma}}\ and\ \bibinfo {author} {\bibfnamefont {J.-W.}\ \bibnamefont {Qiu}},\
  }\href {\doibase 10.1103/PhysRevD.98.074021} {\bibfield  {journal} {\bibinfo
  {journal} {Phys. Rev. D}\ }\textbf {\bibinfo {volume} {98}},\ \bibinfo
  {pages} {074021} (\bibinfo {year} {2018}{\natexlab{a}})},\ \Eprint
  {http://arxiv.org/abs/1404.6860} {arXiv:1404.6860 [hep-ph]} \BibitemShut
  {NoStop}%
\bibitem [{\citenamefont {Ma}\ and\ \citenamefont
  {Qiu}(2018{\natexlab{b}})}]{Ma:2017pxb}%
  \BibitemOpen
  \bibfield  {author} {\bibinfo {author} {\bibfnamefont {Y.-Q.}\ \bibnamefont
  {Ma}}\ and\ \bibinfo {author} {\bibfnamefont {J.-W.}\ \bibnamefont {Qiu}},\
  }\href {\doibase 10.1103/PhysRevLett.120.022003} {\bibfield  {journal}
  {\bibinfo  {journal} {Phys. Rev. Lett.}\ }\textbf {\bibinfo {volume} {120}},\
  \bibinfo {pages} {022003} (\bibinfo {year} {2018}{\natexlab{b}})},\ \Eprint
  {http://arxiv.org/abs/1709.03018} {arXiv:1709.03018 [hep-ph]} \BibitemShut
  {NoStop}%
\bibitem [{\citenamefont {Fan}\ \emph {et~al.}(2018)\citenamefont {Fan},
  \citenamefont {Yang}, \citenamefont {Anthony}, \citenamefont {Lin},\ and\
  \citenamefont {Liu}}]{Fan:2018dxu}%
  \BibitemOpen
  \bibfield  {author} {\bibinfo {author} {\bibfnamefont {Z.-Y.}\ \bibnamefont
  {Fan}}, \bibinfo {author} {\bibfnamefont {Y.-B.}\ \bibnamefont {Yang}},
  \bibinfo {author} {\bibfnamefont {A.}~\bibnamefont {Anthony}}, \bibinfo
  {author} {\bibfnamefont {H.-W.}\ \bibnamefont {Lin}}, \ and\ \bibinfo
  {author} {\bibfnamefont {K.-F.}\ \bibnamefont {Liu}},\ }\href {\doibase
  10.1103/PhysRevLett.121.242001} {\bibfield  {journal} {\bibinfo  {journal}
  {Phys. Rev. Lett.}\ }\textbf {\bibinfo {volume} {121}},\ \bibinfo {pages}
  {242001} (\bibinfo {year} {2018})},\ \Eprint
  {http://arxiv.org/abs/1808.02077} {arXiv:1808.02077 [hep-lat]} \BibitemShut
  {NoStop}%
\bibitem [{\citenamefont {Fan}\ \emph {et~al.}(2021)\citenamefont {Fan},
  \citenamefont {Zhang},\ and\ \citenamefont {Lin}}]{Fan:2020cpa}%
  \BibitemOpen
  \bibfield  {author} {\bibinfo {author} {\bibfnamefont {Z.}~\bibnamefont
  {Fan}}, \bibinfo {author} {\bibfnamefont {R.}~\bibnamefont {Zhang}}, \ and\
  \bibinfo {author} {\bibfnamefont {H.-W.}\ \bibnamefont {Lin}},\ }\href
  {\doibase 10.1142/S0217751X21500809} {\bibfield  {journal} {\bibinfo
  {journal} {Int. J. Mod. Phys. A}\ }\textbf {\bibinfo {volume} {36}},\
  \bibinfo {pages} {2150080} (\bibinfo {year} {2021})},\ \Eprint
  {http://arxiv.org/abs/2007.16113} {arXiv:2007.16113 [hep-lat]} \BibitemShut
  {NoStop}%
\bibitem [{\citenamefont {Khan}\ \emph
  {et~al.}(2021{\natexlab{a}})\citenamefont {Khan} \emph
  {et~al.}}]{HadStruc:2021wmh}%
  \BibitemOpen
  \bibfield  {author} {\bibinfo {author} {\bibfnamefont {T.}~\bibnamefont
  {Khan}} \emph {et~al.} (\bibinfo {collaboration} {HadStruc}),\ }\href
  {\doibase 10.1103/PhysRevD.104.094516} {\bibfield  {journal} {\bibinfo
  {journal} {Phys. Rev. D}\ }\textbf {\bibinfo {volume} {104}},\ \bibinfo
  {pages} {094516} (\bibinfo {year} {2021}{\natexlab{a}})},\ \Eprint
  {http://arxiv.org/abs/2107.08960} {arXiv:2107.08960 [hep-lat]} \BibitemShut
  {NoStop}%
\bibitem [{\citenamefont {Fan}\ and\ \citenamefont {Lin}(2021)}]{Fan:2021bcr}%
  \BibitemOpen
  \bibfield  {author} {\bibinfo {author} {\bibfnamefont {Z.}~\bibnamefont
  {Fan}}\ and\ \bibinfo {author} {\bibfnamefont {H.-W.}\ \bibnamefont {Lin}},\
  }\href {\doibase 10.1016/j.physletb.2021.136778} {\bibfield  {journal}
  {\bibinfo  {journal} {Phys. Lett. B}\ }\textbf {\bibinfo {volume} {823}},\
  \bibinfo {pages} {136778} (\bibinfo {year} {2021})},\ \Eprint
  {http://arxiv.org/abs/2104.06372} {arXiv:2104.06372 [hep-lat]} \BibitemShut
  {NoStop}%
\bibitem [{\citenamefont {Salas-Chavira}\ \emph {et~al.}(2021)\citenamefont
  {Salas-Chavira}, \citenamefont {Fan},\ and\ \citenamefont
  {Lin}}]{Salas-Chavira:2021wui}%
  \BibitemOpen
  \bibfield  {author} {\bibinfo {author} {\bibfnamefont {A.}~\bibnamefont
  {Salas-Chavira}}, \bibinfo {author} {\bibfnamefont {Z.}~\bibnamefont {Fan}},
  \ and\ \bibinfo {author} {\bibfnamefont {H.-W.}\ \bibnamefont {Lin}},\
  }\href@noop {} {\  (\bibinfo {year} {2021})},\ \Eprint
  {http://arxiv.org/abs/2112.03124} {arXiv:2112.03124 [hep-lat]} \BibitemShut
  {NoStop}%
\bibitem [{\citenamefont {Cichy}\ and\ \citenamefont
  {Constantinou}(2019)}]{Cichy:2018mum}%
  \BibitemOpen
  \bibfield  {author} {\bibinfo {author} {\bibfnamefont {K.}~\bibnamefont
  {Cichy}}\ and\ \bibinfo {author} {\bibfnamefont {M.}~\bibnamefont
  {Constantinou}},\ }\href {\doibase 10.1155/2019/3036904} {\bibfield
  {journal} {\bibinfo  {journal} {Adv. High Energy Phys.}\ }\textbf {\bibinfo
  {volume} {2019}},\ \bibinfo {pages} {3036904} (\bibinfo {year} {2019})},\
  \Eprint {http://arxiv.org/abs/1811.07248} {arXiv:1811.07248 [hep-lat]}
  \BibitemShut {NoStop}%
\bibitem [{\citenamefont {Constantinou}\ \emph {et~al.}(2021)\citenamefont
  {Constantinou} \emph {et~al.}}]{Constantinou:2020hdm}%
  \BibitemOpen
  \bibfield  {author} {\bibinfo {author} {\bibfnamefont {M.}~\bibnamefont
  {Constantinou}} \emph {et~al.},\ }\href {\doibase 10.1016/j.ppnp.2021.103908}
  {\bibfield  {journal} {\bibinfo  {journal} {Prog. Part. Nucl. Phys.}\
  }\textbf {\bibinfo {volume} {121}},\ \bibinfo {pages} {103908} (\bibinfo
  {year} {2021})},\ \Eprint {http://arxiv.org/abs/2006.08636} {arXiv:2006.08636
  [hep-ph]} \BibitemShut {NoStop}%
\bibitem [{\citenamefont {Ji}\ \emph {et~al.}(2021)\citenamefont {Ji},
  \citenamefont {Liu}, \citenamefont {Liu}, \citenamefont {Zhang},\ and\
  \citenamefont {Zhao}}]{Ji:2020ect}%
  \BibitemOpen
  \bibfield  {author} {\bibinfo {author} {\bibfnamefont {X.}~\bibnamefont
  {Ji}}, \bibinfo {author} {\bibfnamefont {Y.-S.}\ \bibnamefont {Liu}},
  \bibinfo {author} {\bibfnamefont {Y.}~\bibnamefont {Liu}}, \bibinfo {author}
  {\bibfnamefont {J.-H.}\ \bibnamefont {Zhang}}, \ and\ \bibinfo {author}
  {\bibfnamefont {Y.}~\bibnamefont {Zhao}},\ }\href {\doibase
  10.1103/RevModPhys.93.035005} {\bibfield  {journal} {\bibinfo  {journal}
  {Rev. Mod. Phys.}\ }\textbf {\bibinfo {volume} {93}},\ \bibinfo {pages}
  {035005} (\bibinfo {year} {2021})},\ \Eprint
  {http://arxiv.org/abs/2004.03543} {arXiv:2004.03543 [hep-ph]} \BibitemShut
  {NoStop}%
\bibitem [{\citenamefont {Constantinou}\ \emph {et~al.}(2022)\citenamefont
  {Constantinou} \emph {et~al.}}]{Constantinou:2022yye}%
  \BibitemOpen
  \bibfield  {author} {\bibinfo {author} {\bibfnamefont {M.}~\bibnamefont
  {Constantinou}} \emph {et~al.},\ }\href@noop {} {\  (\bibinfo {year}
  {2022})},\ \Eprint {http://arxiv.org/abs/2202.07193} {arXiv:2202.07193
  [hep-lat]} \BibitemShut {NoStop}%
\bibitem [{\citenamefont {Izubuchi}\ \emph {et~al.}(2018)\citenamefont
  {Izubuchi}, \citenamefont {Ji}, \citenamefont {Jin}, \citenamefont
  {Stewart},\ and\ \citenamefont {Zhao}}]{Izubuchi:2018srq}%
  \BibitemOpen
  \bibfield  {author} {\bibinfo {author} {\bibfnamefont {T.}~\bibnamefont
  {Izubuchi}}, \bibinfo {author} {\bibfnamefont {X.}~\bibnamefont {Ji}},
  \bibinfo {author} {\bibfnamefont {L.}~\bibnamefont {Jin}}, \bibinfo {author}
  {\bibfnamefont {I.~W.}\ \bibnamefont {Stewart}}, \ and\ \bibinfo {author}
  {\bibfnamefont {Y.}~\bibnamefont {Zhao}},\ }\href {\doibase
  10.1103/PhysRevD.98.056004} {\bibfield  {journal} {\bibinfo  {journal} {Phys.
  Rev. D}\ }\textbf {\bibinfo {volume} {98}},\ \bibinfo {pages} {056004}
  (\bibinfo {year} {2018})},\ \Eprint {http://arxiv.org/abs/1801.03917}
  {arXiv:1801.03917 [hep-ph]} \BibitemShut {NoStop}%
\bibitem [{\citenamefont {Ji}\ \emph {et~al.}(2018)\citenamefont {Ji},
  \citenamefont {Zhang},\ and\ \citenamefont {Zhao}}]{Ji:2017oey}%
  \BibitemOpen
  \bibfield  {author} {\bibinfo {author} {\bibfnamefont {X.}~\bibnamefont
  {Ji}}, \bibinfo {author} {\bibfnamefont {J.-H.}\ \bibnamefont {Zhang}}, \
  and\ \bibinfo {author} {\bibfnamefont {Y.}~\bibnamefont {Zhao}},\ }\href
  {\doibase 10.1103/PhysRevLett.120.112001} {\bibfield  {journal} {\bibinfo
  {journal} {Phys. Rev. Lett.}\ }\textbf {\bibinfo {volume} {120}},\ \bibinfo
  {pages} {112001} (\bibinfo {year} {2018})},\ \Eprint
  {http://arxiv.org/abs/1706.08962} {arXiv:1706.08962 [hep-ph]} \BibitemShut
  {NoStop}%
\bibitem [{\citenamefont {Green}\ \emph {et~al.}(2018)\citenamefont {Green},
  \citenamefont {Jansen},\ and\ \citenamefont {Steffens}}]{Green:2017xeu}%
  \BibitemOpen
  \bibfield  {author} {\bibinfo {author} {\bibfnamefont {J.}~\bibnamefont
  {Green}}, \bibinfo {author} {\bibfnamefont {K.}~\bibnamefont {Jansen}}, \
  and\ \bibinfo {author} {\bibfnamefont {F.}~\bibnamefont {Steffens}},\ }\href
  {\doibase 10.1103/PhysRevLett.121.022004} {\bibfield  {journal} {\bibinfo
  {journal} {Phys. Rev. Lett.}\ }\textbf {\bibinfo {volume} {121}},\ \bibinfo
  {pages} {022004} (\bibinfo {year} {2018})},\ \Eprint
  {http://arxiv.org/abs/1707.07152} {arXiv:1707.07152 [hep-lat]} \BibitemShut
  {NoStop}%
\bibitem [{\citenamefont {Zhang}\ \emph {et~al.}(2019)\citenamefont {Zhang},
  \citenamefont {Ji}, \citenamefont {Sch\"afer}, \citenamefont {Wang},\ and\
  \citenamefont {Zhao}}]{Zhang:2018diq}%
  \BibitemOpen
  \bibfield  {author} {\bibinfo {author} {\bibfnamefont {J.-H.}\ \bibnamefont
  {Zhang}}, \bibinfo {author} {\bibfnamefont {X.}~\bibnamefont {Ji}}, \bibinfo
  {author} {\bibfnamefont {A.}~\bibnamefont {Sch\"afer}}, \bibinfo {author}
  {\bibfnamefont {W.}~\bibnamefont {Wang}}, \ and\ \bibinfo {author}
  {\bibfnamefont {S.}~\bibnamefont {Zhao}},\ }\href {\doibase
  10.1103/PhysRevLett.122.142001} {\bibfield  {journal} {\bibinfo  {journal}
  {Phys. Rev. Lett.}\ }\textbf {\bibinfo {volume} {122}},\ \bibinfo {pages}
  {142001} (\bibinfo {year} {2019})},\ \Eprint
  {http://arxiv.org/abs/1808.10824} {arXiv:1808.10824 [hep-ph]} \BibitemShut
  {NoStop}%
\bibitem [{\citenamefont {Li}\ \emph {et~al.}(2019)\citenamefont {Li},
  \citenamefont {Ma},\ and\ \citenamefont {Qiu}}]{Li:2018tpe}%
  \BibitemOpen
  \bibfield  {author} {\bibinfo {author} {\bibfnamefont {Z.-Y.}\ \bibnamefont
  {Li}}, \bibinfo {author} {\bibfnamefont {Y.-Q.}\ \bibnamefont {Ma}}, \ and\
  \bibinfo {author} {\bibfnamefont {J.-W.}\ \bibnamefont {Qiu}},\ }\href
  {\doibase 10.1103/PhysRevLett.122.062002} {\bibfield  {journal} {\bibinfo
  {journal} {Phys. Rev. Lett.}\ }\textbf {\bibinfo {volume} {122}},\ \bibinfo
  {pages} {062002} (\bibinfo {year} {2019})},\ \Eprint
  {http://arxiv.org/abs/1809.01836} {arXiv:1809.01836 [hep-ph]} \BibitemShut
  {NoStop}%
\bibitem [{\citenamefont {Balitsky}\ \emph {et~al.}(2020)\citenamefont
  {Balitsky}, \citenamefont {Morris},\ and\ \citenamefont
  {Radyushkin}}]{Balitsky:2019krf}%
  \BibitemOpen
  \bibfield  {author} {\bibinfo {author} {\bibfnamefont {I.}~\bibnamefont
  {Balitsky}}, \bibinfo {author} {\bibfnamefont {W.}~\bibnamefont {Morris}}, \
  and\ \bibinfo {author} {\bibfnamefont {A.}~\bibnamefont {Radyushkin}},\
  }\href {\doibase 10.1016/j.physletb.2020.135621} {\bibfield  {journal}
  {\bibinfo  {journal} {Phys. Lett. B}\ }\textbf {\bibinfo {volume} {808}},\
  \bibinfo {pages} {135621} (\bibinfo {year} {2020})},\ \Eprint
  {http://arxiv.org/abs/1910.13963} {arXiv:1910.13963 [hep-ph]} \BibitemShut
  {NoStop}%
\bibitem [{\citenamefont {Gribov}\ and\ \citenamefont
  {Lipatov}(1972)}]{Gribov:1972ri}%
  \BibitemOpen
  \bibfield  {author} {\bibinfo {author} {\bibfnamefont {V.~N.}\ \bibnamefont
  {Gribov}}\ and\ \bibinfo {author} {\bibfnamefont {L.~N.}\ \bibnamefont
  {Lipatov}},\ }\href@noop {} {\bibfield  {journal} {\bibinfo  {journal} {Sov.
  J. Nucl. Phys.}\ }\textbf {\bibinfo {volume} {15}},\ \bibinfo {pages} {438}
  (\bibinfo {year} {1972})}\BibitemShut {NoStop}%
\bibitem [{\citenamefont {Altarelli}\ and\ \citenamefont
  {Parisi}(1977)}]{Altarelli:1977zs}%
  \BibitemOpen
  \bibfield  {author} {\bibinfo {author} {\bibfnamefont {G.}~\bibnamefont
  {Altarelli}}\ and\ \bibinfo {author} {\bibfnamefont {G.}~\bibnamefont
  {Parisi}},\ }\href {\doibase 10.1016/0550-3213(77)90384-4} {\bibfield
  {journal} {\bibinfo  {journal} {Nucl. Phys. B}\ }\textbf {\bibinfo {volume}
  {126}},\ \bibinfo {pages} {298} (\bibinfo {year} {1977})}\BibitemShut
  {NoStop}%
\bibitem [{\citenamefont {Dokshitzer}(1977)}]{Dokshitzer:1977sg}%
  \BibitemOpen
  \bibfield  {author} {\bibinfo {author} {\bibfnamefont {Y.~L.}\ \bibnamefont
  {Dokshitzer}},\ }\href@noop {} {\bibfield  {journal} {\bibinfo  {journal}
  {Sov. Phys. JETP}\ }\textbf {\bibinfo {volume} {46}},\ \bibinfo {pages} {641}
  (\bibinfo {year} {1977})}\BibitemShut {NoStop}%
\bibitem [{\citenamefont {Morningstar}\ and\ \citenamefont
  {Peardon}(2004)}]{Morningstar:2003gk}%
  \BibitemOpen
  \bibfield  {author} {\bibinfo {author} {\bibfnamefont {C.}~\bibnamefont
  {Morningstar}}\ and\ \bibinfo {author} {\bibfnamefont {M.~J.}\ \bibnamefont
  {Peardon}},\ }\href {\doibase 10.1103/PhysRevD.69.054501} {\bibfield
  {journal} {\bibinfo  {journal} {Phys. Rev. D}\ }\textbf {\bibinfo {volume}
  {69}},\ \bibinfo {pages} {054501} (\bibinfo {year} {2004})},\ \Eprint
  {http://arxiv.org/abs/hep-lat/0311018} {arXiv:hep-lat/0311018} \BibitemShut
  {NoStop}%
\bibitem [{\citenamefont {Edwards}\ \emph {et~al.}(2016)\citenamefont
  {Edwards}, \citenamefont {Jo\'o}, \citenamefont {Orginos}, \citenamefont
  {Richards},\ and\ \citenamefont {Winter}}]{lattices}%
  \BibitemOpen
  \bibfield  {author} {\bibinfo {author} {\bibfnamefont {R.}~\bibnamefont
  {Edwards}}, \bibinfo {author} {\bibfnamefont {B.}~\bibnamefont {Jo\'o}},
  \bibinfo {author} {\bibfnamefont {K.}~\bibnamefont {Orginos}}, \bibinfo
  {author} {\bibfnamefont {D.}~\bibnamefont {Richards}}, \ and\ \bibinfo
  {author} {\bibfnamefont {F.}~\bibnamefont {Winter}},\ }\href@noop {}
  {\bibfield  {journal} {\bibinfo  {journal} {U.S. 2+1 flavor clover lattice
  generation program}\ } (\bibinfo {year} {2016})},\ \Eprint
  {http://arxiv.org/abs/unpublished} {unpublished} \BibitemShut {NoStop}%
\bibitem [{\citenamefont {Duane}\ \emph {et~al.}(1987)\citenamefont {Duane},
  \citenamefont {Kennedy}, \citenamefont {Pendleton},\ and\ \citenamefont
  {Roweth}}]{Duane:1987de}%
  \BibitemOpen
  \bibfield  {author} {\bibinfo {author} {\bibfnamefont {S.}~\bibnamefont
  {Duane}}, \bibinfo {author} {\bibfnamefont {A.~D.}\ \bibnamefont {Kennedy}},
  \bibinfo {author} {\bibfnamefont {B.~J.}\ \bibnamefont {Pendleton}}, \ and\
  \bibinfo {author} {\bibfnamefont {D.}~\bibnamefont {Roweth}},\ }\href
  {\doibase 10.1016/0370-2693(87)91197-X} {\bibfield  {journal} {\bibinfo
  {journal} {Phys. Lett. B}\ }\textbf {\bibinfo {volume} {195}},\ \bibinfo
  {pages} {216} (\bibinfo {year} {1987})}\BibitemShut {NoStop}%
\bibitem [{\citenamefont {Borsanyi}(2012)}]{Borsanyi:2012zs}%
  \BibitemOpen
  \bibfield  {author} {\bibinfo {author} {\bibfnamefont {S.~t.}\ \bibnamefont
  {Borsanyi}},\ }\href {\doibase 10.1007/JHEP09(2012)010} {\bibfield  {journal}
  {\bibinfo  {journal} {JHEP}\ }\textbf {\bibinfo {volume} {09}},\ \bibinfo
  {pages} {010} (\bibinfo {year} {2012})},\ \Eprint
  {http://arxiv.org/abs/1203.4469} {arXiv:1203.4469 [hep-lat]} \BibitemShut
  {NoStop}%
\bibitem [{\citenamefont {L\"uscher}(2010)}]{Luscher:2010iy}%
  \BibitemOpen
  \bibfield  {author} {\bibinfo {author} {\bibfnamefont {M.}~\bibnamefont
  {L\"uscher}},\ }\href {\doibase 10.1007/JHEP08(2010)071} {\bibfield
  {journal} {\bibinfo  {journal} {JHEP}\ }\textbf {\bibinfo {volume} {08}},\
  \bibinfo {pages} {071} (\bibinfo {year} {2010})},\ \bibinfo {note} {[Erratum:
  JHEP 03, 092 (2014)]},\ \Eprint {http://arxiv.org/abs/1006.4518}
  {arXiv:1006.4518 [hep-lat]} \BibitemShut {NoStop}%
\bibitem [{\citenamefont {Luscher}\ and\ \citenamefont
  {Weisz}(2011)}]{Luscher:2011bx}%
  \BibitemOpen
  \bibfield  {author} {\bibinfo {author} {\bibfnamefont {M.}~\bibnamefont
  {Luscher}}\ and\ \bibinfo {author} {\bibfnamefont {P.}~\bibnamefont
  {Weisz}},\ }\href {\doibase 10.1007/JHEP02(2011)051} {\bibfield  {journal}
  {\bibinfo  {journal} {JHEP}\ }\textbf {\bibinfo {volume} {02}},\ \bibinfo
  {pages} {051} (\bibinfo {year} {2011})},\ \Eprint
  {http://arxiv.org/abs/1101.0963} {arXiv:1101.0963 [hep-th]} \BibitemShut
  {NoStop}%
\bibitem [{\citenamefont {Luscher}(2013)}]{Luscher:2013cpa}%
  \BibitemOpen
  \bibfield  {author} {\bibinfo {author} {\bibfnamefont {M.}~\bibnamefont
  {Luscher}},\ }\href {\doibase 10.1007/JHEP04(2013)123} {\bibfield  {journal}
  {\bibinfo  {journal} {JHEP}\ }\textbf {\bibinfo {volume} {04}},\ \bibinfo
  {pages} {123} (\bibinfo {year} {2013})},\ \Eprint
  {http://arxiv.org/abs/1302.5246} {arXiv:1302.5246 [hep-lat]} \BibitemShut
  {NoStop}%
\bibitem [{\citenamefont {Peardon}\ \emph {et~al.}(2009)\citenamefont
  {Peardon}, \citenamefont {Bulava}, \citenamefont {Foley}, \citenamefont
  {Morningstar}, \citenamefont {Dudek}, \citenamefont {Edwards}, \citenamefont
  {Joo}, \citenamefont {Lin}, \citenamefont {Richards},\ and\ \citenamefont
  {Juge}}]{Peardon:2009gh}%
  \BibitemOpen
  \bibfield  {author} {\bibinfo {author} {\bibfnamefont {M.}~\bibnamefont
  {Peardon}}, \bibinfo {author} {\bibfnamefont {J.}~\bibnamefont {Bulava}},
  \bibinfo {author} {\bibfnamefont {J.}~\bibnamefont {Foley}}, \bibinfo
  {author} {\bibfnamefont {C.}~\bibnamefont {Morningstar}}, \bibinfo {author}
  {\bibfnamefont {J.}~\bibnamefont {Dudek}}, \bibinfo {author} {\bibfnamefont
  {R.~G.}\ \bibnamefont {Edwards}}, \bibinfo {author} {\bibfnamefont
  {B.}~\bibnamefont {Joo}}, \bibinfo {author} {\bibfnamefont {H.-W.}\
  \bibnamefont {Lin}}, \bibinfo {author} {\bibfnamefont {D.~G.}\ \bibnamefont
  {Richards}}, \ and\ \bibinfo {author} {\bibfnamefont {K.~J.}\ \bibnamefont
  {Juge}} (\bibinfo {collaboration} {Hadron Spectrum}),\ }\href {\doibase
  10.1103/PhysRevD.80.054506} {\bibfield  {journal} {\bibinfo  {journal} {Phys.
  Rev. D}\ }\textbf {\bibinfo {volume} {80}},\ \bibinfo {pages} {054506}
  (\bibinfo {year} {2009})},\ \Eprint {http://arxiv.org/abs/0905.2160}
  {arXiv:0905.2160 [hep-lat]} \BibitemShut {NoStop}%
\bibitem [{\citenamefont {Bulava}\ \emph {et~al.}(2012)\citenamefont {Bulava},
  \citenamefont {Donnellan},\ and\ \citenamefont {Sommer}}]{Bulava:2011yz}%
  \BibitemOpen
  \bibfield  {author} {\bibinfo {author} {\bibfnamefont {J.}~\bibnamefont
  {Bulava}}, \bibinfo {author} {\bibfnamefont {M.}~\bibnamefont {Donnellan}}, \
  and\ \bibinfo {author} {\bibfnamefont {R.}~\bibnamefont {Sommer}},\ }\href
  {\doibase 10.1007/JHEP01(2012)140} {\bibfield  {journal} {\bibinfo  {journal}
  {JHEP}\ }\textbf {\bibinfo {volume} {01}},\ \bibinfo {pages} {140} (\bibinfo
  {year} {2012})},\ \Eprint {http://arxiv.org/abs/1108.3774} {arXiv:1108.3774
  [hep-lat]} \BibitemShut {NoStop}%
\bibitem [{\citenamefont {Edwards}\ \emph {et~al.}(2011)\citenamefont
  {Edwards}, \citenamefont {Dudek}, \citenamefont {Richards},\ and\
  \citenamefont {Wallace}}]{Edwards:2011jj}%
  \BibitemOpen
  \bibfield  {author} {\bibinfo {author} {\bibfnamefont {R.~G.}\ \bibnamefont
  {Edwards}}, \bibinfo {author} {\bibfnamefont {J.~J.}\ \bibnamefont {Dudek}},
  \bibinfo {author} {\bibfnamefont {D.~G.}\ \bibnamefont {Richards}}, \ and\
  \bibinfo {author} {\bibfnamefont {S.~J.}\ \bibnamefont {Wallace}},\ }\href
  {\doibase 10.1103/PhysRevD.84.074508} {\bibfield  {journal} {\bibinfo
  {journal} {Phys. Rev. D}\ }\textbf {\bibinfo {volume} {84}},\ \bibinfo
  {pages} {074508} (\bibinfo {year} {2011})},\ \Eprint
  {http://arxiv.org/abs/1104.5152} {arXiv:1104.5152 [hep-ph]} \BibitemShut
  {NoStop}%
\bibitem [{\citenamefont {Bali}\ \emph {et~al.}(2016)\citenamefont {Bali},
  \citenamefont {Lang}, \citenamefont {Musch},\ and\ \citenamefont
  {Sch\"afer}}]{Bali:2016lva}%
  \BibitemOpen
  \bibfield  {author} {\bibinfo {author} {\bibfnamefont {G.~S.}\ \bibnamefont
  {Bali}}, \bibinfo {author} {\bibfnamefont {B.}~\bibnamefont {Lang}}, \bibinfo
  {author} {\bibfnamefont {B.~U.}\ \bibnamefont {Musch}}, \ and\ \bibinfo
  {author} {\bibfnamefont {A.}~\bibnamefont {Sch\"afer}},\ }\href {\doibase
  10.1103/PhysRevD.93.094515} {\bibfield  {journal} {\bibinfo  {journal} {Phys.
  Rev. D}\ }\textbf {\bibinfo {volume} {93}},\ \bibinfo {pages} {094515}
  (\bibinfo {year} {2016})},\ \Eprint {http://arxiv.org/abs/1602.05525}
  {arXiv:1602.05525 [hep-lat]} \BibitemShut {NoStop}%
\bibitem [{\citenamefont {Egerer}\ \emph
  {et~al.}(2021{\natexlab{a}})\citenamefont {Egerer}, \citenamefont {Edwards},
  \citenamefont {Orginos},\ and\ \citenamefont {Richards}}]{Egerer:2020hnc}%
  \BibitemOpen
  \bibfield  {author} {\bibinfo {author} {\bibfnamefont {C.}~\bibnamefont
  {Egerer}}, \bibinfo {author} {\bibfnamefont {R.~G.}\ \bibnamefont {Edwards}},
  \bibinfo {author} {\bibfnamefont {K.}~\bibnamefont {Orginos}}, \ and\
  \bibinfo {author} {\bibfnamefont {D.~G.}\ \bibnamefont {Richards}},\ }\href
  {\doibase 10.1103/PhysRevD.103.034502} {\bibfield  {journal} {\bibinfo
  {journal} {Phys. Rev. D}\ }\textbf {\bibinfo {volume} {103}},\ \bibinfo
  {pages} {034502} (\bibinfo {year} {2021}{\natexlab{a}})},\ \Eprint
  {http://arxiv.org/abs/2009.10691} {arXiv:2009.10691 [hep-lat]} \BibitemShut
  {NoStop}%
\bibitem [{\citenamefont {Khan}\ \emph
  {et~al.}(2021{\natexlab{b}})\citenamefont {Khan}, \citenamefont {Richards},\
  and\ \citenamefont {Winter}}]{Khan:2020ahz}%
  \BibitemOpen
  \bibfield  {author} {\bibinfo {author} {\bibfnamefont {T.}~\bibnamefont
  {Khan}}, \bibinfo {author} {\bibfnamefont {D.}~\bibnamefont {Richards}}, \
  and\ \bibinfo {author} {\bibfnamefont {F.}~\bibnamefont {Winter}},\ }\href
  {\doibase 10.1103/PhysRevD.104.034503} {\bibfield  {journal} {\bibinfo
  {journal} {Phys. Rev. D}\ }\textbf {\bibinfo {volume} {104}},\ \bibinfo
  {pages} {034503} (\bibinfo {year} {2021}{\natexlab{b}})},\ \Eprint
  {http://arxiv.org/abs/2010.03052} {arXiv:2010.03052 [hep-lat]} \BibitemShut
  {NoStop}%
\bibitem [{\citenamefont {Blossier}\ \emph {et~al.}(2009)\citenamefont
  {Blossier}, \citenamefont {Della~Morte}, \citenamefont {von Hippel},
  \citenamefont {Mendes},\ and\ \citenamefont {Sommer}}]{Blossier:2009kd}%
  \BibitemOpen
  \bibfield  {author} {\bibinfo {author} {\bibfnamefont {B.}~\bibnamefont
  {Blossier}}, \bibinfo {author} {\bibfnamefont {M.}~\bibnamefont
  {Della~Morte}}, \bibinfo {author} {\bibfnamefont {G.}~\bibnamefont {von
  Hippel}}, \bibinfo {author} {\bibfnamefont {T.}~\bibnamefont {Mendes}}, \
  and\ \bibinfo {author} {\bibfnamefont {R.}~\bibnamefont {Sommer}},\ }\href
  {\doibase 10.1088/1126-6708/2009/04/094} {\bibfield  {journal} {\bibinfo
  {journal} {JHEP}\ }\textbf {\bibinfo {volume} {04}},\ \bibinfo {pages} {094}
  (\bibinfo {year} {2009})},\ \Eprint {http://arxiv.org/abs/0902.1265}
  {arXiv:0902.1265 [hep-lat]} \BibitemShut {NoStop}%
\bibitem [{\citenamefont {Meinel}(2013)}]{XMBF}%
  \BibitemOpen
  \bibfield  {author} {\bibinfo {author} {\bibfnamefont {S.}~\bibnamefont
  {Meinel}},\ }\href@noop {} {\bibfield  {journal} {\bibinfo  {journal} {XMBF
  2.40}\ } (\bibinfo {year} {2013})}\BibitemShut {NoStop}%
\bibitem [{\citenamefont {Egerer}\ \emph {et~al.}(2019)\citenamefont {Egerer},
  \citenamefont {Richards},\ and\ \citenamefont {Winter}}]{Egerer:2018xgu}%
  \BibitemOpen
  \bibfield  {author} {\bibinfo {author} {\bibfnamefont {C.}~\bibnamefont
  {Egerer}}, \bibinfo {author} {\bibfnamefont {D.}~\bibnamefont {Richards}}, \
  and\ \bibinfo {author} {\bibfnamefont {F.}~\bibnamefont {Winter}},\ }\href
  {\doibase 10.1103/PhysRevD.99.034506} {\bibfield  {journal} {\bibinfo
  {journal} {Phys. Rev. D}\ }\textbf {\bibinfo {volume} {99}},\ \bibinfo
  {pages} {034506} (\bibinfo {year} {2019})},\ \Eprint
  {http://arxiv.org/abs/1810.09991} {arXiv:1810.09991 [hep-lat]} \BibitemShut
  {NoStop}%
\bibitem [{\citenamefont {Dudek}\ \emph
  {et~al.}(2012{\natexlab{b}})\citenamefont {Dudek}, \citenamefont {Edwards},\
  and\ \citenamefont {Thomas}}]{Dudek:2012gj}%
  \BibitemOpen
  \bibfield  {author} {\bibinfo {author} {\bibfnamefont {J.~J.}\ \bibnamefont
  {Dudek}}, \bibinfo {author} {\bibfnamefont {R.~G.}\ \bibnamefont {Edwards}},
  \ and\ \bibinfo {author} {\bibfnamefont {C.~E.}\ \bibnamefont {Thomas}},\
  }\href {\doibase 10.1103/PhysRevD.86.034031} {\bibfield  {journal} {\bibinfo
  {journal} {Phys. Rev. D}\ }\textbf {\bibinfo {volume} {86}},\ \bibinfo
  {pages} {034031} (\bibinfo {year} {2012}{\natexlab{b}})},\ \Eprint
  {http://arxiv.org/abs/1203.6041} {arXiv:1203.6041 [hep-ph]} \BibitemShut
  {NoStop}%
\bibitem [{\citenamefont {Karpie}\ \emph {et~al.}(2018)\citenamefont {Karpie},
  \citenamefont {Orginos},\ and\ \citenamefont
  {Zafeiropoulos}}]{Karpie:2018zaz}%
  \BibitemOpen
  \bibfield  {author} {\bibinfo {author} {\bibfnamefont {J.}~\bibnamefont
  {Karpie}}, \bibinfo {author} {\bibfnamefont {K.}~\bibnamefont {Orginos}}, \
  and\ \bibinfo {author} {\bibfnamefont {S.}~\bibnamefont {Zafeiropoulos}},\
  }\href {\doibase 10.1007/JHEP11(2018)178} {\bibfield  {journal} {\bibinfo
  {journal} {JHEP}\ }\textbf {\bibinfo {volume} {11}},\ \bibinfo {pages} {178}
  (\bibinfo {year} {2018})},\ \Eprint {http://arxiv.org/abs/1807.10933}
  {arXiv:1807.10933 [hep-lat]} \BibitemShut {NoStop}%
\bibitem [{\citenamefont {Karpie}\ \emph {et~al.}(2019)\citenamefont {Karpie},
  \citenamefont {Orginos}, \citenamefont {Rothkopf},\ and\ \citenamefont
  {Zafeiropoulos}}]{Karpie:2019eiq}%
  \BibitemOpen
  \bibfield  {author} {\bibinfo {author} {\bibfnamefont {J.}~\bibnamefont
  {Karpie}}, \bibinfo {author} {\bibfnamefont {K.}~\bibnamefont {Orginos}},
  \bibinfo {author} {\bibfnamefont {A.}~\bibnamefont {Rothkopf}}, \ and\
  \bibinfo {author} {\bibfnamefont {S.}~\bibnamefont {Zafeiropoulos}},\ }\href
  {\doibase 10.1007/JHEP04(2019)057} {\bibfield  {journal} {\bibinfo  {journal}
  {JHEP}\ }\textbf {\bibinfo {volume} {04}},\ \bibinfo {pages} {057} (\bibinfo
  {year} {2019})},\ \Eprint {http://arxiv.org/abs/1901.05408} {arXiv:1901.05408
  [hep-lat]} \BibitemShut {NoStop}%
\bibitem [{\citenamefont {Bhat}\ \emph {et~al.}(2021)\citenamefont {Bhat},
  \citenamefont {Cichy}, \citenamefont {Constantinou},\ and\ \citenamefont
  {Scapellato}}]{Bhat:2020ktg}%
  \BibitemOpen
  \bibfield  {author} {\bibinfo {author} {\bibfnamefont {M.}~\bibnamefont
  {Bhat}}, \bibinfo {author} {\bibfnamefont {K.}~\bibnamefont {Cichy}},
  \bibinfo {author} {\bibfnamefont {M.}~\bibnamefont {Constantinou}}, \ and\
  \bibinfo {author} {\bibfnamefont {A.}~\bibnamefont {Scapellato}},\ }\href
  {\doibase 10.1103/PhysRevD.103.034510} {\bibfield  {journal} {\bibinfo
  {journal} {Phys. Rev. D}\ }\textbf {\bibinfo {volume} {103}},\ \bibinfo
  {pages} {034510} (\bibinfo {year} {2021})},\ \Eprint
  {http://arxiv.org/abs/2005.02102} {arXiv:2005.02102 [hep-lat]} \BibitemShut
  {NoStop}%
\bibitem [{\citenamefont {Jo\'o}\ \emph {et~al.}(2020)\citenamefont {Jo\'o},
  \citenamefont {Karpie}, \citenamefont {Orginos}, \citenamefont {Radyushkin},
  \citenamefont {Richards},\ and\ \citenamefont {Zafeiropoulos}}]{Joo:2020spy}%
  \BibitemOpen
  \bibfield  {author} {\bibinfo {author} {\bibfnamefont {B.}~\bibnamefont
  {Jo\'o}}, \bibinfo {author} {\bibfnamefont {J.}~\bibnamefont {Karpie}},
  \bibinfo {author} {\bibfnamefont {K.}~\bibnamefont {Orginos}}, \bibinfo
  {author} {\bibfnamefont {A.~V.}\ \bibnamefont {Radyushkin}}, \bibinfo
  {author} {\bibfnamefont {D.~G.}\ \bibnamefont {Richards}}, \ and\ \bibinfo
  {author} {\bibfnamefont {S.}~\bibnamefont {Zafeiropoulos}},\ }\href {\doibase
  10.1103/PhysRevLett.125.232003} {\bibfield  {journal} {\bibinfo  {journal}
  {Phys. Rev. Lett.}\ }\textbf {\bibinfo {volume} {125}},\ \bibinfo {pages}
  {232003} (\bibinfo {year} {2020})},\ \Eprint
  {http://arxiv.org/abs/2004.01687} {arXiv:2004.01687 [hep-lat]} \BibitemShut
  {NoStop}%
\bibitem [{\citenamefont {Egerer}\ \emph
  {et~al.}(2021{\natexlab{b}})\citenamefont {Egerer}, \citenamefont {Edwards},
  \citenamefont {Kallidonis}, \citenamefont {Orginos}, \citenamefont
  {Radyushkin}, \citenamefont {Richards}, \citenamefont {Romero},\ and\
  \citenamefont {Zafeiropoulos}}]{Egerer:2021ymv}%
  \BibitemOpen
  \bibfield  {author} {\bibinfo {author} {\bibfnamefont {C.}~\bibnamefont
  {Egerer}}, \bibinfo {author} {\bibfnamefont {R.~G.}\ \bibnamefont {Edwards}},
  \bibinfo {author} {\bibfnamefont {C.}~\bibnamefont {Kallidonis}}, \bibinfo
  {author} {\bibfnamefont {K.}~\bibnamefont {Orginos}}, \bibinfo {author}
  {\bibfnamefont {A.~V.}\ \bibnamefont {Radyushkin}}, \bibinfo {author}
  {\bibfnamefont {D.~G.}\ \bibnamefont {Richards}}, \bibinfo {author}
  {\bibfnamefont {E.}~\bibnamefont {Romero}}, \ and\ \bibinfo {author}
  {\bibfnamefont {S.}~\bibnamefont {Zafeiropoulos}} (\bibinfo {collaboration}
  {HadStruc}),\ }\href {\doibase 10.1007/JHEP11(2021)148} {\bibfield  {journal}
  {\bibinfo  {journal} {JHEP}\ }\textbf {\bibinfo {volume} {11}},\ \bibinfo
  {pages} {148} (\bibinfo {year} {2021}{\natexlab{b}})},\ \Eprint
  {http://arxiv.org/abs/2107.05199} {arXiv:2107.05199 [hep-lat]} \BibitemShut
  {NoStop}%
\bibitem [{\citenamefont {Cichy}\ \emph {et~al.}(2019)\citenamefont {Cichy},
  \citenamefont {Del~Debbio},\ and\ \citenamefont {Giani}}]{Cichy:2019ebf}%
  \BibitemOpen
  \bibfield  {author} {\bibinfo {author} {\bibfnamefont {K.}~\bibnamefont
  {Cichy}}, \bibinfo {author} {\bibfnamefont {L.}~\bibnamefont {Del~Debbio}}, \
  and\ \bibinfo {author} {\bibfnamefont {T.}~\bibnamefont {Giani}},\ }\href
  {\doibase 10.1007/JHEP10(2019)137} {\bibfield  {journal} {\bibinfo  {journal}
  {JHEP}\ }\textbf {\bibinfo {volume} {10}},\ \bibinfo {pages} {137} (\bibinfo
  {year} {2019})},\ \Eprint {http://arxiv.org/abs/1907.06037} {arXiv:1907.06037
  [hep-ph]} \BibitemShut {NoStop}%
\bibitem [{\citenamefont {Del~Debbio}\ \emph {et~al.}(2021)\citenamefont
  {Del~Debbio}, \citenamefont {Giani}, \citenamefont {Karpie}, \citenamefont
  {Orginos}, \citenamefont {Radyushkin},\ and\ \citenamefont
  {Zafeiropoulos}}]{DelDebbio:2020rgv}%
  \BibitemOpen
  \bibfield  {author} {\bibinfo {author} {\bibfnamefont {L.}~\bibnamefont
  {Del~Debbio}}, \bibinfo {author} {\bibfnamefont {T.}~\bibnamefont {Giani}},
  \bibinfo {author} {\bibfnamefont {J.}~\bibnamefont {Karpie}}, \bibinfo
  {author} {\bibfnamefont {K.}~\bibnamefont {Orginos}}, \bibinfo {author}
  {\bibfnamefont {A.}~\bibnamefont {Radyushkin}}, \ and\ \bibinfo {author}
  {\bibfnamefont {S.}~\bibnamefont {Zafeiropoulos}},\ }\href {\doibase
  10.1007/JHEP02(2021)138} {\bibfield  {journal} {\bibinfo  {journal} {JHEP}\
  }\textbf {\bibinfo {volume} {02}},\ \bibinfo {pages} {138} (\bibinfo {year}
  {2021})},\ \Eprint {http://arxiv.org/abs/2010.03996} {arXiv:2010.03996
  [hep-ph]} \BibitemShut {NoStop}%
\bibitem [{\citenamefont {Alexandrou}\ \emph
  {et~al.}(2021{\natexlab{d}})\citenamefont {Alexandrou}, \citenamefont
  {Bacchio}, \citenamefont {Clo\"et}, \citenamefont {Constantinou},
  \citenamefont {Hadjiyiannakou}, \citenamefont {Koutsou},\ and\ \citenamefont
  {Lauer}}]{Alexandrou:2021mmi}%
  \BibitemOpen
  \bibfield  {author} {\bibinfo {author} {\bibfnamefont {C.}~\bibnamefont
  {Alexandrou}}, \bibinfo {author} {\bibfnamefont {S.}~\bibnamefont {Bacchio}},
  \bibinfo {author} {\bibfnamefont {I.}~\bibnamefont {Clo\"et}}, \bibinfo
  {author} {\bibfnamefont {M.}~\bibnamefont {Constantinou}}, \bibinfo {author}
  {\bibfnamefont {K.}~\bibnamefont {Hadjiyiannakou}}, \bibinfo {author}
  {\bibfnamefont {G.}~\bibnamefont {Koutsou}}, \ and\ \bibinfo {author}
  {\bibfnamefont {C.}~\bibnamefont {Lauer}} (\bibinfo {collaboration} {ETM}),\
  }\href {\doibase 10.1103/PhysRevD.104.054504} {\bibfield  {journal} {\bibinfo
   {journal} {Phys. Rev. D}\ }\textbf {\bibinfo {volume} {104}},\ \bibinfo
  {pages} {054504} (\bibinfo {year} {2021}{\natexlab{d}})},\ \Eprint
  {http://arxiv.org/abs/2104.02247} {arXiv:2104.02247 [hep-lat]} \BibitemShut
  {NoStop}%
\bibitem [{\citenamefont {{Ball, Richard D. \textit{et
  al.}}}(2017)}]{Ball:2017nwa}%
  \BibitemOpen
  \bibfield  {author} {\bibinfo {author} {\bibnamefont {{Ball, Richard D.
  \textit{et al.}}}} (\bibinfo {collaboration} {NNPDF}),\ }\href {\doibase
  10.1140/epjc/s10052-017-5199-5} {\bibfield  {journal} {\bibinfo  {journal}
  {Eur. Phys. J. C}\ }\textbf {\bibinfo {volume} {77}},\ \bibinfo {pages} {663}
  (\bibinfo {year} {2017})},\ \Eprint {http://arxiv.org/abs/1706.00428}
  {arXiv:1706.00428 [hep-ph]} \BibitemShut {NoStop}%
\bibitem [{\citenamefont {Sufian}\ \emph {et~al.}(2021)\citenamefont {Sufian},
  \citenamefont {Liu},\ and\ \citenamefont {Paul}}]{Sufian:2020wcv}%
  \BibitemOpen
  \bibfield  {author} {\bibinfo {author} {\bibfnamefont {R.~S.}\ \bibnamefont
  {Sufian}}, \bibinfo {author} {\bibfnamefont {T.}~\bibnamefont {Liu}}, \ and\
  \bibinfo {author} {\bibfnamefont {A.}~\bibnamefont {Paul}},\ }\href {\doibase
  10.1103/PhysRevD.103.036007} {\bibfield  {journal} {\bibinfo  {journal}
  {Phys. Rev. D}\ }\textbf {\bibinfo {volume} {103}},\ \bibinfo {pages}
  {036007} (\bibinfo {year} {2021})},\ \Eprint
  {http://arxiv.org/abs/2012.01532} {arXiv:2012.01532 [hep-ph]} \BibitemShut
  {NoStop}%
\bibitem [{\citenamefont {Lin}\ \emph {et~al.}(2018{\natexlab{b}})\citenamefont
  {Lin}, \citenamefont {Melnitchouk}, \citenamefont {Prokudin}, \citenamefont
  {Sato},\ and\ \citenamefont {Shows}}]{Lin:2017stx}%
  \BibitemOpen
  \bibfield  {author} {\bibinfo {author} {\bibfnamefont {H.-W.}\ \bibnamefont
  {Lin}}, \bibinfo {author} {\bibfnamefont {W.}~\bibnamefont {Melnitchouk}},
  \bibinfo {author} {\bibfnamefont {A.}~\bibnamefont {Prokudin}}, \bibinfo
  {author} {\bibfnamefont {N.}~\bibnamefont {Sato}}, \ and\ \bibinfo {author}
  {\bibfnamefont {H.}~\bibnamefont {Shows}},\ }\href {\doibase
  10.1103/PhysRevLett.120.152502} {\bibfield  {journal} {\bibinfo  {journal}
  {Phys. Rev. Lett.}\ }\textbf {\bibinfo {volume} {120}},\ \bibinfo {pages}
  {152502} (\bibinfo {year} {2018}{\natexlab{b}})},\ \Eprint
  {http://arxiv.org/abs/1710.09858} {arXiv:1710.09858 [hep-ph]} \BibitemShut
  {NoStop}%
\bibitem [{\citenamefont {Bringewatt}\ \emph {et~al.}(2021)\citenamefont
  {Bringewatt}, \citenamefont {Sato}, \citenamefont {Melnitchouk},
  \citenamefont {Qiu}, \citenamefont {Steffens},\ and\ \citenamefont
  {Constantinou}}]{Bringewatt:2020ixn}%
  \BibitemOpen
  \bibfield  {author} {\bibinfo {author} {\bibfnamefont {J.}~\bibnamefont
  {Bringewatt}}, \bibinfo {author} {\bibfnamefont {N.}~\bibnamefont {Sato}},
  \bibinfo {author} {\bibfnamefont {W.}~\bibnamefont {Melnitchouk}}, \bibinfo
  {author} {\bibfnamefont {J.-W.}\ \bibnamefont {Qiu}}, \bibinfo {author}
  {\bibfnamefont {F.}~\bibnamefont {Steffens}}, \ and\ \bibinfo {author}
  {\bibfnamefont {M.}~\bibnamefont {Constantinou}},\ }\href {\doibase
  10.1103/PhysRevD.103.016003} {\bibfield  {journal} {\bibinfo  {journal}
  {Phys. Rev. D}\ }\textbf {\bibinfo {volume} {103}},\ \bibinfo {pages}
  {016003} (\bibinfo {year} {2021})},\ \Eprint
  {http://arxiv.org/abs/2010.00548} {arXiv:2010.00548 [hep-ph]} \BibitemShut
  {NoStop}%
\bibitem [{\citenamefont {Barry}\ \emph {et~al.}(2022)\citenamefont {Barry}
  \emph {et~al.}}]{Barry:2022itu}%
  \BibitemOpen
  \bibfield  {author} {\bibinfo {author} {\bibfnamefont {P.~C.}\ \bibnamefont
  {Barry}} \emph {et~al.},\ }\href@noop {} {\  (\bibinfo {year} {2022})},\
  \Eprint {http://arxiv.org/abs/2204.00543} {arXiv:2204.00543 [hep-ph]}
  \BibitemShut {NoStop}%
\bibitem [{\citenamefont {Hou}\ \emph {et~al.}(2022{\natexlab{a}})\citenamefont
  {Hou}, \citenamefont {Lin}, \citenamefont {Yan},\ and\ \citenamefont
  {Yuan}}]{Hou:2022sdf}%
  \BibitemOpen
  \bibfield  {author} {\bibinfo {author} {\bibfnamefont {T.-J.}\ \bibnamefont
  {Hou}}, \bibinfo {author} {\bibfnamefont {H.-W.}\ \bibnamefont {Lin}},
  \bibinfo {author} {\bibfnamefont {M.}~\bibnamefont {Yan}}, \ and\ \bibinfo
  {author} {\bibfnamefont {C.~P.}\ \bibnamefont {Yuan}},\ }\href@noop {} {\
  (\bibinfo {year} {2022}{\natexlab{a}})},\ \Eprint
  {http://arxiv.org/abs/2204.07944} {arXiv:2204.07944 [hep-ph]} \BibitemShut
  {NoStop}%
\bibitem [{\citenamefont {Hou}\ \emph {et~al.}(2022{\natexlab{b}})\citenamefont
  {Hou}, \citenamefont {Yan}, \citenamefont {Liang}, \citenamefont {Liu},\ and\
  \citenamefont {Yuan}}]{Hou:2022ajg}%
  \BibitemOpen
  \bibfield  {author} {\bibinfo {author} {\bibfnamefont {T.-J.}\ \bibnamefont
  {Hou}}, \bibinfo {author} {\bibfnamefont {M.}~\bibnamefont {Yan}}, \bibinfo
  {author} {\bibfnamefont {J.}~\bibnamefont {Liang}}, \bibinfo {author}
  {\bibfnamefont {K.-F.}\ \bibnamefont {Liu}}, \ and\ \bibinfo {author}
  {\bibfnamefont {C.~P.}\ \bibnamefont {Yuan}},\ }\href@noop {} {\  (\bibinfo
  {year} {2022}{\natexlab{b}})},\ \Eprint {http://arxiv.org/abs/2206.02431}
  {arXiv:2206.02431 [hep-ph]} \BibitemShut {NoStop}%
\bibitem [{\citenamefont {Towns}\ \emph {et~al.}(2014)\citenamefont {Towns},
  \citenamefont {Cockerill}, \citenamefont {Dahan}, \citenamefont {Foster},
  \citenamefont {Gaither}, \citenamefont {Grimshaw}, \citenamefont {Hazlewood},
  \citenamefont {Lathrop}, \citenamefont {Lifka}, \citenamefont {Peterson},
  \citenamefont {Roskies}, \citenamefont {Scott},\ and\ \citenamefont
  {Wilkins-Diehr}}]{6866038}%
  \BibitemOpen
  \bibfield  {author} {\bibinfo {author} {\bibfnamefont {J.}~\bibnamefont
  {Towns}}, \bibinfo {author} {\bibfnamefont {T.}~\bibnamefont {Cockerill}},
  \bibinfo {author} {\bibfnamefont {M.}~\bibnamefont {Dahan}}, \bibinfo
  {author} {\bibfnamefont {I.}~\bibnamefont {Foster}}, \bibinfo {author}
  {\bibfnamefont {K.}~\bibnamefont {Gaither}}, \bibinfo {author} {\bibfnamefont
  {A.}~\bibnamefont {Grimshaw}}, \bibinfo {author} {\bibfnamefont
  {V.}~\bibnamefont {Hazlewood}}, \bibinfo {author} {\bibfnamefont
  {S.}~\bibnamefont {Lathrop}}, \bibinfo {author} {\bibfnamefont
  {D.}~\bibnamefont {Lifka}}, \bibinfo {author} {\bibfnamefont {G.~D.}\
  \bibnamefont {Peterson}}, \bibinfo {author} {\bibfnamefont {R.}~\bibnamefont
  {Roskies}}, \bibinfo {author} {\bibfnamefont {J.}~\bibnamefont {Scott}}, \
  and\ \bibinfo {author} {\bibfnamefont {N.}~\bibnamefont {Wilkins-Diehr}},\
  }\href {\doibase 10.1109/MCSE.2014.80} {\bibfield  {journal} {\bibinfo
  {journal} {Computing in Science \& Engineering}\ }\textbf {\bibinfo {volume}
  {16}},\ \bibinfo {pages} {62} (\bibinfo {year} {2014})}\BibitemShut {NoStop}%
\bibitem [{\citenamefont {Nystrom}\ \emph {et~al.}(2015)\citenamefont
  {Nystrom}, \citenamefont {Levine}, \citenamefont {Roskies},\ and\
  \citenamefont {Scott}}]{Nystrom:2015:BUF:2792745.2792775}%
  \BibitemOpen
  \bibfield  {author} {\bibinfo {author} {\bibfnamefont {N.~A.}\ \bibnamefont
  {Nystrom}}, \bibinfo {author} {\bibfnamefont {M.~J.}\ \bibnamefont {Levine}},
  \bibinfo {author} {\bibfnamefont {R.~Z.}\ \bibnamefont {Roskies}}, \ and\
  \bibinfo {author} {\bibfnamefont {J.~R.}\ \bibnamefont {Scott}},\ }\href
  {\doibase 10.1145/2792745.2792775} {\ \bibinfo {series} {XSEDE '15},\
  \bibinfo {pages} {30:1} (\bibinfo {year} {2015})}\BibitemShut {NoStop}%
\bibitem [{\citenamefont {Edwards}\ and\ \citenamefont
  {Joo}(2005)}]{Edwards:2004sx}%
  \BibitemOpen
  \bibfield  {author} {\bibinfo {author} {\bibfnamefont {R.~G.}\ \bibnamefont
  {Edwards}}\ and\ \bibinfo {author} {\bibfnamefont {B.}~\bibnamefont {Joo}}
  (\bibinfo {collaboration} {SciDAC, LHPC, UKQCD}),\ }\href {\doibase
  10.1016/j.nuclphysbps.2004.11.254} {\bibfield  {journal} {\bibinfo  {journal}
  {Nucl. Phys. Proc. Suppl.}\ }\textbf {\bibinfo {volume} {140}},\ \bibinfo
  {pages} {832} (\bibinfo {year} {2005})},\ \bibinfo {note} {[,832(2004)]},\
  \Eprint {http://arxiv.org/abs/hep-lat/0409003} {arXiv:hep-lat/0409003
  [hep-lat]} \BibitemShut {NoStop}%
\bibitem [{\citenamefont {Clark}\ \emph {et~al.}(2010)\citenamefont {Clark},
  \citenamefont {Babich}, \citenamefont {Barros}, \citenamefont {Brower},\ and\
  \citenamefont {Rebbi}}]{Clark:2009wm}%
  \BibitemOpen
  \bibfield  {author} {\bibinfo {author} {\bibfnamefont {M.~A.}\ \bibnamefont
  {Clark}}, \bibinfo {author} {\bibfnamefont {R.}~\bibnamefont {Babich}},
  \bibinfo {author} {\bibfnamefont {K.}~\bibnamefont {Barros}}, \bibinfo
  {author} {\bibfnamefont {R.~C.}\ \bibnamefont {Brower}}, \ and\ \bibinfo
  {author} {\bibfnamefont {C.}~\bibnamefont {Rebbi}},\ }\href {\doibase
  10.1016/j.cpc.2010.05.002} {\bibfield  {journal} {\bibinfo  {journal}
  {Comput. Phys. Commun.}\ }\textbf {\bibinfo {volume} {181}},\ \bibinfo
  {pages} {1517} (\bibinfo {year} {2010})},\ \Eprint
  {http://arxiv.org/abs/0911.3191} {arXiv:0911.3191 [hep-lat]} \BibitemShut
  {NoStop}%
\bibitem [{\citenamefont {Babich}\ \emph {et~al.}(2010)\citenamefont {Babich},
  \citenamefont {Clark},\ and\ \citenamefont {Joo}}]{Babich:2010mu}%
  \BibitemOpen
  \bibfield  {author} {\bibinfo {author} {\bibfnamefont {R.}~\bibnamefont
  {Babich}}, \bibinfo {author} {\bibfnamefont {M.~A.}\ \bibnamefont {Clark}}, \
  and\ \bibinfo {author} {\bibfnamefont {B.}~\bibnamefont {Joo}},\ }in\
  \href@noop {} {\emph {\bibinfo {booktitle} {{SC 10 (Supercomputing 2010)}}}}\
  (\bibinfo {year} {2010})\ \Eprint {http://arxiv.org/abs/1011.0024}
  {arXiv:1011.0024 [hep-lat]} \BibitemShut {NoStop}%
\bibitem [{\citenamefont {Jo{\'o}}\ \emph {et~al.}(2016)\citenamefont
  {Jo{\'o}}, \citenamefont {Kalamkar}, \citenamefont {Kurth}, \citenamefont
  {Vaidyanathan},\ and\ \citenamefont {Walden}}]{QPhiX2}%
  \BibitemOpen
  \bibfield  {author} {\bibinfo {author} {\bibfnamefont {B.}~\bibnamefont
  {Jo{\'o}}}, \bibinfo {author} {\bibfnamefont {D.~D.}\ \bibnamefont
  {Kalamkar}}, \bibinfo {author} {\bibfnamefont {T.}~\bibnamefont {Kurth}},
  \bibinfo {author} {\bibfnamefont {K.}~\bibnamefont {Vaidyanathan}}, \ and\
  \bibinfo {author} {\bibfnamefont {A.}~\bibnamefont {Walden}}\ }(\bibinfo
  {publisher} {Springer International Publishing},\ \bibinfo {address} {Cham},\
  \bibinfo {year} {2016})\ pp.\ \bibinfo {pages} {415--427}\BibitemShut
  {NoStop}%
\bibitem [{\citenamefont {Stanzione}\ \emph {et~al.}(2020)\citenamefont
  {Stanzione}, \citenamefont {West}, \citenamefont {Evans}, \citenamefont
  {Minyard}, \citenamefont {Ghattas},\ and\ \citenamefont
  {Panda}}]{10.1145/3311790.3396656}%
  \BibitemOpen
  \bibfield  {author} {\bibinfo {author} {\bibfnamefont {D.}~\bibnamefont
  {Stanzione}}, \bibinfo {author} {\bibfnamefont {J.}~\bibnamefont {West}},
  \bibinfo {author} {\bibfnamefont {R.~T.}\ \bibnamefont {Evans}}, \bibinfo
  {author} {\bibfnamefont {T.}~\bibnamefont {Minyard}}, \bibinfo {author}
  {\bibfnamefont {O.}~\bibnamefont {Ghattas}}, \ and\ \bibinfo {author}
  {\bibfnamefont {D.~K.}\ \bibnamefont {Panda}},\ }in\ \href {\doibase
  10.1145/3311790.3396656} {\emph {\bibinfo {booktitle} {Practice and
  Experience in Advanced Research Computing}}},\ \bibinfo {series and number}
  {PEARC '20}\ (\bibinfo  {publisher} {Association for Computing Machinery},\
  \bibinfo {address} {New York, NY, USA},\ \bibinfo {year} {2020})\ p.\
  \bibinfo {pages} {106–111}\BibitemShut {NoStop}%
\bibitem [{\citenamefont {Wolff}(2004)}]{Wolff:2003sm}%
  \BibitemOpen
  \bibfield  {author} {\bibinfo {author} {\bibfnamefont {U.}~\bibnamefont
  {Wolff}} (\bibinfo {collaboration} {ALPHA}),\ }\href {\doibase
  10.1016/S0010-4655(03)00467-3} {\bibfield  {journal} {\bibinfo  {journal}
  {Comput. Phys. Commun.}\ }\textbf {\bibinfo {volume} {156}},\ \bibinfo
  {pages} {143} (\bibinfo {year} {2004})},\ \bibinfo {note} {[Erratum:
  Comput.Phys.Commun. 176, 383 (2007)]},\ \Eprint
  {http://arxiv.org/abs/hep-lat/0306017} {arXiv:hep-lat/0306017} \BibitemShut
  {NoStop}%
\end{thebibliography}%

\appendix

\section{Choice of basis of interpolators} \label{sec:interpol}
The comparison plots in Fig.~\ref{fig:opcomp} illustrate the case for $p=0.41$ GeV and the first momentum with the momentum smearing, $p=1.64$ GeV at the smallest and largest flow times $\tau/a^2=1.0$ and $3.8$ used in this work.

\befs 
\centering

\includegraphics[scale=0.6]{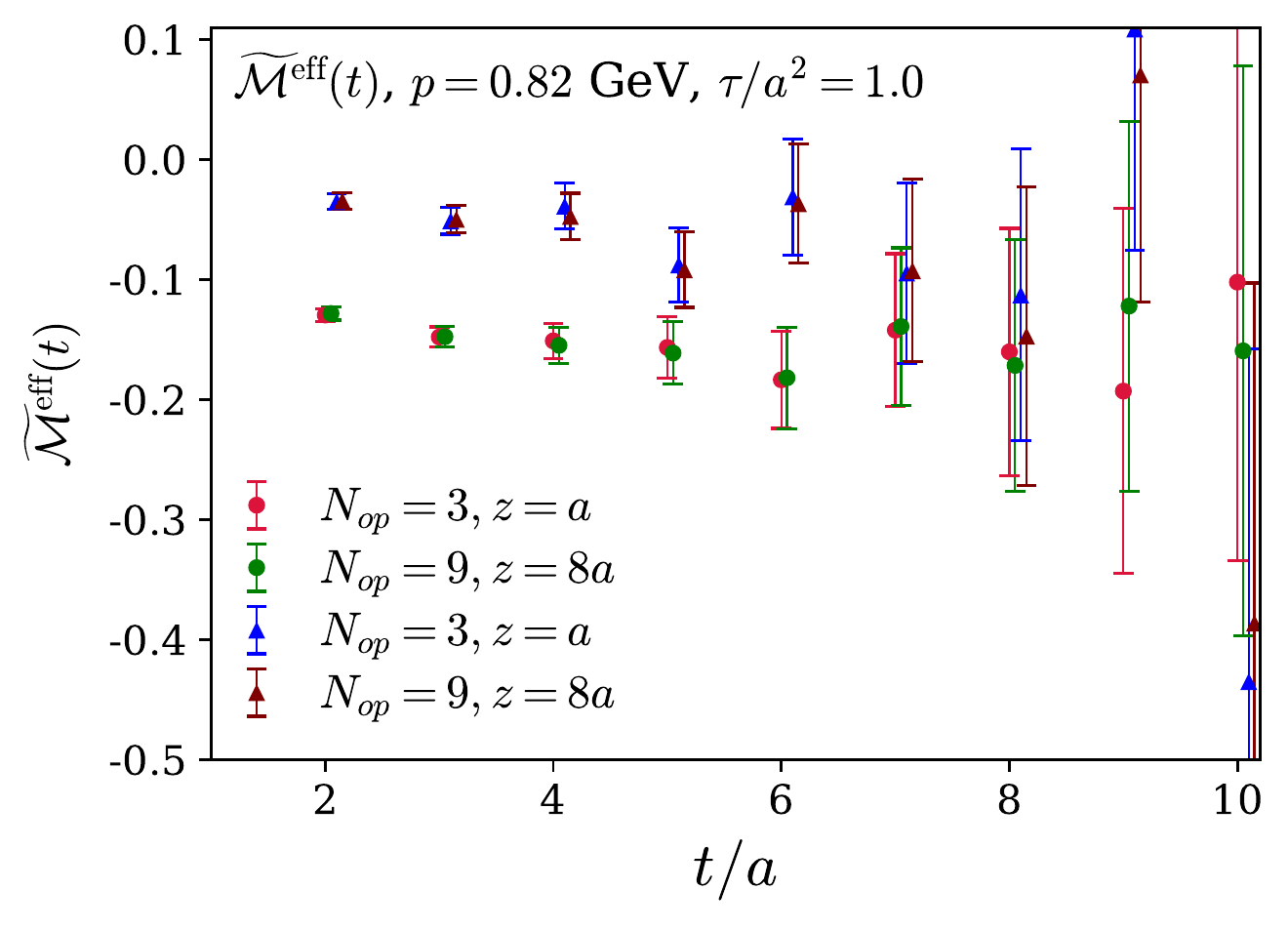}
\includegraphics[scale=0.6]{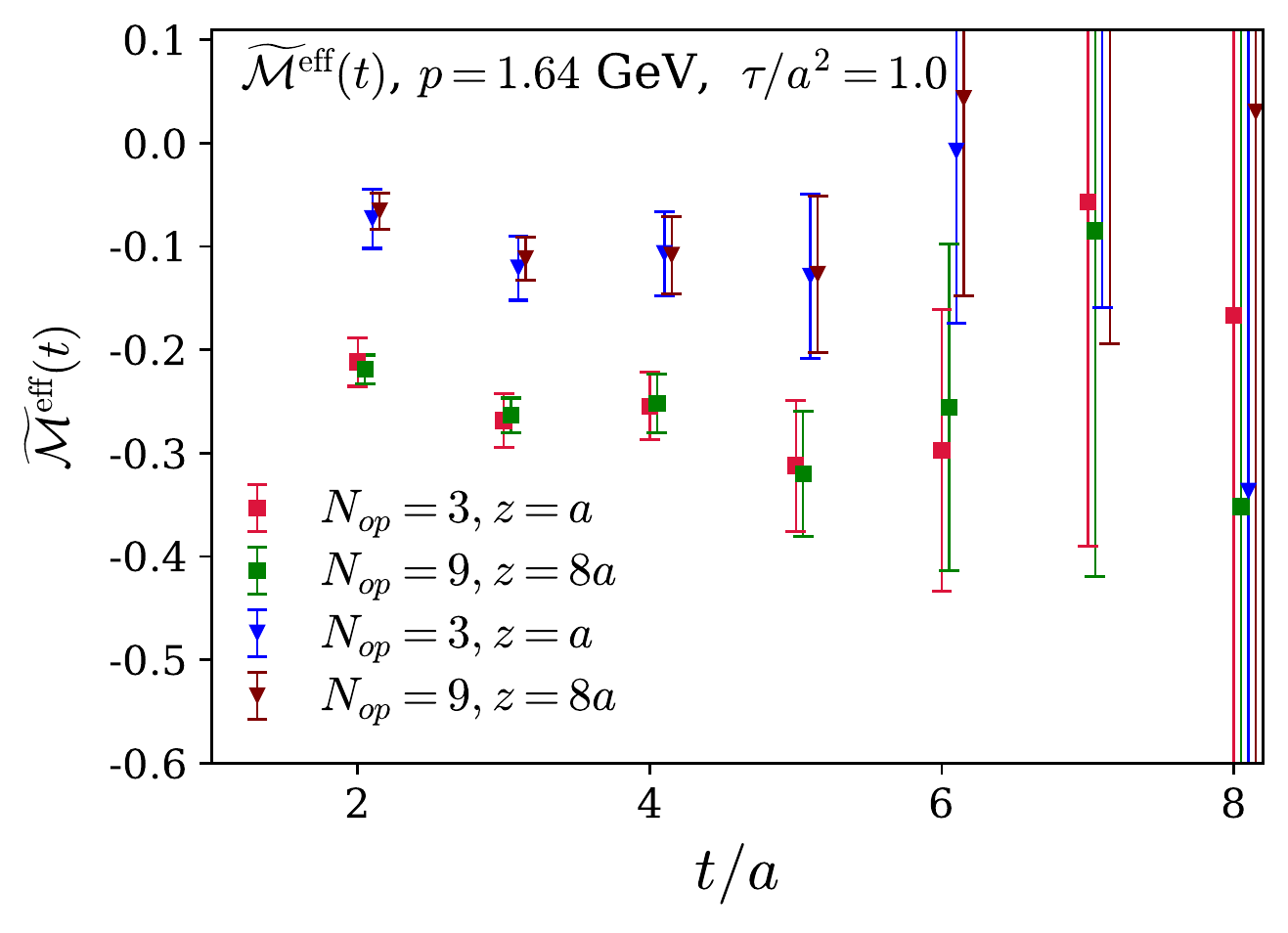}

\includegraphics[scale=0.6]{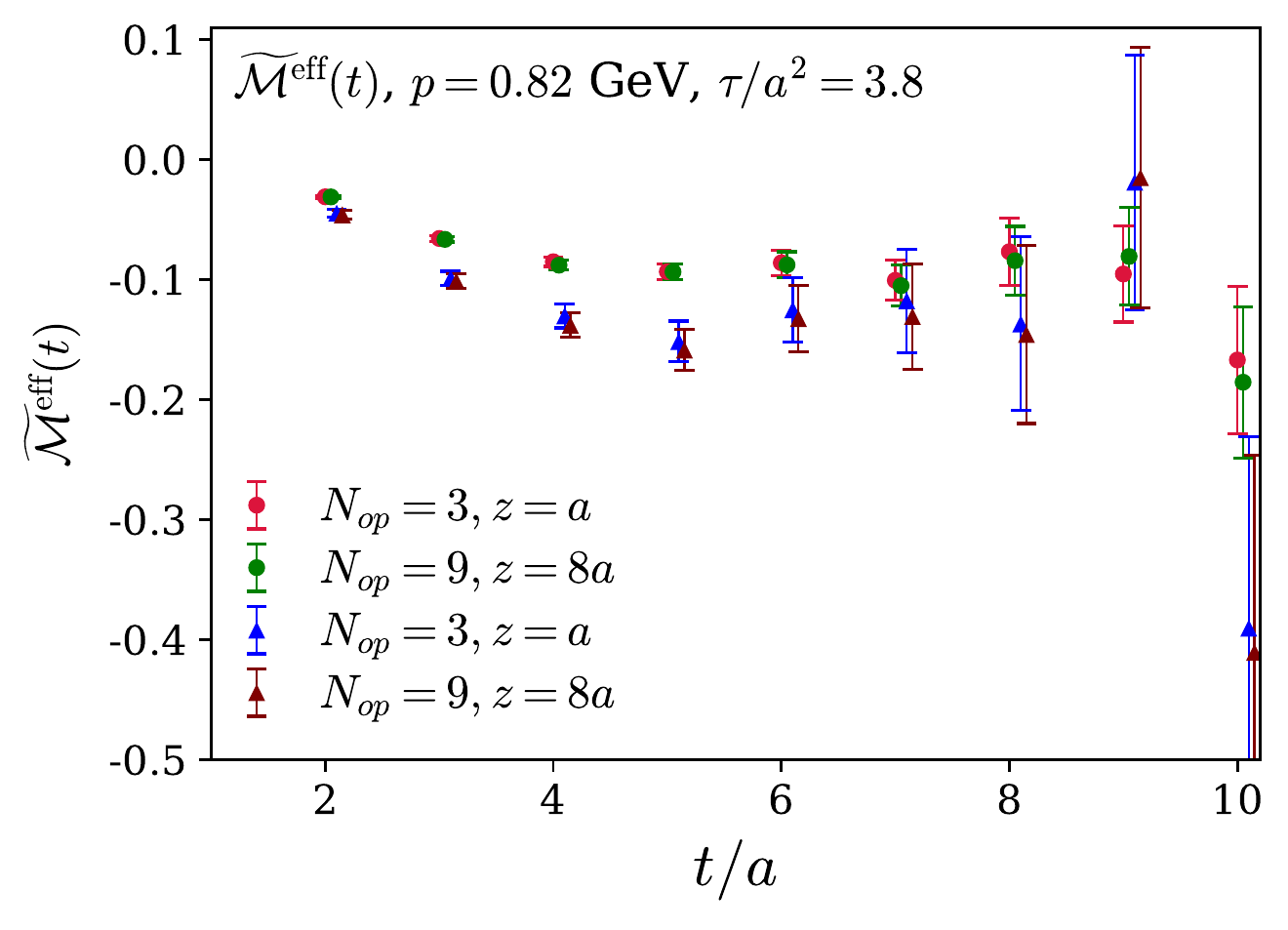}
\includegraphics[scale=0.6]{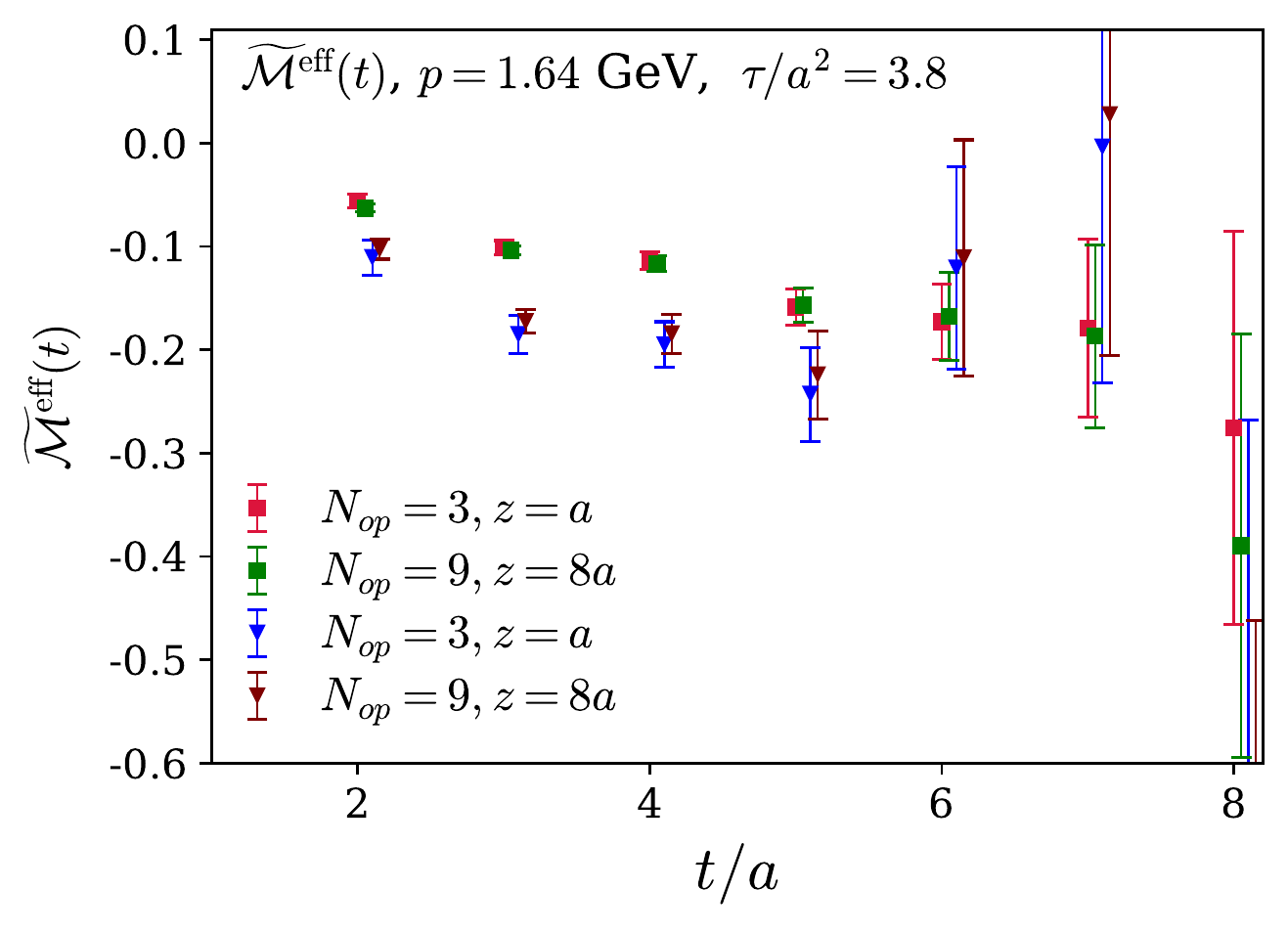}

\caption{\label{fig:opcomp}   Comparison plots of a few polarized gluon matrix elements $\wt{M}^{\rm eff}(t)$ between 3 and 9 different number of interpolators as illustrated in Table~\ref{tab:interpolator}. The smaller basis of interpolators is used in this work based on the observation that the smaller basis of interpolators reproduces the matrix elements constructed with the larger basis of interpolators through the sGEVP analysis. }
\eefs{mockdemocn}

\section{Example fit results for unpolarized and polarized gluonic matrix elements}\label{appendix:fitparams}

\begin{table*}
  \centering
  \setlength{\tabcolsep}{10pt}
  \renewcommand{\arraystretch}{1.5}
  \begin{tabular}{ccccccccc}
  \toprule
    $\tau/a^2$ &  $p \,$(GeV) & $z \,(a)$ & $\nu$ &  $A$ & $B$ & $\Delta E$ & $\chi^2/{\rm d.o.f.}$ \\
    \midrule
    $1.4$ & $0.82$ & $2$ & 0.79 & -0.261(21) & 0.484(256) & 1.18(13) & $0.66(23)$ \\
    $1.4$ & $0.82$ & $8$ & 3.14 & -0.097(19) & 0.204(97) & 1.18(13) & $0.66(23)$ \\
    $1.4$ & $1.64$ & $2$ & 1.57 & -0.409(38) & 3.92(60) & 1.40(13) & $0.37(18)$ \\
    $1.4$ & $1.64$ & $8$ & 6.28 & -0.155(14) & 2.15(55) & 2.14(15) & $0.37(18)$ \\
    $1.4$ & $2.46$ & $2$ & 2.36 & -0.62(14) & 1.31(83) & 1.27(20) & $0.18(14)$ \\
    $1.4$ & $2.46$ & $8$ & 9.42 & -0.197(48) & 0.46(25) & 1.27(20) & $0.18(14)$ \\
    $3.0$ & $0.82$ & $2$ & 0.79 & -0.216(17) & 0.474(28) & 1.24(6) & $0.60(23)$ \\
    $3.0$ & $0.82$ & $8$ & 3.14 & -0.151(11) & 0.618(74) & 1.24(6) & $0.60(23)$ \\
    $3.0$ & $1.64$ & $2$ & 1.57 & -0.288(30) & 1.62(81) & 1.66(20) & $0.60(28)$ \\
    $3.0$ & $1.64$ & $8$ & 6.28 & -0.200(15) & 1.42(50) & 1.66(20) & $0.60(28)$ \\
    $3.0$ & $2.46$ & $2$ & 2.36 & -0.56(17) & 0.97(17) & 0.89(20) & $0.23(18)$ \\
    $3.0$ & $2.46$ & $8$ & 9.42 & -0.35(10) & 0.44(25) & 0.89(20) & $0.23(18)$ \\
    \bottomrule
  \end{tabular}
\caption{ The fitted parameters and the goodness of the fits for the matrix elements $\wt{\mathcal M}^{\rm eff}(t)$ shown in Fig.~\ref{fig:Mefftau}. For a particular flow time and nucleon momentum, we first fit the matrix elements at $z=2a$; the information regarding the fit parameter $\Delta E$ from this fit is used to set the prior for $\Delta E$ in a simultaneous correlated fit for the matrix elements of all the non-zero separations.}\label{tab:fitparams}
\end{table*}

From Fig.~\ref{fig:Mefftau} and the corresponding fit parameters in Table~\ref{tab:fitparams} we see that the lattice data are described well by our fit procedure. The $\chi^2/{\rm d.o.f.}$ shows that the choice of prior-width for $\Delta E$ at $z > 0$ is an appropriate one. We notice from Fig.~\ref{fig:Mefftau} that the matrix elements for $z = 6a $ = 0.564 fm, have a flat behavior as a function of the source-sink separations. This can also be understood from the smallness of $B$-parameters listed in Table~\ref{tab:fitparams}, with relatively larger uncertainties.

\section{Zero Flow Time Extrapolated Reduced Matrix Elements}\label{zero_flow_time_reduced_mtx_elem}

For each nucleon momentum and each field separation, the flowed reduced matrix elements for different flow times are fit to a linear expression: $\wt{\mathfrak{ M}}(\tau) = c_0 + c_1 \tau$, where the fit parameter, $c_0$ gives the reduced pseudo-ITD at zero flow time limit. The fit parameters, $c_0$ and $c_1$ and the covariance between are tabulated in Table~\ref{tab:zft_reduced_mtx_elem}.

 \begin{table*}
  \centering
  \renewcommand{\arraystretch}{1.2}
  \setlength{\tabcolsep}{20pt}

  \begin{tabular}{ccccccccc}
    \toprule
    $p$ (GeV)  & $z \, (a)$ & $\nu$ & $c_0$ & $c_1$ & ${\rm cov}[c_0,c_1]$ \\
    \midrule
    $0.41$ & $1$ & $0.20$ & 0.697(67) & -0.091(19) & -0.00116 \\
    $0.41$ & $2$ & $0.39$ & 1.294(122) & -0.174(33) & -0.00373 \\
    $0.41$ & $3$ & $0.59$ & 1.859(165) & -0.249(44) & -0.00690 \\
    $0.41$ & $4$ & $0.79$ & 2.348(244) & -0.315(68) & -0.01554 \\
    $0.41$ & $5$ & $0.98$ & 2.682(345) & -0.314(96) & -0.03155 \\
    $0.41$ & $6$ & $1.18$ & 2.783(414) & -0.275(120) & -0.04657 \\
    $0.41$ & $7$ & $1.37$ & 2.891(504) & -0.255(135) & -0.06358 \\
    $0.41$ & $8$ & $1.57$ & 3.209(591) & -0.325(136) & -0.07627 \\
    $0.82$ & $1$ & $0.39$ & 0.386(39) & -0.057(12) & -0.00042 \\
    $0.82$ & $2$ & $0.79$ & 0.672(61) & -0.094(17) & -0.00096 \\
    $0.82$ & $3$ & $1.18$ & 1.0(1) & -0.143(26) & -0.00233 \\
    $0.82$ & $4$ & $1.57$ & 1.147(119) & -0.138(33) & -0.0036 \\
    $0.82$ & $5$ & $1.96$ & 1.445(185) & -0.191(52) & -0.0091 \\
    $0.82$ & $6$ & $2.36$ & 1.335(219) & -0.119(66) & -0.0134 \\
    $0.82$ & $7$ & $2.75$ & 1.470(258) & -0.143(74) & -0.0177 \\
    $0.82$ & $8$ & $3.14$ & 1.501(271) & -0.139(64) & -0.0161 \\
    $1.23$ & $1$ & $0.59$ & 0.272(29) & -0.040(9) & -0.00025 \\
    $1.23$ & $2$ & $1.18$ & 0.473(50) & -0.067(15) & -0.00066 \\
    $1.23$ & $3$ & $1.77$ & 0.644(59) & -0.084(17) & -0.00090 \\
    $1.23$ & $4$ & $2.36$ & 0.744(87) & -0.087(27) & -0.00216 \\
    $1.23$ & $5$ & $2.95$ & 0.936(121) & -0.108(37) & -0.00414 \\
    $1.23$ & $6$ & $3.53$ & 0.953(154) & -0.099(47) & -0.00669 \\
    $1.23$ & $7$ & $4.12$ & 0.984(185) & -0.086(52) & -0.00893 \\
    $1.23$ & $8$ & $4.71$ & 0.977(213) & -0.082(53) & -0.01058 \\
    $1.64$ & $1$ & $0.79$ & 0.215(22) & -0.035(8) &   -0.00015 \\
    $1.64$ & $2$ & $1.57$ & 0.389(40) & -0.064(12) &  -0.00044 \\
    $1.64$ & $3$ & $2.36$ & 0.512(58) & -0.078(19) & -0.00102 \\
    $1.64$ & $4$ & $3.14$ & 0.635(76) & -0.097(23) & -0.00167 \\
    $1.64$ & $5$ & $3.93$ & 0.685(109) & -0.085(32) & -0.00336 \\
    $1.64$ & $6$ & $4.71$ & 0.760(123) & -0.096(40) & -0.00468 \\
    $1.64$ & $7$ & $5.50$ & 0.798(161) & -0.097(45) & -0.00691 \\
    $1.64$ & $8$ & $6.28$ & 0.740(159) & -0.071(42) & -0.00631 \\
    $2.05$ & $1$ & $0.98$ & 0.188(28) & -0.029(9) & -0.00024 \\
    $2.05$ & $2$ & $1.96$ & 0.315(45) & -0.044(14) & -0.00057 \\
    $2.05$ & $3$ & $2.95$ & 0.456(62) & -0.064(20) & -0.00110 \\
    $2.05$ & $4$ & $3.93$ & 0.539(79) & -0.074(25) & -0.00187 \\
    $2.05$ & $5$ & $4.91$ & 0.646(116) & -0.079(32) & -0.00349 \\
    $2.05$ & $6$ & $5.89$ & 0.649(140) & -0.068(49) & -0.00624 \\
    $2.05$ & $7$ & $6.87$ & 0.817(213) & -0.104(65) & -0.01314 \\
    $2.05$ & $8$ & $7.85$ & 0.852(208) & -0.105(45) & -0.00888 \\
    $2.46$ & $1$ & $1.18$ & 0.175(23) & -0.031(9) & -0.00018 \\
    $2.46$ & $2$ & $2.36$ & 0.279(46) & -0.039(18) & -0.00064 \\
    $2.46$ & $3$ & $3.53$ & 0.419(62) & -0.065(27) & -0.00131 \\
    $2.46$ & $4$ & $4.71$ & 0.463(81) & -0.070(29) & -0.00195 \\
    $2.46$ & $5$ & $5.89$ & 0.549(106) & -0.082(38) & -0.00328 \\
    $2.46$ & $6$ & $7.07$ & 0.424(105) & -0.035(42) & -0.00352 \\
    $2.46$ & $7$ & $8.25$ & 0.560(134) & -0.073(49) & -0.00532 \\
    $2.46$ & $8$ & $9.42$ & 0.454(133) & -0.055(41) & -0.00458 \\
    \bottomrule
  \end{tabular}

  \caption{ Reduced matrix elements extrapolated to zero flow time. The flowed reduced matrix elements are fitted using a linear form: $\wt{\mathfrak{M}}(\tau) = c_0 + c_1 \tau$, where $c_0$ is the reduced matrix elements at the zero flow-time limit. All the $\chi^2/{\rm d.o.f.}$ are smaller than $1$. The fitted parameters and the covariances between them are listed in the final column of the table. \label{tab:zft_reduced_mtx_elem}}
\end{table*}

\section{Effect of perturbative matching kernel on the lattice reduced pseudo-ITD data}\label{appendix:matching}

We choose the result from the 4-parameters fit (labeled as ``Fit-2" in FIG.~\ref{fig:mpcfit}) and investigate the effect of matching for two different values of $z=a$ and $2a$ in the  $\overline{\rm MS}$  renormalization  scheme  at  $\mu=2$ GeV and show the results in FIG.~\ref{fig:matching_plot}. We have ignored the  fraction of the hadron momentum carried by the singlet quarks and also the mixing
with the quark singlet sector in the matching relation~\eqref{eq:matching}. As we can see, the effect of matching is  smaller compared to the significantly larger uncertainty in our data while the systematic uncertainties associated with the removal of the contamination term are unknown and can be much larger. 

\begin{figure}
\center{\includegraphics[scale=0.7]{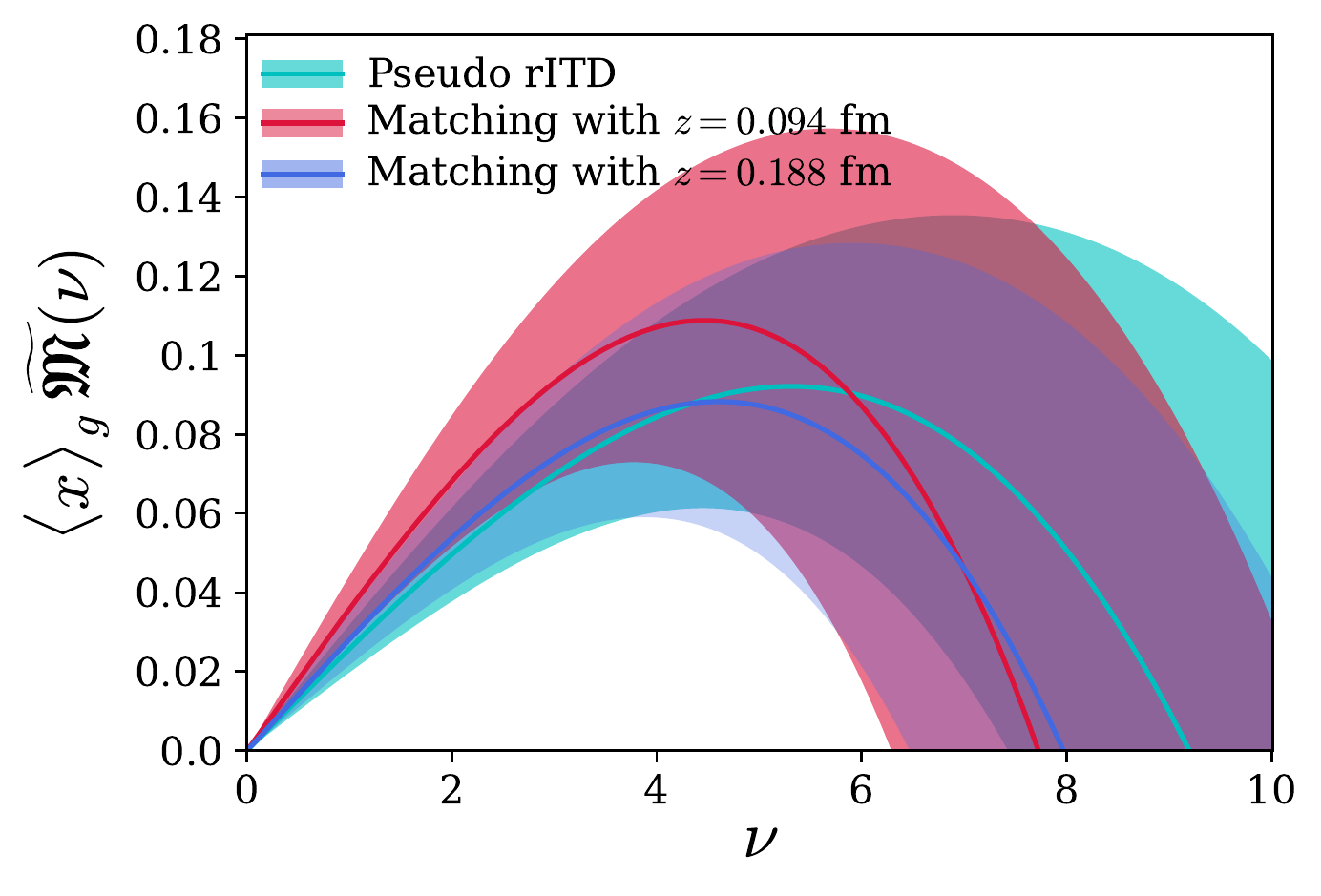}}\caption{Ioffe-time distributions after the implementation of the perturbative matching kernel on the lattice reduced pseudo-ITD  data in  the $\overline{\rm MS}$  renormalization  scheme  at  2 GeV. For the  overall normalization of the ITDs, 
we take  $\langle x \rangle_g$=0.427(92) from~\cite{Alexandrou:2020sml} in the $\overline{\rm MS}$ scheme at renormalization scale $\mu = 2$ GeV.
\label{fig:matching_plot}}
\end{figure}

\section{Autocorrelation of the bare matrix elements}\label{appendix:autocorr}

To check whether the bare matrix elements extracted from the different gauge configurations are correlated or not, the associated integrated autocorrelation time, $\tau_{\mathrm{int}}$ is calculated for each source-sink separation of the nucleon two-point correlators. The integrated autocorrelation time is a measure of how efficiently the Monte Carlo algorithm can be implemented to calculate the desired quantity. In this calculation, each gauge configuration is separated from the next gauge configuration by 10 hybrid Monte Carlo trajectories. Calculation of $\tau_{\mathrm{int}}$ for a given quantity will indicate whether this separation of 10 hybrid Monte Carlo trajectories is sufficient to remove the autocorrelation between the gauge configurations for that quantity. The method of the calculation follows the procedure described in~\cite{Wolff:2003sm}. According to this method, for the autocorrelation to vanish, $\tau_{\mathrm{int}}$ needs to be 0.5.

\befs 
\centering
\includegraphics[scale=0.6]{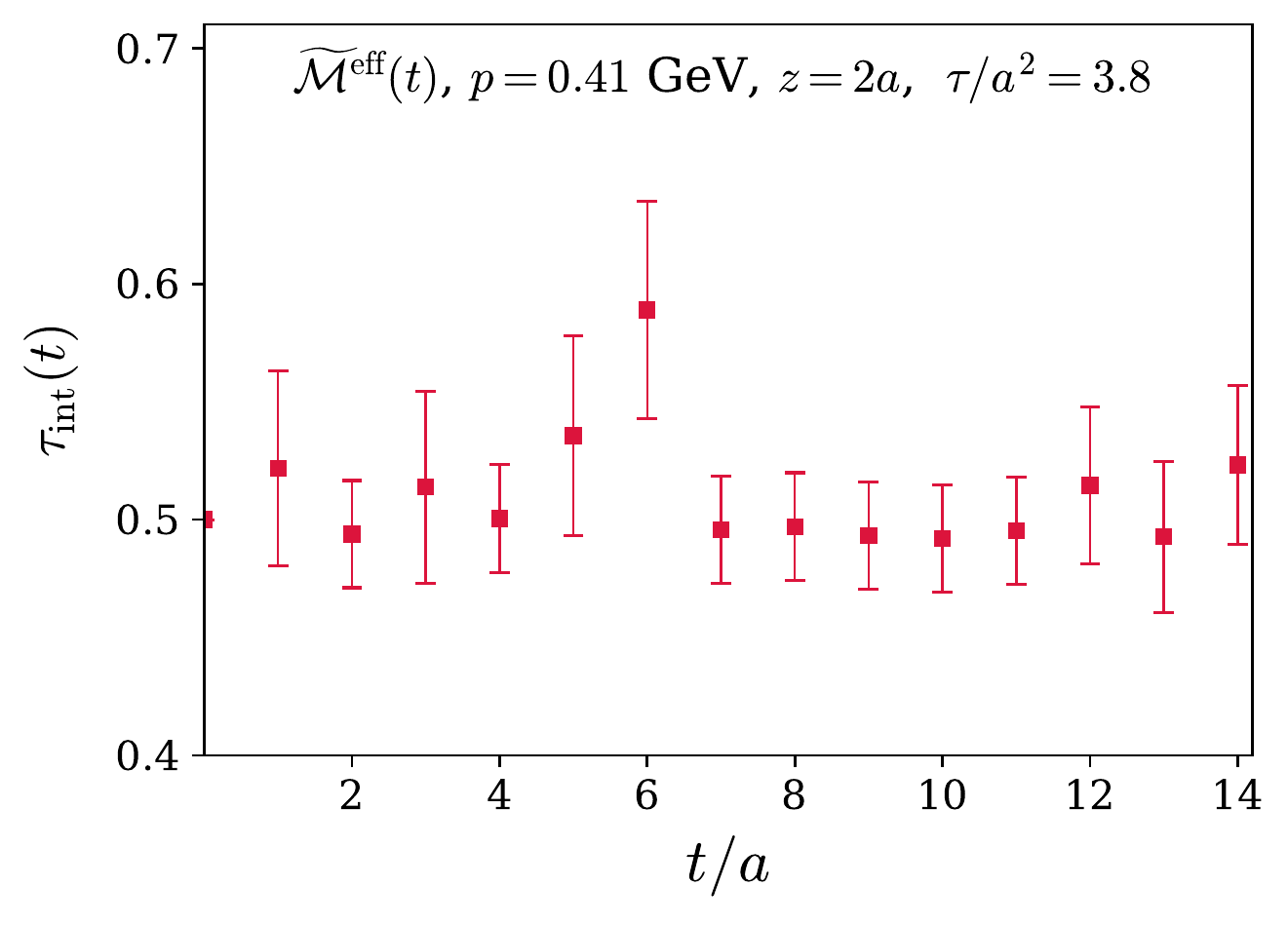}
\includegraphics[scale=0.6]{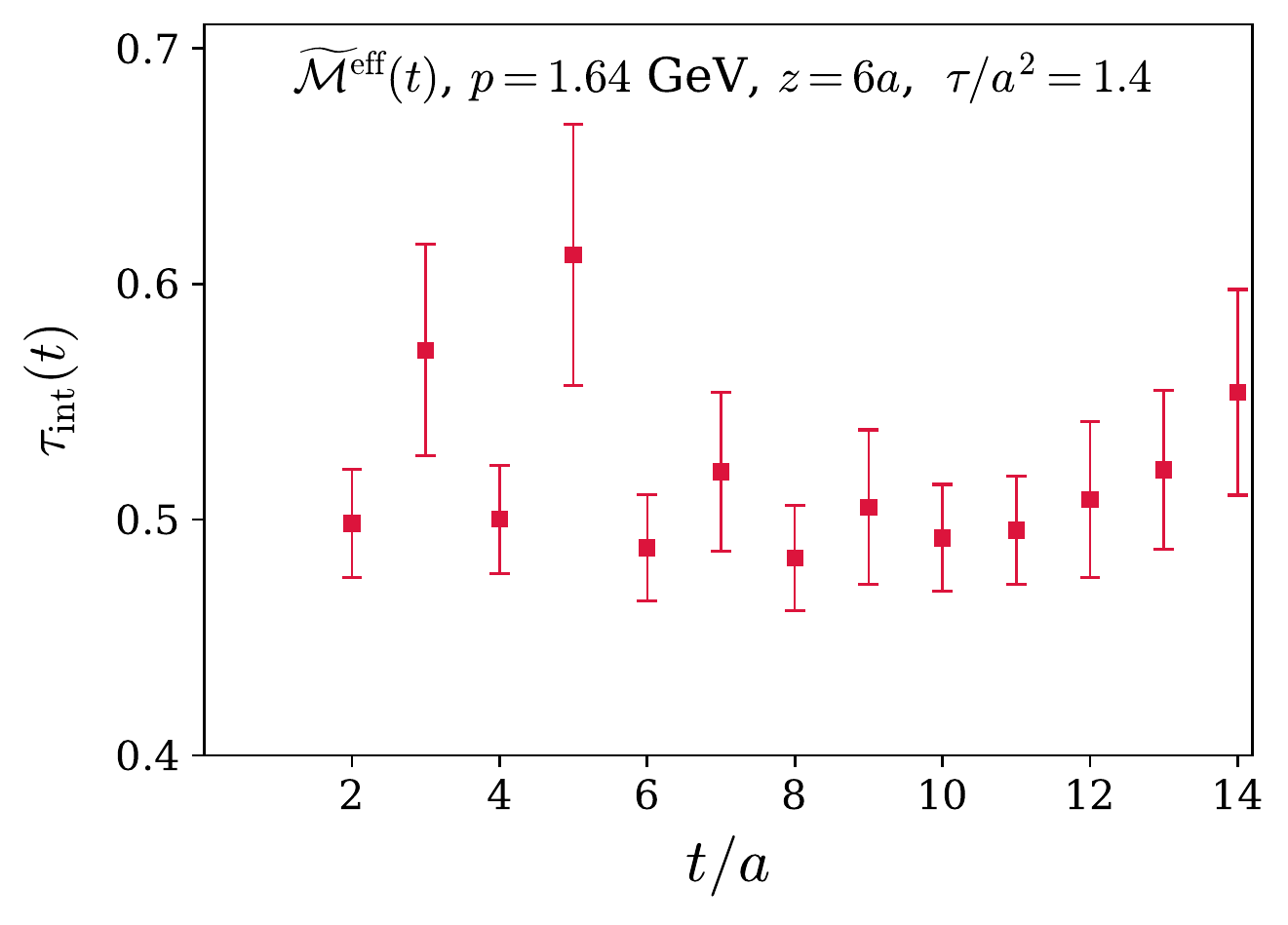}
\caption{The integrated autocorrelation time of the gluonic bare  effective matrix elements. The left panel has $\tau_{\mathrm{int}}$ for the nucleon momentum, $p = \frac{2 \pi}{a L}$ = 0.41 GeV, the field separation, $z = 2a $ = 0.188 fm, at the flow time, $\tau/a^2$ = 3.8, plotted with respect to the source-sink separation of the nucleonic two-point correlators and the right panel has $\tau_{\mathrm{int}}$ for the nucleon momentum, $p = 4 \times \frac{2 \pi}{a L}$ = 1.64 GeV, the field separation, $z = 6a $ = 0.564 fm, at the flow time, $\tau/a^2$ = 1.4. \label{fig:integrated_autocorrelation_time}}
\eefs{mockdemocn}

In Figure~\ref{fig:integrated_autocorrelation_time}, the integrated autocorrelation times of the gluonic bare matrix elements for the nucleon momentum, $p = \frac{2 \pi}{a L}$ = 0.41 GeV, the field separation, $z = 2a $ = 0.188 fm, at the flow time, $\tau/a^2$ = 3.8, and for the nucleon momentum, $p = 4 \times \frac{2 \pi}{a L}$ = 1.64 GeV, the field separation, $z = 6a $ = 0.564 fm, at the flow time, $\tau/a^2$ = 1.4 are shown. It can be seen that $\tau_{\mathrm{int}}$ stays close to the value of 0.5 for all the source-sink separations indicating that the autocorrelation among the gluonic bare matrix elements calculated on the gauge configurations separated from the next configuration by 10 hybrid Monte Carlo trajectories is negligible. Similar values of $\tau_{\mathrm{int}}$ are found for the gluonic bare matrix elements with other nucleon momenta, field separations and flow times.

\end{document}